\documentclass[journal]{IEEEtran}

\usepackage[utf8]{inputenc}
\usepackage[T1]{fontenc}
\usepackage{graphicx}
\graphicspath{{figures/}{./}}
\usepackage{amsmath,amssymb}
\usepackage{booktabs}
\usepackage{tabularx}
\usepackage{longtable}
\usepackage{array}
\usepackage{makecell}
\usepackage[table]{xcolor}
\usepackage{pgfplots}
\pgfplotsset{compat=1.16}
\usepackage{tikz}
\usetikzlibrary{shapes.geometric,arrows.meta,positioning,calc,patterns}
\usepackage{placeins}
\usepackage{multirow}
\usepackage{url}
\definecolor{linkblue}{HTML}{1A4F8B}
\usepackage[colorlinks=true,linkcolor=linkblue,citecolor=linkblue,urlcolor=linkblue,breaklinks=true]{hyperref}
\renewcommand{\arraystretch}{1.12}

\newcommand{\flown}{\emph{Flight-validated}}
\newcommand{\demonstrated}{\emph{Ground-demonstrated}}
\newcommand{\announced}{\emph{Announced}}
\newcommand{\concept}{\emph{Conceptual}}
\newcommand{\notyet}{\emph{No flight evidence}}
\newcommand{\partialtag}{\emph{Partially demonstrated}}
\newcommand{\est}{(est.)}

\definecolor{cFlown}{HTML}{F4C57E}
\definecolor{cDemo}{HTML}{F8DDB0}
\definecolor{cPartial}{HTML}{D1D5DB}
\definecolor{cConcept}{HTML}{E5E7EB}
\definecolor{cNotFlown}{HTML}{F1F5F9}
\newcommand{\cFlown}{\cellcolor{cFlown}\emph{Flight-validated}}
\newcommand{\cDemo}{\cellcolor{cDemo}\emph{Ground-demonstrated}}
\newcommand{\cPartial}{\cellcolor{cPartial}\emph{Partially demonstrated}}
\newcommand{\cConcept}{\cellcolor{cConcept}\emph{Conceptual}}
\newcommand{\cNotFlown}{\cellcolor{cNotFlown}\emph{No flight evidence}}

\begin{document}

\title{HAPS through the Lens of Satellites and UAVs:\\
A Function-Level Perspective on the Emerging High Altitude Economy}
% \title{HAPS Through the Function-Level Lens:\\
% Sensing, Navigation, and Communications in a Multi-Tier NTN}

\author{Mukhtiar~Ahmad
        and~Mohamed-Slim~Alouini,~\IEEEmembership{Fellow,~IEEE}%
\thanks{\emph{Corresponding author: Mukhtiar Ahmad.}}
\thanks{The authors are with the Computer, Electrical and Mathematical Sciences and Engineering (CEMSE) Division, King Abdullah University of Science and Technology (KAUST), Thuwal 23955-6900, Kingdom of Saudi Arabia (e-mail: mukhtiar.ahmad@kaust.edu.sa; slim.alouini@kaust.edu.sa).}%
}

\maketitle

\begin{abstract}
High-Altitude Platform Stations (HAPS) operate in the lower stratosphere at 17--27~km, between satellites and Unmanned Aerial Vehicles (UAVs). For this third tier the architectural case has long outpaced the flight evidence, but a wave of 2020--2026 stratospheric flights now permits a direct comparison. We evaluate HAPS function by function across sensing, navigation, and communication, taking operational satellite and UAV implementations as the reference. We define a strict evidence rule, counting a function as flight-validated only on operationally relevant stratospheric data return at or above 18~km, and apply it to nineteen functions. The resulting count is lower than the literature implies: five functions have credibly crossed over (optical Earth observation, hyperspectral imaging, methane imaging, RF/SIGINT, and broadband relay), yet these rest on only four flight programs, with at most one carrying peer-reviewed flight evidence. One function is ground-demonstrated, two are partially demonstrated, three are conceptual, and eight remain unflown. Four engineering domains (size, weight, and power; station-keeping; aperture; and viewing geometry), bounded by an operational envelope of platform stability and payload operability, explain the pattern. The governing advantage is persistence at close range, not altitude. Eight use cases, supported by same-sensor forward simulations, translate the pattern into missions, led by resilient public-safety mission-critical services (MCX). On this evidence, HAPS fits as a persistent regional tier in a multi-tier non-terrestrial network and as the seed of an emerging \emph{High Altitude Economy}. Carrier-grade service, station-keeping precision, and regulation remain the principal open problems, and we pose the persistent-tier reading as a testable hypothesis with dated 2030 markers.
\end{abstract}

\begin{IEEEkeywords}
HAPS, high-altitude platform, stratosphere, High Altitude Economy, Low Altitude Economy, Space Economy, remote sensing, Earth observation, navigation, satellites, UAV, persistence, SAR, hyperspectral, ISAC, NTN, FSO, ISR, DOAS, public safety, mission-critical services (MCX).
\end{IEEEkeywords}
% =============================================================
%  Body sections (split into sections/ for organization)
% =============================================================

% ----------------------------------------------------------------------------
\section{Introduction}\label{sec:perspective}
% ----------------------------------------------------------------------------

\begin{figure*}[!t]
\centering
\includegraphics[width=0.86\textwidth,keepaspectratio]{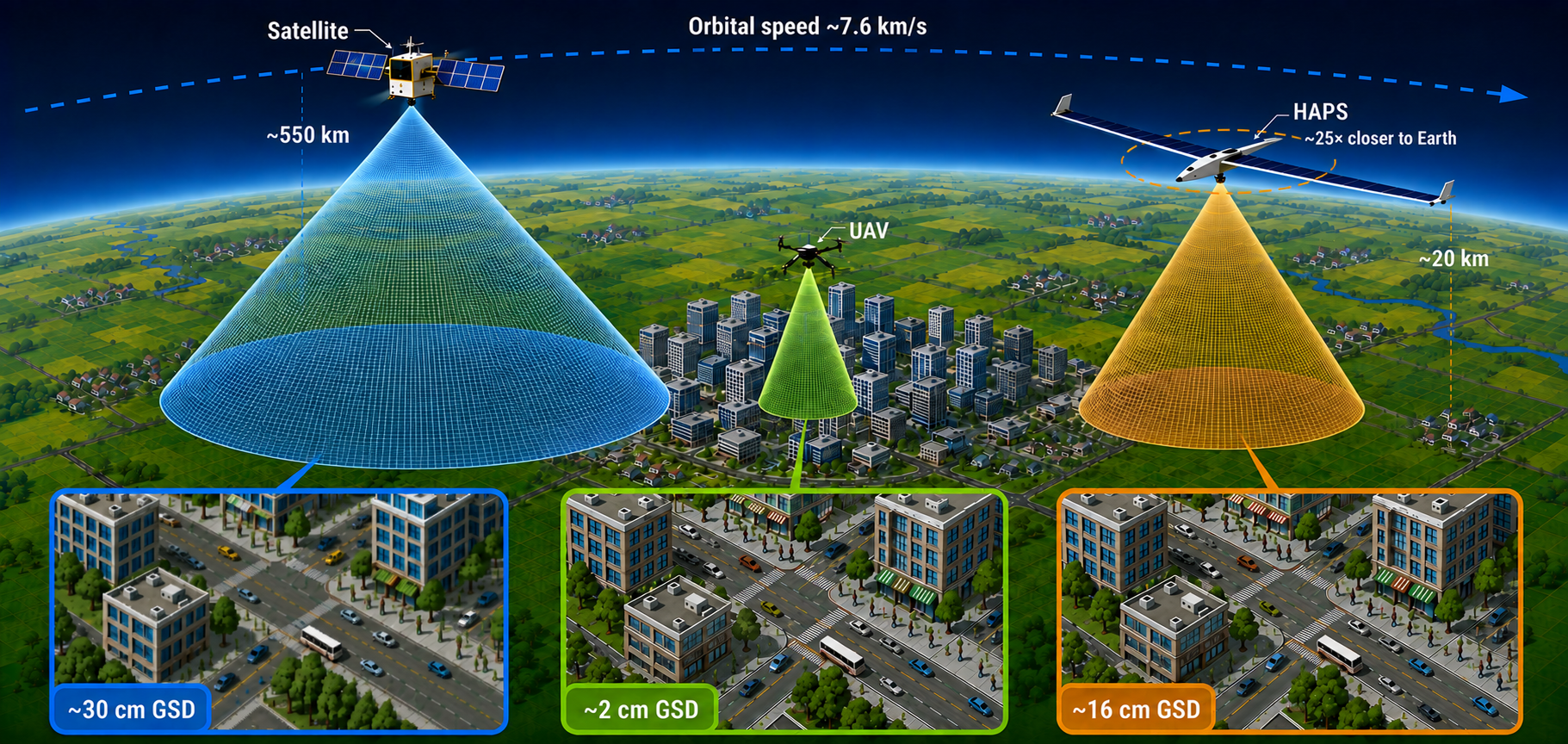}
\caption{The three-tier stack evaluated in this paper. A LEO satellite at $\sim$500~km (blue) sweeps a wide footprint in minutes; a HAPS at $\sim$20~km (orange) station-keeps over the city for weeks; a UAV (green) hovers over a single block at fine resolution. The insets show the representative ground sample of each tier. The paper asks, function by function, which regimes of coverage, persistence, and resolution HAPS can credibly cross over to.}
\label{fig:bigpicture}
\end{figure*}

\IEEEPARstart{H}{IGH}-Altitude Platform Stations (HAPS) are uncrewed aircraft, airships, or super-pressure balloons that hold station in the lower stratosphere, above weather and commercial air traffic, for weeks to months, rather than the hours typical of aircraft or the seconds-to-minutes typical of orbital overflights. Their defining property is \emph{persistence at close range}. A HAPS sits some twenty-five times closer to the ground than a low-Earth-orbit satellite (for a $\sim$500~km reference orbit), yet stays over a chosen target area continuously instead of sweeping past it in minutes (Fig.~\ref{fig:bigpicture}). The platform class has been studied for two decades. However, until 2020 the academic literature outpaced the flight evidence. A wave of 2020--2026 stratospheric demonstrations has now reversed that balance.

This reversal rests on a concrete flight record that the foundational HAPS literature of the early 2000s~\cite{vazquez-castro2002channel,karapantazis2005haps} anticipated but could not yet demonstrate: a 67-day continuous solar high-altitude long-endurance (HALE) flight, the first publicly reported stratospheric flight of a methane-tuned Short-Wave Infrared (SWIR) hyperspectral imager, Software-Defined Radio (SDR) ISR above 20~km, a 336-day balloon mission, and a ground-tested stratospheric Synthetic Aperture Radar (SAR), inventoried program by program in \S\ref{sec:background}. The pivot draws on three converging technology enablers (lithium-ion battery density, III-V solar conversion efficiency, and edge-compute power efficiency for on-board ISR) that together bring autonomous flight control and useful sensing within the solar HALE payload envelope cataloged in \S\ref{sec:why}. This record reflects an August~2026 snapshot. Several of the 2024--2026 milestones noted above are operator- or third-party-attested rather than peer-reviewed at the time of writing, and \S\ref{sec:lens} weights each source accordingly. The open question is therefore which remote-sensing, navigation, and communication \emph{functions} this third tier can actually deliver beyond what satellites and Unmanned Aerial Vehicles (UAVs) already do.\footnote{We adopt a platform-class scope of 17--27~km for this paper. The ITU Radio Regulations (Article~1.66A) place a HAPS between 20 and 50~km, and the WRC-23 HIBS framework (Resolutions~213, 218, and 221) narrows this to 18--25~km for IMT use~\cite{hapsAllianceRegPos2024}. Our band reaches slightly lower so that operational programs flying just below the ITU floor are included. For the function-by-function \emph{Flight-validated} evidence rule of \S\ref{sec:lens} we apply a stricter $\geq$18~km threshold, so recent sub-18~km flights are counted as \emph{near-flight} rather than \emph{Flight-validated} (Table~\ref{tab:jurisdictions}).}

Satellite and UAV portfolios fall into two complementary regimes. Low Earth Orbit (LEO) and Medium Earth Orbit (MEO) satellites give global coverage at the cost of revisit over any one target, since each platform sweeps past a fixed point in minutes. Geostationary Earth Orbit (GEO) satellites stare continuously at a hemisphere, but at $\sim$36{,}000~km slant range this precludes the meter-scale Ground Sample Distance (GSD) and low-latency two-way link that downward-viewing missions and broadband service typically require. In contrast, UAVs give sub-centimeter GSD and real-time latency, at the cost of small footprint and short endurance. The dimension neither class addresses well is \emph{a persistent presence over a target area}: continuous methane attribution to a specific facility, sub-minute fire-line tracking, persistent radio-frequency (RF) signal intelligence (SIGINT), or Direct-to-Unmodified-Smartphone (D2US) and mission-critical public-safety services (MCX) in a disaster zone. These are the cases for which HAPS (solar HALE aircraft, super-pressure balloons, and stratospheric airships) appear well suited.

Relative to a LEO satellite at approximately 500~km at the same carrier frequency and aperture, a 20~km HAPS gives roughly $28$~dB lower one-way RF free-space path loss, a $25\times$ shorter slant range, and (for diffraction-limited optics at the same aperture) a $25\times$ finer GSD. The corresponding $1/R^{2}$ irradiance ratio is $625\times$ at the entrance pupil. This is a geometric upper bound, not a usable Signal-to-Noise Ratio (SNR) gain: in the shot-noise-limited regime in which several flagship LEO sensors operate, it collapses toward $\sqrt{625}=25\times$. We derive this read-noise versus shot-noise distinction once in \S\ref{sec:tutorial} and reference it thereafter. Revisit over a fixed area improves by 1--3 orders of magnitude through continuous station-kept observation. Against UAVs, HAPS trades sub-cm GSD and true real-time tasking for an instantaneous footprint two to three orders of magnitude larger and an endurance measured in months rather than hours.

HAPS fit the cases where the \emph{persistence gain} (sustained time on station at a useful range) outweighs the \emph{stratospheric penalties} (SWaP envelope, station-keeping precision, aperture, and platform stability). When persistence does not dominate the mission profile, for example global wide-area screening, gravimetry, and ocean altimetry, HAPS adds little over the established satellite tier. We therefore treat HAPS not as a satellite replacement but as one layer in a multi-tier Non-Terrestrial Network (NTN) composed with satellites and UAVs, an integrated satellite-airborne-terrestrial architecture explored explicitly for global-connectivity gains~\cite{alsharoa2020satAirTerr}.

Platform-centric comparisons (Zephyr vs.\ PHASA-35 vs.\ Sceye) tend to track marketing maturity rather than capability migration. In contrast, functions that are already operational on satellites or UAVs (optical Earth Observation (EO), SAR, AIS, IoT relay, etc.) provide a fixed reference set that any HAPS implementation must match or beat. We therefore catalog nineteen sensing, navigation, and communication functions, take the satellite and UAV implementations as the operational reference, and ask of each: \emph{has it credibly transferred to a HAPS, and if not, what limits it?} The status of each function is tabulated in \S\ref{sec:lens} and explained by four engineering domains (SWaP, station-keeping, aperture, and above-cloud viewing geometry) bounded by a two-fold operational envelope of platform stability and payload operability (\S\ref{sec:why}). The analysis favors breadth across the three pillars over exhaustive per-function depth; for mission-level detail we point to the operational catalogs of \S\ref{sec:functions}.

The central claim of this paper is that the rate at which functions have actually transferred to HAPS is substantially lower than the surrounding literature implies, and the engineering envelope tighter than assumed. First, five of nineteen functions have credibly transferred, and they rest on only four flight programs. Second, the above-cloud advantage survives scrutiny only for upward- and limb-viewing geometries. Third, the binding constraints are SWaP, station-keeping, and aperture rather than altitude. The multi-tier view itself is established ground, held by the HAPS Alliance and the 6G NTN literature; our contribution is the evidence basis beneath it, with an explicit account of which eight functions remain open and what limits each. We also identify a HAPS-native frontier beyond the satellite-referenced catalog, in-situ and acoustic sensing, that has no orbital analogue (\S\ref{sec:functions}).

Our contributions are fourfold. (i) We apply a function-by-function evidence rule to HAPS against satellite and UAV reference systems, with an explicit definition of \emph{Flight-validated} as operationally relevant stratospheric data return at $\geq$~18~km, not merely payload carriage. (ii) We propose a four-domain engineering taxonomy (SWaP, station-keeping, aperture, above-cloud viewing) plus a two-fold operational envelope (platform stability, payload operability) that together explain which functions transfer and which do not, and we map this taxonomy onto the HAPS Alliance nine-level Technology Readiness Level scale in \S\ref{sec:forward}. (iii) We place HAPS in the multi-tier 3GPP NTN stack as both a relay node and an emerging edge-compute and sensing node. (iv) We map the value chain, market scale, engineering-to-economic differentiations, and gating constraints of the stratospheric tier (\S\ref{sec:hae}) as an emerging \emph{High Altitude Economy} (HAE) between the Low Altitude Economy~\cite{stateCouncilLAE2024} and the established Space Economy~\cite{wefMckinseySpace2024}. To make the comparison concrete, we accompany seven of the use cases with scaling-law visualizations generated by controlled same-sensor forward simulation (\S\ref{sec:gains}; models and parameters in the \hyperref[sec:appendix-sim]{Appendix}), placing an identical instrument on a 500~km satellite and a 20~km HAPS so that the satellite-to-HAPS difference is a geometry-and-physics best case rather than a measured result.
A complementary resilience argument (developed in \S\ref{sec:layer}, \emph{Infrastructure Resilience}) supports the multi-tier framing: a stratospheric tier is uncorrelated with the dominant LEO failure modes.

Relative to prior surveys, this perspective differs in three concrete ways. The classical channel-modeling and broadband-communication treatments of V\'azquez-Castro et al.~\cite{vazquez-castro2002channel} and Karapantazis and Pavlidou~\cite{karapantazis2005haps}, and the comprehensive 6G NTN survey of Kurt et al.~\cite{kurt2021haps}, all treat HAPS primarily as a communication node and catalog the platform class in general architectural terms. The two older treatments predate the 2020--2026 flight wave, so we also engage the most recent literature directly. The 2024--2026 6G HAPS roadmap of Abbasi et al.~\cite{abbasi2024haps6G} and the HAPS Alliance reference-architecture and 6G white papers~\cite{hapsAllianceRefArch2024,hapsAlliance6G2026} refresh the standards and architecture picture, yet they remain communication- and architecture-centric. The most recent 2025 surveys preserve this orientation: Svistunov et al.~\cite{svistunov2025hapsSurvey} consolidate the 6G-NTN role of HAPS, and Elkhazraji et al.~\cite{elkhazraji2025hapsOptical} couple gigabit free-space-optical backhaul with a single atmospheric-sensing modality (DIAL) over shared optical links. None of them applies a per-function evidence rule across sensing, navigation, and communication, and none reports a validated-versus-open cross-over count, which is the gap this perspective fills. \emph{First}, we apply a strict, falsifiable per-function evidence rule across nineteen sensing, navigation, and communication functions, in which each function's status is overturned by a single counterexample of stratospheric data return, producing the headline count (5/1/2/3/8) that prior surveys do not, and we read it conservatively: the five flown functions rest on four flight programs, amounting to three to four genuinely independent demonstrations, and at most one of the five is supported by peer-reviewed flight evidence (\S\ref{sec:lens}). \emph{Second}, we give the three pillars equal weight in the analysis, treating remote sensing as a first-class pillar rather than as a downstream payload of the communication link. The use cases of \S\ref{sec:gains} include methane plume attribution, sub-decimeter optical EO, photon-budget LiDAR, and thermal-IR fire detection on equal footing with broadband relay and SIGINT. \emph{Third}, we frame NTN integration as the architectural question and engage with the HAPS Alliance's own reference architecture, regulatory positions, technology-readiness scale, and operational-risk analysis~\cite{hapsAllianceRefArch2024,hapsAllianceRegPos2024,haps-defense-trl-2025,hapsAllianceRisk2024} so that the perspective speaks to both the academic literature and the industrial stratospheric standardization effort.

\begin{figure}[!t]
\centering
\includegraphics[width=0.7\linewidth]{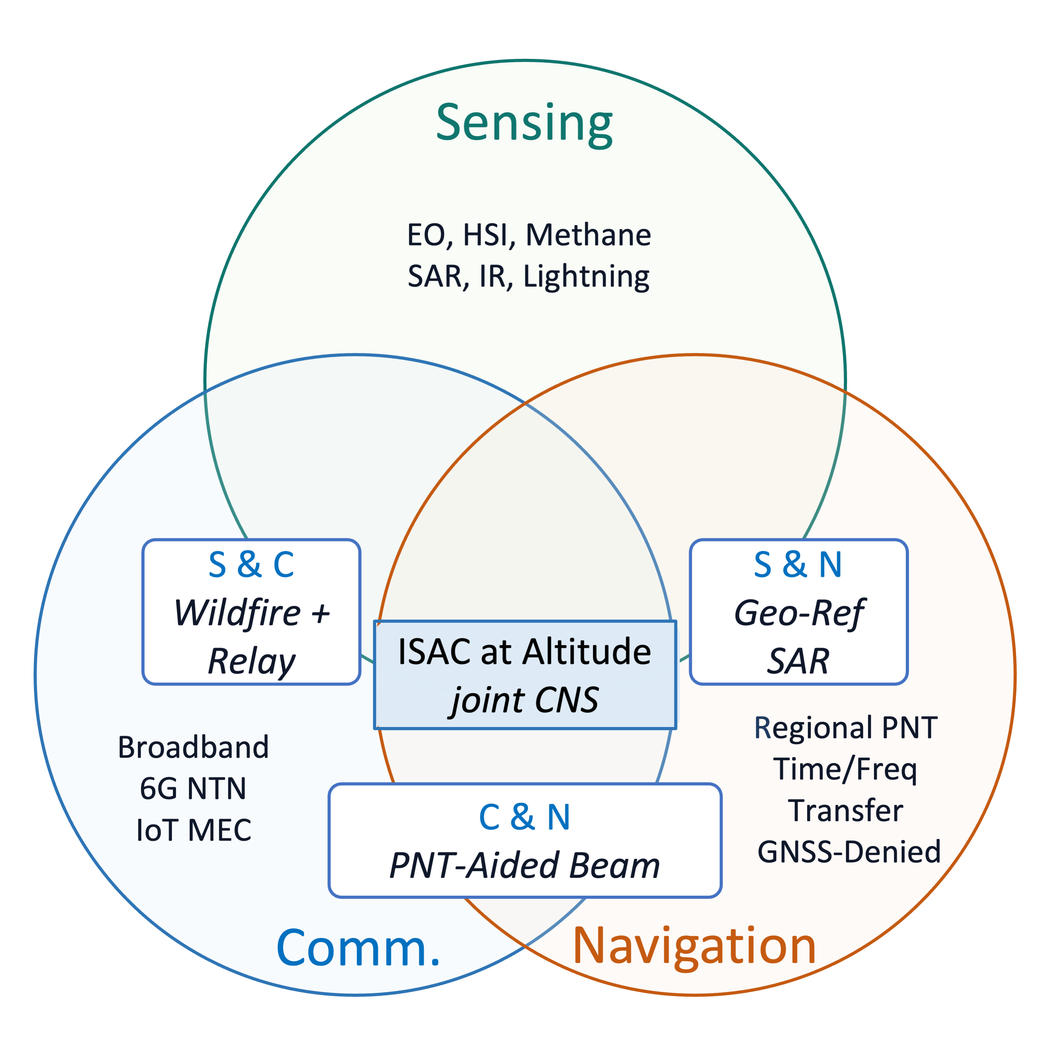}
\caption{Three-pillar convergence on a HAPS: pillar lobes (sensing, navigation, communication), pairwise joint services (S\,\&\,C, S\,\&\,N, C\,\&\,N), and the central integrated stratospheric services layer.}
\label{fig:trefoil}
\end{figure}

Fig.~\ref{fig:trefoil} organizes the rest of the paper. Each lobe of the trefoil lists one pillar's primary functions. The pairwise overlaps host joint services that depend on two pillars co-resident on the same airframe: wildfire detection plus first-responder relay (S\,\&\,C), geo-referenced SAR (S\,\&\,N), and PNT-aided beamforming (C\,\&\,N). The center is the integrated stratospheric services layer where persistence, a single timing source, a single inertial frame, and on-board edge compute jointly enable Integrated Sensing, Communication, and Navigation (also called ISAC at altitude). The remainder of the paper follows this structure: Sections \ref{sec:functions}--\ref{sec:lens} catalog the per-pillar functions, \S\ref{sec:why} and \S\ref{sec:gains} explain the transfer pattern, and \S\ref{sec:layer} returns to the center as the multi-tier architectural result. Section~\ref{sec:conclusion} closes with time-bounded predictions on carrier-grade KPIs and fleet scale whose outcome over the coming years will support or weaken the persistent-tier reading.

% ----------------------------------------------------------------------------
\section{Three Platform Classes}\label{sec:background}
% ----------------------------------------------------------------------------

\begin{figure*}[!t]
\centering
{\footnotesize\textbf{Heavier-than-air} (aerodynamic / rotor lift)}\\[4pt]
\begin{minipage}[t]{0.32\textwidth}\centering
  \includegraphics[width=\linewidth,height=2.35cm,keepaspectratio]{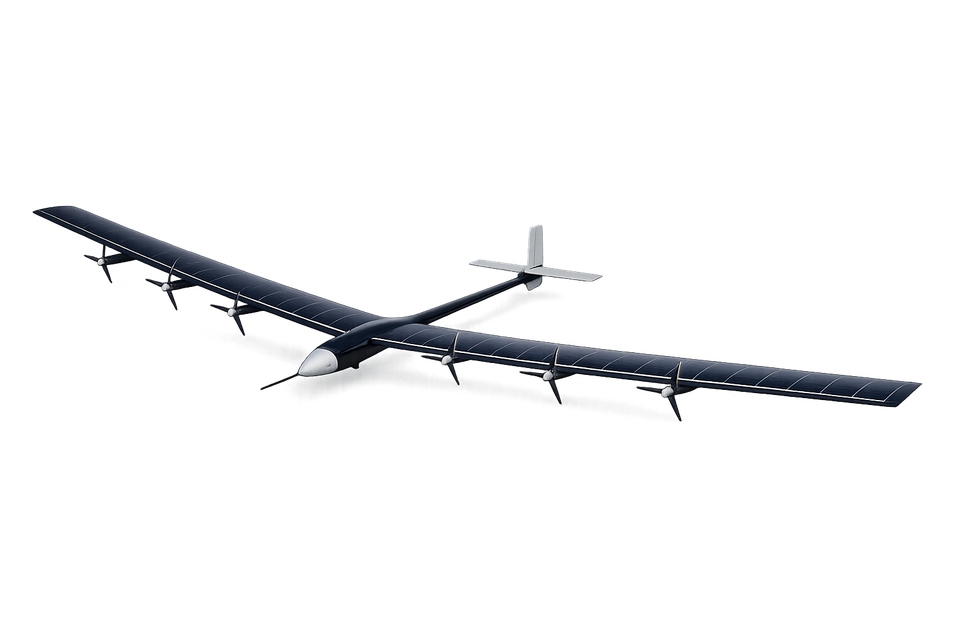}\\[2pt]
  {\footnotesize (a) Solar HALE aircraft}
\end{minipage}\hfill
\begin{minipage}[t]{0.32\textwidth}\centering
  \includegraphics[width=\linewidth,height=2.35cm,keepaspectratio]{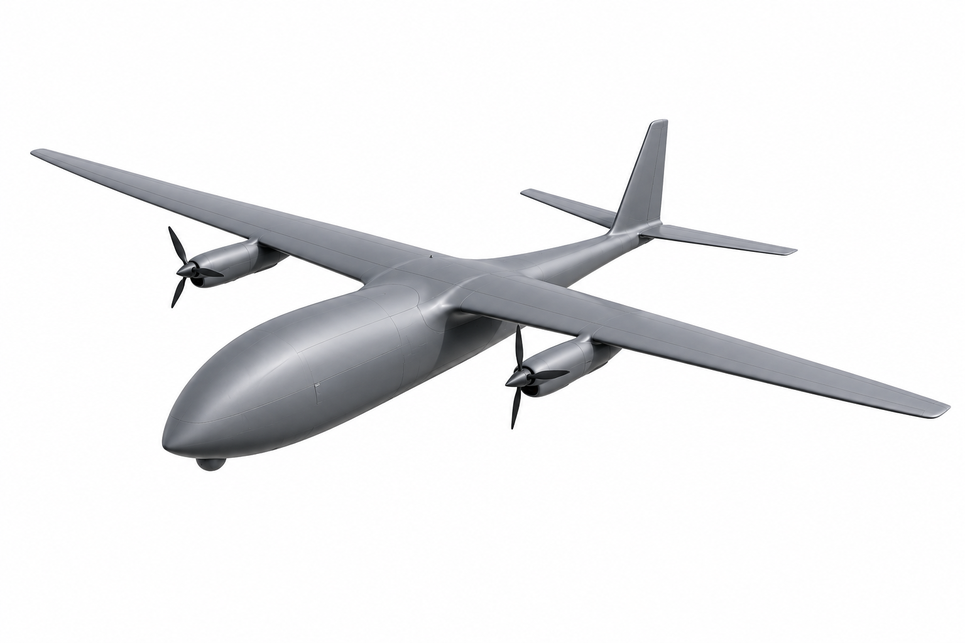}\\[2pt]
  {\footnotesize (b) Hydrogen-electric fixed-wing}
\end{minipage}\hfill
\begin{minipage}[t]{0.32\textwidth}\centering
  \includegraphics[width=\linewidth,height=2.35cm,keepaspectratio]{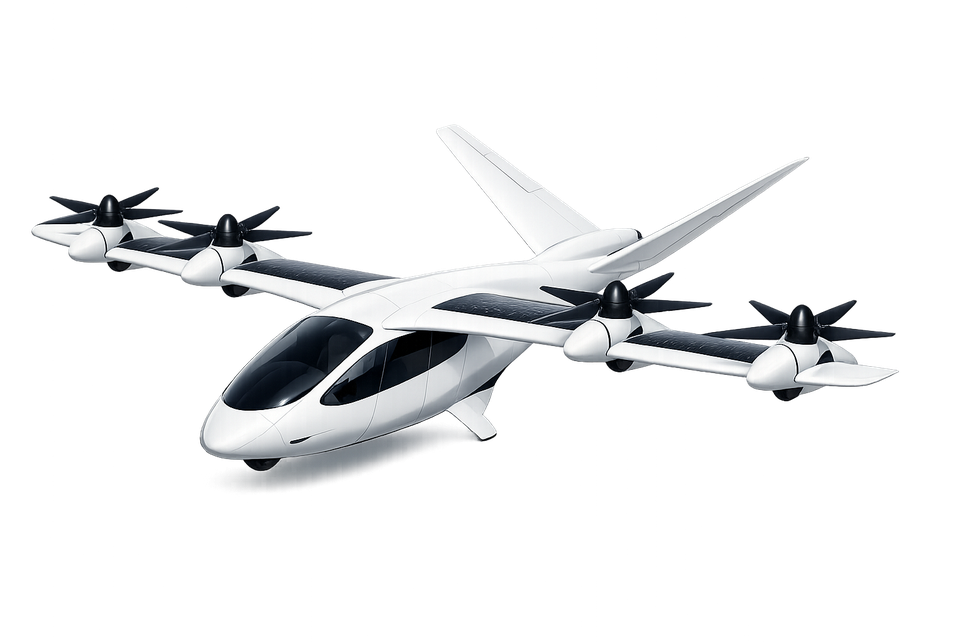}\\[2pt]
  {\footnotesize (c) eVTOL (emerging)}
\end{minipage}

\vspace{9pt}
{\footnotesize\textbf{Lighter-than-air} (buoyant)}\\[4pt]
\begin{minipage}[t]{0.32\textwidth}\centering
  \includegraphics[width=\linewidth,height=2.35cm,keepaspectratio]{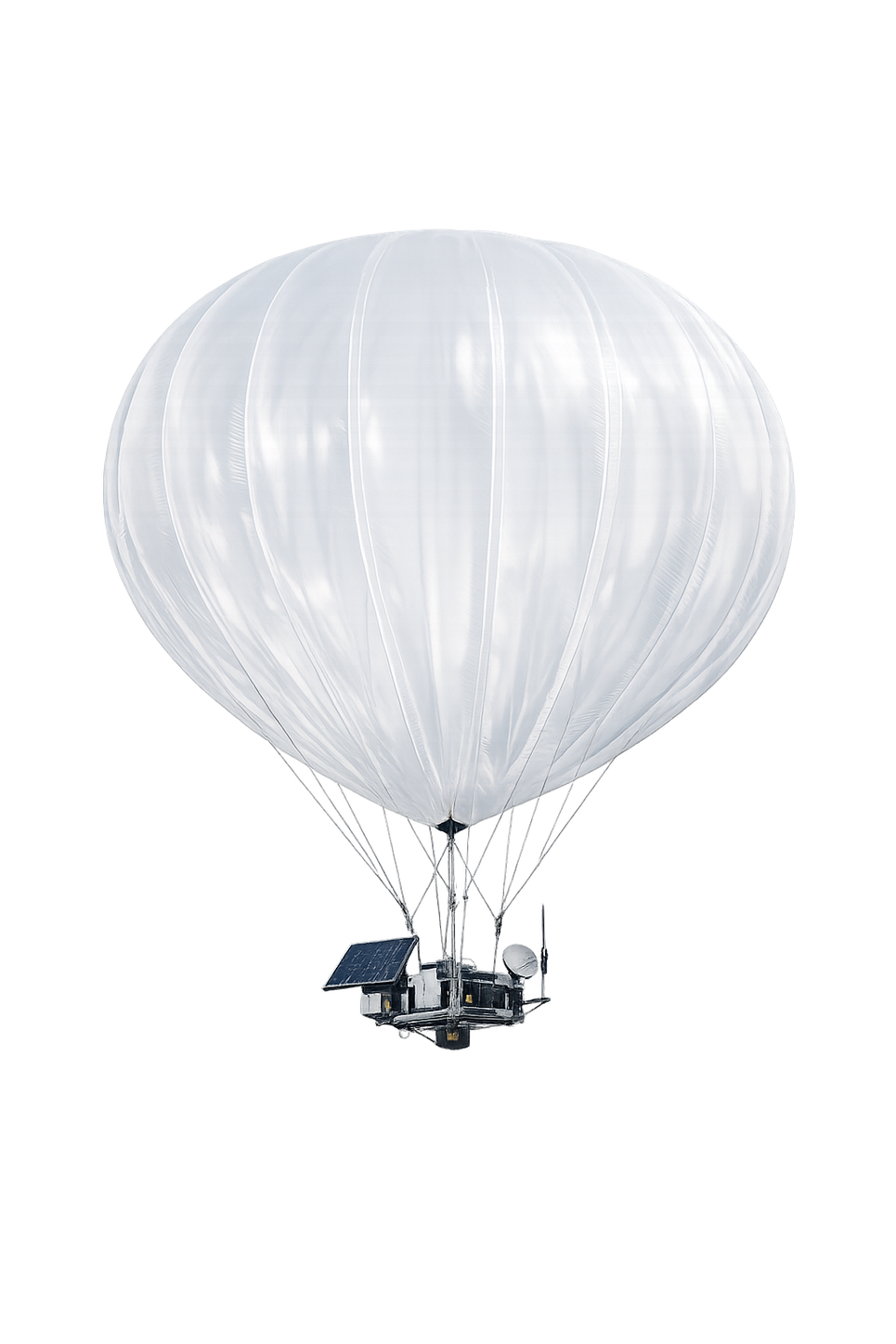}\\[2pt]
  {\footnotesize (d) Super-pressure balloon}
\end{minipage}\hspace{0.04\textwidth}
\begin{minipage}[t]{0.32\textwidth}\centering
  \includegraphics[width=\linewidth,height=2.35cm,keepaspectratio]{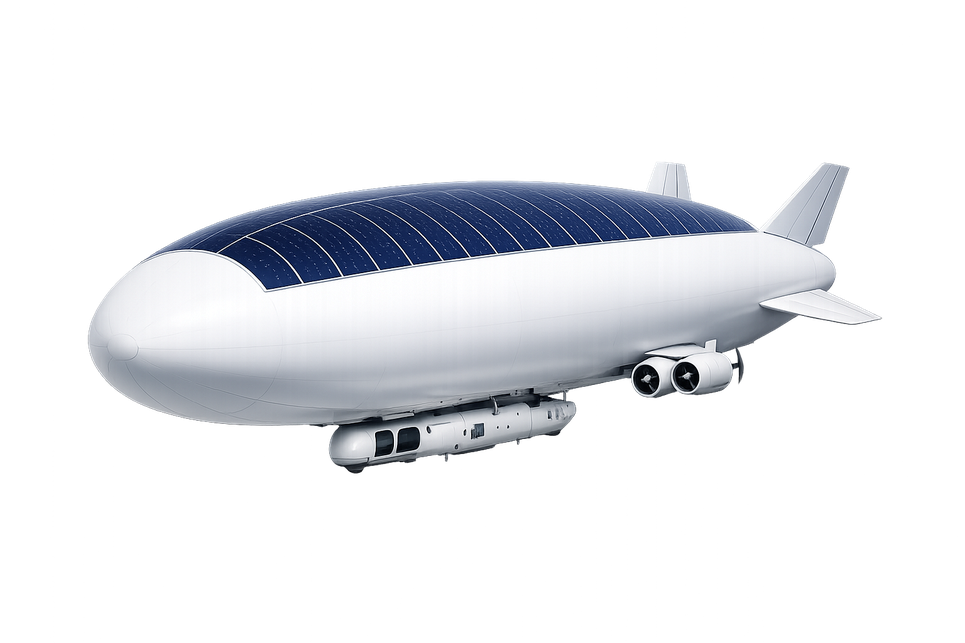}\\[2pt]
  {\footnotesize (e) Stratospheric airship}
\end{minipage}

\caption{The HAPS platform design space, grouped by flight principle. \emph{Heavier-than-air} (aerodynamically or rotor-lifted): (a)~a solar high-altitude long-endurance (HALE) aircraft (e.g., Zephyr, PHASA-35), (b)~a hydrogen-electric fixed-wing aircraft (e.g., Global Observer, Stratospheric Platforms), and (c)~a hydrogen-electric eVTOL, an emerging runway-independent variant. \emph{Lighter-than-air} (buoyant): (d)~a super-pressure balloon (e.g., Stratollite) and (e)~a solar stratospheric airship (e.g., Sceye, Stratobus). The three platform classes compared in Table~\ref{tab:bg} (solar aircraft, balloon, airship) span both categories, whereas the two emerging variants developed in \S\ref{sec:why} (hydrogen fixed-wing and eVTOL) are both heavier-than-air. Illustrative renderings, not to scale.}
\label{fig:platforms}
\end{figure*}

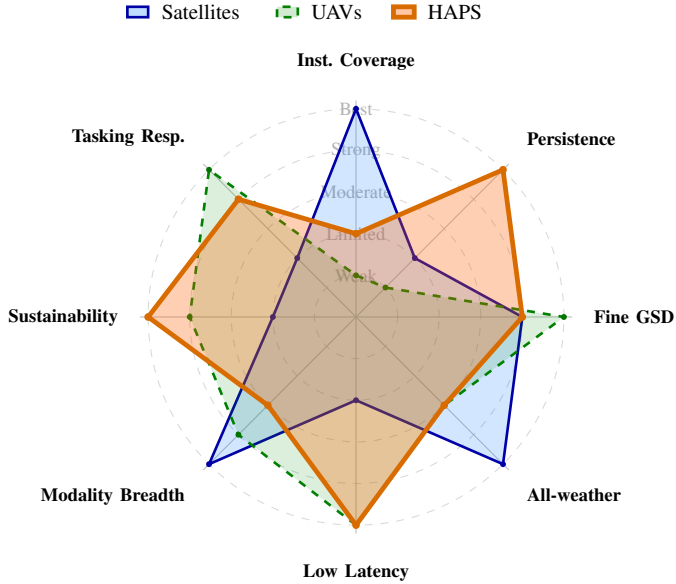
\begin{figure}[!t]
\centering
\begin{tikzpicture}[scale=0.55,every node/.style={font=\scriptsize}]

% Concentric reference rings (1=Weak ... 5=Best)
\foreach \r in {1,2,3,4,5} {
  \draw[gray!30,dashed,thin] (0,0) circle (\r);
}
% Inner-ring level labels along the top axis
\foreach \r/\lab in {1/Weak,2/Limited,3/Moderate,4/Strong,5/Best} {
  \node[gray!60,font=\scriptsize,inner sep=1pt] at (90:\r) {\lab};
}

% Eight axes at 45-degree spacing
\foreach \angle in {90,45,0,-45,-90,-135,180,135} {
  \draw[gray!50,thin] (0,0) -- (\angle:5.2);
}

% Outer dimension labels
\node[anchor=south]      at (90:5.7)   {\textbf{Inst.\ Coverage}};
\node[anchor=south west] at (45:5.5)   {\textbf{Persistence}};
\node[anchor=west]       at (0:5.5)    {\textbf{Fine GSD}};
\node[anchor=north west] at (-45:5.5)  {\textbf{All-weather}};
\node[anchor=north]      at (-90:5.7)  {\textbf{Low Latency}};
\node[anchor=north east] at (-135:5.5) {\textbf{Modality Breadth}};
\node[anchor=east]       at (180:5.5)  {\textbf{Sustainability}};
\node[anchor=south east] at (135:5.5)  {\textbf{Tasking Resp.}};

% Satellites polygon (BLUE, solid thin line): Coverage=5, Persistence=2, GSD=4, All-weather=5, Latency=2, Breadth=5, Sust=2, TaskResp=2
\fill[blue!55!cyan,fill opacity=0.18] (90:5) -- (45:2) -- (0:4) -- (-45:5) -- (-90:2) -- (-135:5) -- (180:2) -- (135:2) -- cycle;
\draw[draw=blue!65!black,line width=1.0pt]
  (90:5) -- (45:2) -- (0:4) -- (-45:5) -- (-90:2) -- (-135:5) -- (180:2) -- (135:2) -- cycle;
\foreach \a/\v in {90/5,45/2,0/4,-45/5,-90/2,-135/5,180/2,135/2} {\fill[blue!65!black] (\a:\v) circle (2pt);}

% UAVs polygon (GREEN, DASHED line): Coverage=1, Persistence=1, GSD=5, All-weather=3, Latency=5, Breadth=4, Sust=4, TaskResp=5
\fill[green!55!black,fill opacity=0.14] (90:1) -- (45:1) -- (0:5) -- (-45:3) -- (-90:5) -- (-135:4) -- (180:4) -- (135:5) -- cycle;
\draw[draw=green!50!black,line width=1.0pt,dashed]
  (90:1) -- (45:1) -- (0:5) -- (-45:3) -- (-90:5) -- (-135:4) -- (180:4) -- (135:5) -- cycle;
\foreach \a/\v in {90/1,45/1,0/5,-45/3,-90/5,-135/4,180/4,135/5} {\fill[green!45!black] (\a:\v) circle (2pt);}

% HAPS polygon (ORANGE, THICK solid line --- the focus): Coverage=2, Persistence=5, GSD=4, All-weather=3, Latency=5, Breadth=3, Sust=5, TaskResp=4
\fill[orange!75!red,fill opacity=0.30] (90:2) -- (45:5) -- (0:4) -- (-45:3) -- (-90:5) -- (-135:3) -- (180:5) -- (135:4) -- cycle;
\draw[draw=orange!85!black,line width=1.8pt]
  (90:2) -- (45:5) -- (0:4) -- (-45:3) -- (-90:5) -- (-135:3) -- (180:5) -- (135:4) -- cycle;
\foreach \a/\v in {90/2,45/5,0/4,-45/3,-90/5,-135/3,180/5,135/4} {\fill[orange!85!black] (\a:\v) circle (2.6pt);}

% Manual legend at top
\begin{scope}[shift={(-5.5,7.2)},every node/.style={font=\footnotesize,anchor=west}]
  \draw[draw=blue!65!black,line width=1.0pt,fill=blue!55!cyan,fill opacity=0.25] (0,0) rectangle ++(0.5,0.25);
  \node at (0.6,0.125) {Satellites};
  \draw[draw=green!50!black,line width=1.0pt,dashed,fill=green!55!black,fill opacity=0.18] (3.6,0) rectangle ++(0.5,0.25);
  \node at (4.2,0.125) {UAVs};
  \draw[draw=orange!85!black,line width=1.8pt,fill=orange!75!red,fill opacity=0.35] (6.4,0) rectangle ++(0.5,0.25);
  \node at (7.0,0.125) {HAPS};
\end{scope}

\end{tikzpicture}
\caption{Capability fingerprints of satellites (blue, solid), UAVs (green, dashed), and HAPS (orange, bold) across eight mission-design dimensions. Axes are ordinal positions, not normalized scores; the figure is a structural mnemonic and should not be read as a quantitative ranking.}
\label{fig:radar}
\end{figure}

Table~\ref{tab:bg} gives a compact platform-class comparison on the parameters most relevant to mission design, and Fig.~\ref{fig:radar} renders the same picture as capability fingerprints across eight mission-design dimensions. Fig.~\ref{fig:platforms} shows the physical morphology of the platform types, grouped into the two classical flight-principle categories of heavier-than-air (aerodynamically or rotor-lifted) and lighter-than-air (buoyant) vehicles. Each axis runs from \emph{Weak} (center) to \emph{Best} (perimeter). The polygon is oriented so a longer footprint is more desirable on every axis (longer ``Fine GSD'' = finer pixel size, longer ``Low Latency'' = shorter delay). Each axis is anchored by the quantitative entries of Table~\ref{tab:bg}. The persistence axis reflects the days-to-months on-station endurance of a HAPS against the minutes-per-pass of a LEO satellite, and the Fine-GSD and Low-Latency axes reflect the resolution and delay values tabulated there. The axes are ordinal positions rather than normalized scores, so the figure is a structural mnemonic for those measured differences, not a numerical ranking. Satellites win on instantaneous coverage and modality breadth, UAVs win on Ground Sample Distance (GSD) and latency, and HAPS (where flight-validated) wins decisively on persistence and matches UAVs on latency. Acronyms and notation that recur in this paper are listed in Table~\ref{tab:acronyms}. The function-level analysis that this paper argues for begins in \S\ref{sec:functions}, where the qualitative picture in Fig.~\ref{fig:radar} is quantified function-by-function.

The 2020--2026 flight record anchors the comparison. AALTO Zephyr completed a 67-day continuous flight in 2025, the longest continuous aircraft flight on record, though the airframe was lost at the end of the mission. Sceye flew a methane-tuned SWIR hyperspectral imager in the stratosphere in August~2024, the first publicly reported flight of its kind. BAE PHASA-35, which had completed its first stratospheric flight from Spaceport America in June~2023 (announced July~2023)~\cite{baeSpaceportJul2023}, carried an SDR ISR sensor above 20~km in stratospheric trials announced in December~2024. Aerostar Thunderhead remained controllable for 336~days (HBAL684~\cite{aerostarHBAL684}, a navigated long-duration traverse rather than a fixed-target station-keep). DLR has ground-tested HAPSAR, the first dedicated stratospheric S-band SAR, alongside MACS-HAP, a 15~cm-GSD optical camera designed for the 20~km altitude band with a 90~mm aperture and an environmental-test campaign across the $-30^{\circ}$C to $+50^{\circ}$C stratospheric envelope~\cite{dlr-macshap}.

\begin{table}[!htbp]
\centering
\caption{Acronyms and notation used in this paper.}
\label{tab:acronyms}
\footnotesize
\rowcolors{2}{gray!8}{white}
\renewcommand{\arraystretch}{1.15}
\setlength{\tabcolsep}{4pt}
\begin{tabular}{@{}>{\raggedright\arraybackslash}p{1.4cm} >{\raggedright\arraybackslash}p{6.5cm}@{}}
\toprule
\textbf{Acronym} & \textbf{Definition} \\
\midrule
ADS-B & Automatic Dependent Surveillance-Broadcast \\
AIS & Automatic Identification System (maritime) \\
ALARP & As Low As Reasonably Practicable \\
APNT & Alternative Position, Navigation, and Timing \\
ATM & Air Traffic Management \\
BLER & Block Error Rate \\
$C_{n}^{2}$ & Refractive-index structure constant (turbulence) \\
Carrier-grade KPI & Throughput / BLER / mobility / MTBF metrics required for commercial-grade service (see \S\ref{sec:cgkpi}) \\
D2US & Direct-to-Unmodified-Smartphone \\
DIAL & Differential Absorption LiDAR \\
DOAS & Differential Optical Absorption Spectroscopy \\
ELINT & Electronic Intelligence \\
EO & Earth Observation \\
FSO & Free-Space Optical (communication) \\
FSPL & Free-Space Path Loss \\
Fleet multiplier & Number of platforms needed to maintain coverage through scheduled and unscheduled outages \\
GSD & Ground Sample Distance \\
HALE & High-Altitude Long-Endurance (aircraft) \\
HAPS & High-Altitude Platform Station \\
HIBS & High-altitude IMT Base Station \\
HSI & Hyperspectral Imaging \\
ISAC & Integrated Sensing and Communication \\
ISR & Intelligence, Surveillance, Reconnaissance \\
LEO/\allowbreak MEO/\allowbreak GEO & Low/Medium/Geostationary Earth Orbit \\
MCX & Mission-critical services (3GPP MCPTT / MCData / MCVideo profiles) \\
MEC & Multi-access Edge Computing \\
MTBPL & Mean Time Between Platform Losses \\
NE$\sigma^{0}$ & Noise-equivalent radar backscatter coefficient \\
NTN & Non-Terrestrial Network \\
PNT & Position, Navigation, and Timing \\
Regenerative payload & Onboard gNB-DU terminates the air interface (decode-and-forward) \\
SAR & Synthetic Aperture Radar \\
SBAS & Satellite-Based Augmentation System \\
SIGINT & Signal Intelligence \\
SWaP & Size, Weight, and Power \\
TRL & Technology Readiness Level \\
Transparent payload & Bent-pipe relay: amplify and forward without decoding \\
UAV & Unmanned Aerial Vehicle \\
\bottomrule
\end{tabular}
\end{table}

\begin{table*}[!tbp]
\caption{Quick comparison of the three platform classes across the parameters most relevant to mission design.}
\label{tab:bg}
\centering
\footnotesize
\setlength{\tabcolsep}{4pt}
\renewcommand{\arraystretch}{1.06}
\rowcolors{2}{gray!8}{white}
\begin{tabular}{@{}>{\raggedright\arraybackslash}p{3.0cm} >{\raggedright\arraybackslash}p{4.5cm} >{\raggedright\arraybackslash}p{4.3cm} >{\raggedright\arraybackslash}p{5.6cm}@{}}
\toprule
\textbf{Parameter} & \textbf{Satellites} & \textbf{UAVs} & \textbf{HAPS} \\
\midrule
Altitude & 400--36{,}000~km (LEO/MEO/GEO/HEO) & 0.03--18~km AGL, mostly $<$8~km & 17--27~km (lower stratosphere) \\
Platform classes & Cubesat, smallsat, large EO sat, GEO, MEO/GNSS & Multirotor, fixed-wing, VTOL hybrid, MALE, HALE & Solar- or hydrogen-electric HALE aircraft, super-pressure balloon, stratospheric airship \\
Endurance & 5--20+ years on-orbit & 0.5--34~h per flight & Days to months per flight \\
Best operational GSD & 16~cm (Umbra SAR spotlight), 31~cm (WV-3 EO pan) & sub-cm at 120~m AGL & 18~cm EO~\est{} (Zephyr OPAZ); 15~cm EO ground-test (DLR MACS-HAP~\cite{dlr-macshap}); $\sim$3--5~m SWIR HSI flown (Sceye + HySpex SWIR-640, Aug~2024)~\cite{hyspexSceye2024} \\
Revisit over target & 5~min--16~d; sub-hourly with LEO constellation & Mission-bounded ($\sim$1~h to days) & \textbf{Persistent station-keeping (days--months) over fixed area} \\
Sensor footprint and swept area & 10--100~km$^2$ (EO scene); 25{,}000--60{,}000~km$^2$ (LEO SAR swath); $\sim$$1.5\times10^{8}$~km$^2$ (GEO full disk) & $<$1~km$^2$ (multirotor); few hundred~km$^2$ (MALE EO/IR); 10{,}000--50{,}000~km$^2$ (HALE SAR swath) & $\sim$419~km$^2$ swept area at $\pm 30^{\circ}$ gimbal swing from 20~km, $\pi(20\tan30^{\circ})^{2}$~\cite{dlr-macshap}; $\sim$7{,}850--31{,}400~km$^2$ wide-FOV service area (50--100~km radius, up to $\sim$200$\times$ a terrestrial cell; \emph{industry-target}: no station-kept HAPS has yet sustained service over this area)~\cite{hapsAlliance6G2026,hapsAllianceRegPos2024}; LOS to $\sim$$8\times10^5$~km$^2$ horizon \\
End-to-end latency~\cite{hapsAlliance6G2026,tgpp38811} & LEO 20--60~ms; MEO 60--200~ms; GEO 480--560~ms & Real-time to seconds (LOS) & End-to-end target 1--10~ms (Alliance KPI); propagation-only $\approx$0.07~ms at nadir (geometric lower bound) \\
Atmospheric path of view & Full atmosphere & Sub-PBL to lower-troposphere & Lower-stratospheric (sensor above 95\,\% of mass; downward water-vapor column same as LEO since $>$99\,\% of H$_{2}$O sits below 12~km, see \S\ref{sec:why}) \\
All-weather, day/night & Yes via flown SAR and IR/MW sounders; not optical & UAV-SAR limited; UAVs fly below cloud directly & Moderate: RF/SIGINT and atmospheric LiDAR all-weather; HAPS optical/HSI still view through 0--13~km cloud column; SAR hardware-mature but not yet flight-validated~\cite{dlr-hapsar} \\
Tasking responsiveness & Hours--days (programmed) & Minutes (operator) & Hours (re-position) to minutes (payload pointing) \\
Regulatory regime & Frequency and orbital slots; NOAA CRSRA licensing and EAR for commercial high-res imagery; ITAR for defense exports & FAA Part~107 / EASA Open/Specific/Certified & ITU-R Res.~165~\cite{itu_Res165}; 3GPP TR~38.811~\cite{tgpp38811}; FAA/EASA HAPS category nascent \\
Modality breadth (active functions of 19) & 19 (see Table~\ref{tab:functions}; atmospheric LiDAR via EarthCARE ATLID, public data from Jan~2025) & 12 (see Table~\ref{tab:functions}) & 5 flight-validated; 1 ground-demonstrated, 2 partial, 3 conceptual, 8 unflown (tags and qualifications in Table~\ref{tab:lens} and \S\ref{sec:lens}) \\
\bottomrule
\end{tabular}
\end{table*}

The number of national-jurisdiction HAPS programs with public flight, near-flight, or active stratospheric-development evidence has grown to eight as of August~2026~\cite{inflection-haps-economics} (Table~\ref{tab:jurisdictions}). Applying the strict-flight evidence rule alone yields five flown programs across three jurisdictions (United Kingdom, United States, Japan). The broader count includes near-flight and active development. The shift to an early-commercial regime in 2024--26 rests on three converging technology enablers. Lithium-ion cell energy density has crossed the $\sim$350~Wh/kg threshold needed for multi-month flight (Amprius silicon-anode at 450~Wh/kg for Zephyr~\cite{amprius2025Aalto67d}). Single-junction III-V solar cells (single-cell devices built from III-V semiconductors such as the Alta Devices GaAs cell) now reach $\sim$29\,\% NREL-certified efficiency, while stacked multi-junction III-V cells, in which several junctions are layered to absorb different parts of the solar spectrum, exceed 30\,\% one-sun~\cite{nrel-celleff}, up from $\sim$22\,\% a decade ago. Edge processors deliver several TOPS/W (Jetson AGX Orin, Hailo-8~\cite{nvidia-orin,hailo-8}), bringing on-board ISR inference and autonomous flight control within the solar HALE envelope (\S\ref{sec:why}). The framing aligns with ITU-R IMT-2030~\cite{itu_M2160}, the HAPS Alliance multi-layered NTN architecture~\cite{hapsAlliance6G2026}, and the academic 6G HAPS roadmap of~\cite{abbasi2024haps6G,kurt2021haps}, all of which position HAPS as a stratospheric tier between terrestrial and satellite layers and identify \emph{Integrated Sensing and Communication (ISAC)} as 6G-native. We adopt this framing with one caveat: no HAPS demonstrator has yet flown a shared-aperture, shared-waveform ISAC payload, and we therefore treat ISAC on HAPS as a forward research direction rather than a flown capability.

% ----------------------------------------------------------------------------
\section{NTN Integration}\label{sec:ntn-integration}
% ----------------------------------------------------------------------------

\subsection{NTN Context} HAPS carries one of two NTN payload classes: a \emph{transparent} (bent-pipe) relay that amplifies and forwards an analogue carrier without decoding, or a \emph{regenerative} payload that terminates the air interface onboard. The HAPS Alliance reference architecture splits the system into Aviation subsystems (flight vehicle, energy, fleet/air-traffic management) and Service subsystems (service link, feeder link, core network)~\cite{hapsAllianceRefArch2024,hapsAllianceAdvantages2025}. The regenerative-versus-transparent choice is a Service-subsystem decision. In the transparent case the air interface terminates at a terrestrial base station and HAPS only forwards the analogue carrier. In the regenerative case the gNodeB-DU (and increasingly an MEC instance for AI-RAN inference~\cite{hapsAlliance6G2026,itu_M2160}) sits onboard the platform, terminating user-plane processing in the stratosphere and sending only an aggregated feeder link to ground. The platform is then a base station, not a relay. The regenerative split changes the feeder-link constraint: a transparent HAPS spends spectrum on every active user, while a regenerative HAPS spends feeder spectrum only on backhaul, allowing larger user populations per platform but pushing onboard SWaP up. Mobility management is structurally easier than for LEO NTN because a station-kept HAPS produces a fixed cell footprint, eliminating the high-rate cell-reselection that dominates LEO NTN signaling. The residual mobility load comes from the kilometer-class loiter circle. Doppler shifts and rates are accordingly two-to-three orders of magnitude smaller than for LEO (per the channel-model parameters of 3GPP TR~38.811~\cite{tgpp38811}), removing the pre-/post-compensation timers that the same TR specifies for LEO, but the K/Ka/V-band low-elevation scintillation and rain-rate variability are first-order link-budget concerns. Synchronisation can in principle source from GNSS plus an onboard chip-scale atomic clock, with carrier phase coherent over the on-station period. However, published HAPS-NR demonstrations have not yet validated that phase-coherent reference at carrier-grade precision. Spectrum coexistence with terrestrial 5G/6G in the HAPS bands is the principal active-coordination question, governed by ITU-R Resolutions~122, 150, and 165 (full band-by-band allocations are tabulated in \S\ref{sec:forward}). The HAPS channel itself is well characterized: 3GPP TR~38.811 specifies a Rician line-of-sight model with elevation-angle-dependent shadowing and a tropospheric scintillation tail driven by refractive-index fluctuations~\cite{tgpp38811}, and the academic survey of Kurt et al.~\cite{kurt2021haps} consolidates HAPS-specific rain-attenuation and K/Ka/V-band scintillation statistics. The link-budget consequence is that K/Ka/V-band HAPS gateways at low elevation (below $\sim$20$^{\circ}$) lose 5--10~dB to scintillation plus rain rate even at temperate latitudes, so site diversity and elevation-angle gating are first-order coordination requirements alongside the spectrum allocation itself. Latency decomposes into three distinct terms: the one-way \emph{propagation} latency of $\sim$0.07~ms at nadir is set by geometry, radio-access latency adds the slot/symbol structure of the air interface (sub-ms in NR with mini-slot configurations), and end-to-end latency adds the transport leg through the regenerative or transparent feeder link to the destination application. The 1--10~ms HAPS Alliance target~\cite{hapsAlliance6G2026} is end-to-end and applies under specific assumptions, not to every traffic class.

\begin{table}[!htbp]
\centering
\caption{National-jurisdiction HAPS programs with public flight, near-flight, or active stratospheric-development evidence as of August 2026, ordered by maturity stage.}
\label{tab:jurisdictions}
\footnotesize
\rowcolors{2}{gray!8}{white}
\renewcommand{\arraystretch}{1.06}
\setlength{\tabcolsep}{4pt}
\begin{tabular}{@{}>{\raggedright\arraybackslash}p{1.5cm} >{\raggedright\arraybackslash}p{2.3cm} >{\raggedright\arraybackslash}p{3.9cm}@{}}
\toprule
\textbf{Stage} & \textbf{Country (lead)} & \textbf{Program / platform} \\
\midrule
Flight-validated & United Kingdom & AALTO/Airbus Zephyr; BAE PHASA-35 \\
Flight-validated & United States & Sceye; Aerostar Thunderhead \\
Flight-validated & Japan (AeroVironment) & SoftBank/HAPSMobile Sunglider, LTE demo with Loon payload (flight evidence 2020; HAPSMobile JV absorbed into SoftBank 2023, Table~\ref{tab:lessons}) \\
Near-flight ($<$18~km) & India & DRDO stratospheric airship ($\sim$17~km, May 2025) \\
Near-flight ($<$18~km) & New Zealand & Kea Aerospace ($\sim$17.2~km, Feb 2025) \\
Near-flight ($<$18~km) & UAE & Bayanat--UAVOS Mira Aerospace ApusDuo \\
Active dev.\ (no strat.\ flight) & United States & Radical Evenstar (Seattle) \\
Active dev.\ (ground-tested) & Germany & DLR HAP-alpha \\
Active dev.\ (in development) & France & Thales Stratobus \\
\bottomrule
\end{tabular}
\end{table}

\subsection{Standards} The relevant standardization scaffolding spans 3GPP TR 38.811 (NTN channel modeling)~\cite{tgpp38811}, TR 38.821 (NTN solutions in NR)~\cite{tgpp38821}, Rel-17 NR-NTN normative work, and Rel-19 NTN-NR-IoT enhancements. A complementary review of LEO standardization across the same release window is given in~\cite{darwish2022LEOstandards}. Rel-18 added NTN power-efficiency improvements and IoT-NTN enhancements while remaining transparent-payload-based; Rel-19 introduces regenerative payload definitions, inter-satellite links, and store-and-forward operation (specified for IoT-NTN), plus continued NB-IoT/NTN profiles relevant to HAPS-served low-rate sensors, and is expected to produce the first carrier-grade HAPS NB-IoT trials. Most existing HAPS comms demonstrations remain low-scale, low-user-density technology demos rather than carrier-grade deployments. The AALTO 4G D2US trial~\cite{aaltoFG2025Kenya} is operator-attested, but the user-plane KPIs (sustained throughput, Block Error Rate (BLER) under load, mobility statistics) have not been published.

\subsection{Edge Compute} Beyond air-interface standardization, the architectural placement of HAPS in the 6G stack is now being formalised. A HAPS layer composes with three other tiers in a 6G NTN: a terrestrial 5G/6G layer that handles dense urban traffic, a LEO NTN layer that handles global thin-coverage and inter-satellite mesh, and a UAV layer that handles last-mile high-resolution survey and disaster-zone inspection. The natural HAPS-specific role beyond relay is as a regenerative \emph{edge-compute and sensing node}: the platform terminates the air interface, runs MEC-class inference on locally collected sensor data (methane plume detection, wildfire ignition, ADS-B / AIS, hyperspectral feature extraction), and emits both a service link to UEs and a low-rate downlink of decisions to ground users. This edge layer has been studied as a cloud-enabled HAPS architecture~\cite{mershad2021cloudHaps}, and on-board caching and computation offloading have been shown to reduce latency for vehicular and IoT workloads served from the stratosphere~\cite{ren2022hapsCaching}. The integrated HAPS--terrestrial network proposed in~\cite{shamsabadi2024urbanHaps,alam2021hapsSMBS} positions HAPS as a super-macro base station that absorbs dense-urban capacity through wide-area beams. This co-locates sensing and connectivity at the platform that has both the persistent presence and the close-range geometry to do them. The architecture is the explicit subject of recent inter-platform link studies, including 3GPP-NTN-channel-model simulations of GEO-to-HAP backhaul~\cite{grieco2024sat-hap-link} and HAPS-altitude tuning for downlink capacity under elevation-angle-dependent fading~\cite{jang2025altitude}. The HAPS Alliance numbers cited here are industry-target values rather than independently measured ones.

% ----------------------------------------------------------------------------
\section{Function Reference: Satellites and UAVs}\label{sec:functions}
% ----------------------------------------------------------------------------

We anchor the perspective in a function catalog. Table~\ref{tab:functions} lists the sensing, navigation, and communication functions that satellites and UAVs already perform, each with a representative reference point and a key technical (quality) parameter. The table fixes the reference values that any HAPS implementation of the same function must compete with; exhaustiveness in the system column is the role of mission catalogs such as ESA eoPortal~\cite{eoportal-s2,eoportal-s1} and the WMO OSCAR database. We state the construction rule for the nineteen rows explicitly. A row is included when the function has an operational satellite implementation with civil or dual-use mission heritage, and each row is defined at the level of the measured geophysical or service product. Derived products and signal-of-opportunity variants are folded into their parent rows rather than counted separately (CO$_2$ retrieval and SBAS augmentation, per the table caption). GNSS reflectometry (CYGNSS, Spire GNSS-R) is the most prominent function this rule folds rather than lists. Its products, sea-state and surface soil moisture, are alternative retrievals adjacent to existing rows, and no HAPS GNSS-R flight or program has been published, so a dedicated row would enter with no flight evidence. The catalog is therefore not exhaustive, but the consequence is conservative: any added row could only lower the HAPS cross-over fraction, so the headline count of \S\ref{sec:lens} is an upper bound on coverage breadth. The count is accordingly robust in direction but not exact in magnitude: a different yet equally defensible choice of row granularity could move the cross-over fraction by one or two functions without altering the qualitative finding, because the binding constraint on the headline result is the four-program flight record rather than the row count. We note, however, that the small receiver SWaP of GNSS-R makes it a natural ride-along candidate on the same manifests as the Tier-5 receivers of \S\ref{sec:forward}. A further boundary is one of measurement \emph{principle}. Every row in the catalog senses a distant target through the electromagnetic spectrum, with the gravimetric and magnetic potential-field pair as the only exception, so each function observes from a distance. A fundamentally different class instead measures the medium from within it: persistent in-situ stratospheric sampling of composition, fields, and particles, demonstrated by the NOAA POPS aerosol spectrometer flown on a Stratollite~\cite{noaaPOPSStratollite}, and acoustic (infrasound) sensing, in which the low-noise stratospheric ``quiet zone'' lets a balloon-borne microbarometer record surface explosions and earthquakes more clearly than ground arrays~\cite{bowman2021infrasound}. Because neither has an operational satellite implementation, both fall outside the satellite-referenced cross-over count of \S\ref{sec:lens} rather than adding to it; we read them not as candidate cross-over rows but as a HAPS-native frontier, the limiting case of persistence at close range in which the platform observes by being present in the medium it measures.

A few function-category terms recur and are worth fixing up front. \emph{Synthetic Aperture Radar (SAR)} synthesizes a long antenna along the platform's flight path by coherent processing of successive pulses. The achievable cross-track (range) resolution is set by pulse bandwidth, while the along-track (azimuth) resolution is bounded by half the real-antenna length in the along-track direction (the \emph{minimum azimuth aperture}). \emph{Noise-equivalent sigma-zero (NE$\sigma^0$)} is the radar-cross-section per unit area at which a distributed-target return equals the receiver noise floor (a smaller, more negative number is better). \emph{Differential Optical Absorption Spectroscopy (DOAS)} retrieves trace-gas column amounts from the differential absorption signature of the column against a solar reference spectrum, typically in the UV/visible bands. \emph{Integrated Sensing and Communication (ISAC)} is a 6G-native architecture in which a shared waveform and aperture serve both data transmission and radar-style sensing of the environment. \emph{Ground Sample Distance (GSD)} is the projection of a single detector pixel onto the ground. The \emph{diffraction-limited GSD} is the smallest GSD the optics can in principle resolve at a given wavelength and aperture. These terms appear throughout Tables~\ref{tab:functions}, \ref{tab:lens}, and \ref{tab:gains}. The engineering constraints that bound each are developed in \S\ref{sec:why}.

\subsection{Why HAPS Geometry Helps: Three Numerical Cases}\label{sec:tutorial}
The physical-scaling laws of Table~\ref{tab:gains} follow from first principles in a few lines each. Let $R_{\mathrm{LEO}}=500$~km and $R_{\mathrm{HAPS}}=20$~km be slant ranges at nadir. The ratio $R_{\mathrm{LEO}}/R_{\mathrm{HAPS}}=25$ recurs throughout. (At the $\sim$320~km very-low-Earth-orbit altitude of imagers such as Albedo Clarity-1~\cite{albedoClarity}, this ratio is about 16.)

\textit{Passive optical/IR irradiance ($1/R^{2}$).} For a Lambertian ground patch viewed from above, the radiant flux from that patch through a pupil of fixed area scales as the inverse square of slant range. The HAPS-to-LEO entrance-pupil irradiance ratio for a fixed-size target is therefore
\begin{equation}\label{eq:irradiance}
\frac{E_{\mathrm{HAPS}}}{E_{\mathrm{LEO}}}=\left(\frac{R_{\mathrm{LEO}}}{R_{\mathrm{HAPS}}}\right)^{2}=25^{2}=625\quad(\approx 28~\mathrm{dB}).
\end{equation} Whether this irradiance ratio translates into a usable SNR ratio depends on the dominant noise term: in a read-noise-limited sensor at fixed integration time the SNR gain is $625\,\times$. In a shot-noise-limited sensor the SNR scales as the square root of the signal, and the gain collapses to $\sqrt{625}=25\,\times$. To put numbers on this, consider ICESat-2. NASA's ATLAS instrument flies at $\sim$500~km and transmits 48--172~$\mu$J per strong-beam pulse at 532~nm ($\sim$100~$\mu$J at the energy setting flown for most of the mission) with a 10~kHz pulse-repetition frequency, from which it recovers on the order of a few signal photoelectrons per shot over vegetated land~\cite{icesat2-mission,martino2023atlas}. Re-flying the same pulse, beam divergence, and target from 20~km multiplies the per-shot photoelectron count at a same-aperture receiver by the $1/R^{2}$ factor of $625\,\times$. The more useful corollary is the design statement obtained by inverting the relation: a HAPS LiDAR that only has to match ICESat-2's per-shot photon budget can shrink its pulse energy by the same factor, to $\sim$100/625$\,\approx\,$0.16~$\mu$J. This is precisely the operating point projected for UC5 (\S\ref{sec:uc5}): $\sim$0.16~$\mu$J pulses at a $\sim$1.6~mW average laser power, well within the SWaP envelope of a solar HALE payload bay.

\textit{SAR received power ($1/R^{4}$ point-target).} The monostatic radar equation gives the received power for a single point scatterer of cross-section $\sigma$ as
\begin{equation}\label{eq:radar}
P_{r}=\frac{P_{t}G^{2}\lambda^{2}\sigma}{(4\pi)^{3}R^{4}L},
\end{equation}
where $G$ is the antenna gain, $\lambda$ the wavelength, and $L$ the losses~\cite{moreira2013sar}. Holding $P_{t}$, $G$, $\lambda$, $\sigma$, and $L$ fixed, the HAPS-to-LEO ratio is $(R_{\mathrm{LEO}}/R_{\mathrm{HAPS}})^{4}=25^{4}\approx 3.9\times 10^{5}$ ($\approx$56~dB). To see what this buys in hardware terms: representative X-band LEO SAR satellites transmit several hundred watts to a few kilowatts of peak RF power (Capella $\sim$400--600~W~\cite{capella-stringham}, ICEYE up to $\sim$3~kW~\cite{eoportal-iceye}). The 56~dB free-space relief from a 20~km HAPS therefore implies sub-watt to few-watt peak transmit power for the same point-target return. In practice the budget is partly spent as lower transmit power, partly as shorter antennas at fixed swath, finer resolution at the same SWaP, or longer coherent-integration windows. For distributed targets one factor of $R$ is recovered because the synthetic-aperture integration length, and hence the coherent processing gain, scales linearly with range~\cite{moreira2013sar}, leaving NE$\sigma^{0}$ scaling as $1/R^{3}$ and a HAPS gain of $25^{3}\approx 1.6\times 10^{4}$ ($\approx$42~dB).

\textit{Diffraction-limited GSD ($\propto \lambda R/D$).} The Rayleigh criterion gives the angular resolution of a circular aperture of diameter $D$ at wavelength $\lambda$ as $\theta=1.22\lambda/D$. Projected to the ground at slant range $R$, the diffraction-limited ground sample distance is
\begin{equation}\label{eq:gsd}
\mathrm{GSD}_{\mathrm{diff}}=1.22\,\frac{\lambda R}{D}.
\end{equation} The HAPS-to-LEO GSD ratio at the same aperture is therefore $R_{\mathrm{HAPS}}/R_{\mathrm{LEO}}=1/25$, i.e., $25\,\times$ finer. For instance, a 30~cm aperture at $\lambda=550$~nm (visible green) gives $\text{GSD}_{\mathrm{diff}}=1.22\times 550\times 10^{-9}\times 500\times 10^{3}/0.30\approx 1.12$~m at LEO and $1.22\times 550\times 10^{-9}\times 20\times 10^{3}/0.30\approx 4.5$~cm at HAPS. The diffraction-limited GSD is a best case, the finest GSD the optics can deliver. In practice, sampling pitch, pointing jitter, and atmospheric MTF make the operational GSD coarser (UC4, \S\ref{sec:uc4}).

All three results are gathered in Table~\ref{tab:gains}. The engineering constraints that bound translation of these geometric upper bounds into operational performance are developed in \S\ref{sec:why}.

\begin{table*}[!tbp]
\caption{Function catalog (19 rows): representative satellite and UAV implementations with key quality parameters. Specifications vary by mode and configuration and are indicative reference points, not exhaustive. CO$_2$ retrieval and SBAS augmentation fold into the methane-imaging and GNSS-broadcast rows and are not counted separately in the 5/1/2/3/8 cross-over total of \S\ref{sec:lens}.}
\label{tab:functions}
\centering
\footnotesize
\renewcommand{\arraystretch}{0.9}
\setlength{\tabcolsep}{4pt}
\rowcolors{2}{gray!8}{white}
\begin{tabular}{@{}>{\raggedright\arraybackslash}p{3.0cm} >{\raggedright\arraybackslash}p{5.2cm} >{\raggedright\arraybackslash}p{5.0cm} >{\raggedright\arraybackslash}p{4.0cm}@{}}
\toprule
\textbf{Function} & \textbf{Satellite reference} & \textbf{UAV reference} & \textbf{Key quality parameter} \\
\midrule
Optical EO (pan/MS) & Sentinel-2 10/20/60~m~\cite{esa-s2}; WorldView-3 31~cm pan~\cite{maxar-wv3}; Pl\'eiades Neo 30~cm~\cite{airbus-pneo}; Planet 3~m daily~\cite{planet-ps} & WingtraOne 0.7~cm/px~\cite{wingtra-gen2-datasheet}; eBee~X 2.9~cm/px @120~m~\cite{sensefly-ebeex-spec} & GSD vs.\ revisit; aperture-limited \\
Hyperspectral (VNIR/SWIR) & PRISMA 30~m, 239 bd~\cite{asi-prisma,green1998aviris}; EnMAP 30~m, 224 bd~\cite{enmap-mission}; EMIT 60~m, 285 bd~\cite{nasa-emit} & Headwall Nano-Hyperspec~\cite{headwall-nano-spec}; Specim AFX10~\cite{specim-afx10-datasheet} & Spectral coverage $\times$ SNR \\
Thermal IR & Landsat~8/9 TIRS 100~m native (resampled to 30~m for delivery)~\cite{nasa-landsat9}; ECOSTRESS 38$\times$69~m native~\cite{jpl-ecostress}; SLSTR 1~km SST~\cite{copernicus-slstr} & FLIR Vue/Boson; MicaSense Altum-PT 33.5~cm @120~m~\cite{micasense-altum-pt-spec} & NEdT; integration time \\
Synthetic aperture radar & Sentinel-1 C-band: 5$\times$20~m IW (interferometric wide-swath, default land-acquisition mode), 5$\times$5~m Stripmap, 20$\times$40~m EW, and Wave mode~\cite{esa-s1}; Umbra X-band $\sim$0.16~m spotlight, ICEYE Gen4 X-band $\sim$0.25~m class spotlight (Dwell / persistent), Capella X-band 0.25--0.5~m spotlight (Spotlight-50 and finer modes; stripmap is coarser)~\cite{eoportal-iceye,capella-stringham,umbra-wmo}; NISAR L+S (launched 30~July~2025; declared operational November~2025)~\cite{nasa-nisar}; BIOMASS P-band (launched 29~April~2025; first open data products January~2026)~\cite{esa-biomass} & IMSAR NSP-3 Ku-band 0.1~m on ScanEagle~\cite{imsar-nsp3-spec}; MQ-9 Lynx 0.1~m~\cite{gaasi-lynx-sar} & Resolution $\times$ PRF; antenna SWaP~\cite{moreira2013sar} \\
Topographic LiDAR & ICESat-2 ATLAS (6 beams, 3 strong/weak pairs; measured on-orbit footprint diameter $\sim$11~m against a $\leq$17.4~m requirement)~\cite{icesat2-mission,markus2017icesat2,martino2023atlas}; GEDI 25~m, 8 tracks~\cite{gedi-mission,dubayah2020gedi} & DJI Zenmuse L2 5/4~cm @150~m~\cite{dji-zenmuse-l2-spec}; Riegl VUX-1UAV~\cite{riegl-vux1uav-spec} & Vertical accuracy; pointing stability \\
Atmospheric LiDAR & EarthCARE ATLID (launched May~2024; public data from Jan~2025)~\cite{wehr2023earthcare}; CALIPSO (retired Aug~2023)~\cite{nasa-calipso}; Aeolus (operations ended 2023)~\cite{esa-aeolus} & Lab-scale only & Optical aperture; eye-safety \\
Ocean / cryo radar altimetry & Jason-3 Ku-band cm-class~\cite{nasa-jason3}; Sentinel-6~\cite{esa-sentinel6}; SWOT Ka-band 120~km swath~\cite{nasa-swot}; CryoSat-2~\cite{esa-cryosat2} & Not routine & Range precision; orbit knowledge \\
Precipitation radar & GPM DPR Ku/Ka, 5~km horiz, 250~m vert~\cite{nasa-gpm} & Not routine & Reflectivity sensitivity; vertical resolution \\
Soil moisture (passive + active) & SMAP L-band 36~km~\cite{nasa-smap,entekhabi2010smap}; SMOS 35--50~km~\cite{esa-smos} & Research only & L-band $T_{B}$ NEdT; spatial resolution \\
Methane (CH$_4$) imaging & TROPOMI 5.5$\times$7~km daily~\cite{esa-tropomi,veefkind2012tropomi}; GHGSat-C $\sim$25~m~\cite{ghgsat-mission,jervis2021ghgsat}; MethaneSAT (loss of contact 20~June~2025)~\cite{methanesat}; Carbon Mapper~\cite{carbonmapper}; EMIT 60~m~\cite{nasa-emit} & SeekOps SeekIR~\cite{seekops-seekir-spec}; Bridger GML~\cite{bridger-gml-spec} & Detection limit (kg/h); revisit \\
Atmospheric trace gases (NO$_2$, O$_3$, SO$_2$, HCHO) & TROPOMI 5.5$\times$7~km~\cite{esa-tropomi,veefkind2012tropomi}; TEMPO (GEO N.\ America)~\cite{nasa-tempo}; GEMS (GEO Asia)~\cite{kari-gems}; Sentinel-4 on MTG-S1 (launched 1~July~2025; in calibration/validation through early 2026)~\cite{esa-sentinel4} & Limited UAV DOAS (research) & Slant-column detection limit; spatial resolution \\
Atmospheric sounders (IR/MW) & CrIS~\cite{nesdis-cris}; IASI~\cite{eumetsat-iasi}; MTG-S1 / IRS (GEO, recently launched)~\cite{eumetsat-mtgs}; ATMS~\cite{nesdis-atms} & Not routine & NEdT $<$0.5~K; spectral coverage \\
Gravimetric / magnetic field & GRACE-FO inter-sat ranging~\cite{nasa-gracefo}; Swarm B-field~\cite{esa-swarm} & UAV magnetometer common; gravimeter rare & Inter-asset baseline; B-field NEdT \\
GPS radio occultation & COSMIC-2 (thousands of occultations per day)~\cite{nesdis-cosmic2}; Spire ($\sim$10{,}000 per day constellation capacity)~\cite{spire-product} & Not routine & GNSS geometry; orbital cadence \\
GNSS / PNT broadcast & GPS/GLONASS/Galileo/BeiDou (primary)~\cite{gps-pnt}; WAAS / EGNOS (SBAS augmentation, derived from GNSS broadcast, not counted as a separate function in the cross-over total)~\cite{faa-waas,egnos-system} & Pseudolite ground tx (experimental) & Carrier phase noise; DOP \\
Lightning mapping & GLM on GOES-R 8~km px~\cite{noaa-goes}; LI on MTG~\cite{eumetsat-mtg}; ISS-LIS~\cite{eoportal-isslis} & Not routine & Detection efficiency; flash localization \\
RF / SIGINT / ELINT & HawkEye 360 70~MHz--18~GHz native, 26--40~GHz extension via the December~2023 acquisition of Maxar's RF Solutions business unit (formerly Aurora Insight)~\cite{eoportal-he360,sfl-he360}; multiple ESM operators & RQ-4 ASIP; MQ-9 derivatives & RF sensitivity; geolocation baseline \\
AIS / ADS-B & Spire LEO constellation~\cite{spire-product}; ORBCOMM OG2~\cite{orbcomm-ais} & Saildrone (surface); MQ-9 (limited) & Antenna footprint; collision avoidance \\
Broadband comms relay & Iridium LEO; Starlink mega-constellation (rapidly evolving); Inmarsat L-band GEO & Tactical airborne relays & Throughput; latency; sky availability \\
\bottomrule
\end{tabular}
\end{table*}

\section{HAPS Function Status}\label{sec:lens}

We now apply the evidence rule. \emph{Brochures alone do not promote a payload from \announced\ to \flown.} For each function in Table~\ref{tab:functions}, the question is: what is the strongest publicly attested HAPS evidence as of August~2026?

\subsection{Evidence Rule} The five tags applied per function in Table~\ref{tab:lens} are defined in Table~\ref{tab:evidence}; \announced\ is treated as ``no evidence'' for the purposes of cross-over counting. To see how the rule applies in practice, consider the August~2024 Sceye flight~\cite{hyspexSceye2024,sceyeDiurnalPR2024}: it carried the HySpex SWIR-640 hyperspectral imager into the stratosphere with operator-attested data return: the generic VNIR/SWIR hyperspectral row is therefore \flown, and so is the methane-imaging row because the same flight returned a methane-tuned retrieval against the same band selection used by mission-class GHG sensors. Sceye marketing also lists SAR, IR, and stereo EO as part of an ``advanced payload suite,'' but only the two hyperspectral instruments have publicly attested stratospheric data return. The SAR row therefore remains \demonstrated\ on the strength of the DLR HAPSAR ground-test campaign~\cite{dlr-hapsar} rather than the Sceye announcement, and the IR and stereo-EO rows are tagged \announced.

\begin{table}[!htbp]
\centering
\caption{Evidence-rule tags applied per function in Table~\ref{tab:lens}.}
\label{tab:evidence}
\footnotesize
\rowcolors{2}{gray!8}{white}
\renewcommand{\arraystretch}{1.06}
\setlength{\tabcolsep}{4pt}
\begin{tabular}{@{}>{\raggedright\arraybackslash}p{1.7cm} >{\raggedright\arraybackslash}p{6.0cm}@{}}
\toprule
\textbf{Tag} & \textbf{Requirement} \\
\midrule
\flown & Integration to a stratospheric platform at $\geq$~18~km \emph{plus} operationally relevant data return (calibrated retrieval, imaged scene, or sustained service link) with peer-reviewed or operator-attested support. Payload carriage without data return is not sufficient. \\
\demonstrated & Ground or short-flight test of a complete payload in its target band; no stratospheric data return. \\
\partialtag & Function partially exercised. E.g., a sub-band of the spectral coverage has flown while the operational standard has not, or a sounding rocket / scientific balloon has demonstrated the geometry but not station-kept HAPS operation. \\
\concept & Paper or simulation only. \\
\notyet & No public flight or near-flight evidence. \\
\announced & Brochure manifests without flight evidence; treated as ``no evidence'' for cross-over counting. \\
\bottomrule
\end{tabular}
\end{table}

\subsection{Evidence Weighting} Where multiple sources differ, we weight in descending order: (i) peer-reviewed conference or journal paper with reported flight datum; (ii) operator-attested press release with named platform, altitude, and date; (iii) third-party reporting of a public demonstration; (iv) operator press release without independent confirmation; (v) marketing PDF or company blog. Categories (iv) and (v) are flagged \est\ and called out inline as operator-reported. Several headline metrics in this paper (the Zephyr OPAZ 18~cm GSD, the Sceye projected methane detection limit, and operator-reported HAPS payload-mass and endurance figures aggregated by third parties) sit in category (iv) and remain independently unverified at the time of writing.

\begin{table*}[!tbp]
\caption{HAPS through the lens of the function catalog as of August 2026.}
\label{tab:lens}
\centering
\footnotesize
\setlength{\tabcolsep}{4pt}
\setlength{\emergencystretch}{3em}
\renewcommand{\arraystretch}{1.06}
\rowcolors{2}{gray!8}{white}
\begin{tabular}{@{}>{\raggedright\arraybackslash}p{2.7cm} >{\raggedright\arraybackslash}p{1.9cm} >{\raggedright\arraybackslash}p{7.6cm} >{\raggedright\arraybackslash}p{4.8cm}@{}}
\toprule
\textbf{Function} & \textbf{HAPS status} & \textbf{Strongest HAPS evidence} & \textbf{Quality vs.\ sat / UAV} \\
\midrule
Optical EO (pan/MS) & \cFlown & Zephyr OPAZ 18~cm~\est{} GSD streamed Arizona 2021~\cite{airbusZephyr2021PR}; DLR MACS-HAP 15~cm peer-published, environmental-test complete~\cite{dlr-macshap}; SuperBIT 0.5~m balloon, 39~d 2023~\cite{nasaSuperBIT2023}. & Matches mid-tier sat EO; the commercial reference is moving sub-decimeter: Albedo's VLEO Clarity-1 (launched 2025) was designed for 10~cm imagery, finer than the HAPS 18~cm figure, although the spacecraft was lost before completing that demonstration~\cite{albedoClarity}. \\
Hyperspectral (VNIR/SWIR) & \cFlown & Sceye + HySpex Mjolnir V-1240 + SWIR-640, full diurnal flight Aug~2024~\cite{hyspexSceye2024,sceyeDiurnalPR2024}: first publicly reported methane-tuned stratospheric HSI demo. & Spectral parity with PRISMA / EnMAP and the commercial Pixxel Firefly constellation (5~m, daily)~\cite{pixxelFirefly}; SNR not yet co-temporally validated. \\
Thermal IR & \cPartial & Sceye IR cameras announced~\cite{sceyeNASAUSGS2024}; Stratollite IR reported but specs not public. & Commercial high-resolution thermal is now spaceborne (SatVu HotSat $\sim$3.5~m MWIR~\cite{satvuHotSat}; Albedo 2~m LWIR~\cite{albedoClarity}); UAV/sat parity reachable, integration is the obstacle, not physics. \\
Synthetic aperture radar & \cDemo & DLR HAPSAR S-band ground-tested (5~kg, $<$250~W, $<$0.7~m resolution); HAP-alpha flight planned post-2026~\cite{dlr-hapsar,dlr-hapalpha-tests}. PHASA-35 has Aloft Sensing X-band SAR identified~\cite{av-phasa-aloft}. & \textbf{Hardware mature; flight imagery is the gap.} Commercial LEO SAR already images at $\sim$16--50~cm (ICEYE, Umbra, Capella)~\cite{commercialSAR}; slant range favors HAPS-SAR, antenna SWaP is the residual obstacle. \\
Topographic LiDAR & \cNotFlown & No HAPS-borne nadir topographic / bathymetric LiDAR flown. & UAV LiDAR (Riegl, Zenmuse L2) is reference; pointing-stability gap. \\
Atmospheric LiDAR & \cConcept & No HAPS-borne atmospheric LiDAR has been integrated or flown. Honeywell HALAS is a \emph{ground-based} stratospheric-profiling LiDAR that supports HAPS flight planning~\cite{honeywellHALAS2023}, not a flight payload; DIAL-on-HAPS exists as a design study~\cite{elkhazraji2025hapsOptical}. & EarthCARE ATLID (launched May~2024; commissioning completed late~2024, public data from Jan~2025)~\cite{wehr2023earthcare}, following the CALIPSO~\cite{nasa-calipso} and Aeolus~\cite{esa-aeolus} retirements in 2023; HAPS-borne flight is the next-tier R\&D direction. \\
Ocean / cryo radar altimetry & \cNotFlown & No HAPS-borne radar altimeter flown. & Jason-3 / SWOT~\cite{nasa-jason3,nasa-swot} cm-class; HAPS persistence could enable regional water-level monitoring. \\
Precipitation radar & \cNotFlown & No HAPS-borne precipitation radar flown. & GPM DPR~\cite{nasa-gpm}; HAPS persistence over a basin would be uniquely valuable. \\
Soil moisture (L-band) & \cNotFlown & No HAPS L-band radiometer or SAR flown for soil moisture. & SMAP / SMOS L-band brightness temperature~\cite{nasa-smap,esa-smos}; HAPS-on-basin retrieval is an open application. \\
Methane (CH$_4$) imaging & \cFlown & Methane-tuned channel of the Sceye + HySpex SWIR-640 flight Aug~2024~\cite{hyspexSceye2024} delivered an operator-attested CH$_4$ retrieval. CO$_2$ has not been separately demonstrated on HAPS; HSI bands cover the retrieval but no on-orbit data product. & CH$_4$ detection limit vs.\ GHGSat / MethaneSAT pending; CO$_2$ trails OCO-2/3~\cite{nasa-oco2} and is treated here as an HSI sub-function rather than a separate counted row. \\
Atmospheric trace gases (NO$_2$, O$_3$, SO$_2$) & \cConcept & Sceye HSI VNIR covers DOAS bands, but no HAPS trace-gas retrieval published; band coverage is necessary but not sufficient. & TROPOMI~\cite{esa-tropomi}, TEMPO~\cite{nasa-tempo}; HAPS persistence over a polluted basin would complement geostationary cadence. \\
Atmospheric sounders (IR/MW) & \cNotFlown & No HAPS-borne CrIS / IASI / ATMS analog flown. & Cross-track scanning favors LEO; HAPS is a poor structural match. \\
Gravimetric / magnetic field & \cNotFlown & No HAPS gravimeter / magnetometer mission. Drifting balloon mag (PEGASO~\cite{pegaso-magnetometer}) is not station-kept HAPS. & GRACE-FO inter-sat ranging~\cite{nasa-gracefo} cannot be replicated on a single HAPS; Swarm-class B-field needs higher attitude stability. \\
GPS radio occultation & \cPartial & HASP student GPS-RO experiments~\cite{nasaHASPFactSheet}; no operational HAPS GPS-RO. & COSMIC-2 / Spire~\cite{nesdis-cosmic2,spire-product} are the operational standard; HAPS geometry is sub-optimal vs.\ LEO. \\
GNSS / PNT broadcast (navigation) & \cConcept & APNT proposals~\cite{kurt2021haps,tgpp38811}; AALTO discusses regional PNT augmentation~\cite{aaltoFG2025Kenya}; ITU-R IMT-2030 1--10~cm positioning target~\cite{itu_M2160,hapsAlliance6G2026}. No attested HAPS-PNT broadcast. & Regional augmentation in GNSS-denied scenarios~\cite{gps-pnt}; complement, not replacement, for GNSS. \\
Lightning mapping & \cNotFlown & No HAPS lightning imager flown; wide-FOV optical / VHF would fit easily. & GLM on GOES-R~\cite{noaa-goes}, LI on MTG~\cite{eumetsat-mtg}; lower altitude could improve per-flash localization. \\
RF / SIGINT / ELINT & \cFlown & PHASA-35 SDR ISR Dec~2024~\cite{ustPHASA2024}; Aerostar / Airbus SATCOM relay trial Nov.--Dec.~2024~\cite{aerostarAirbusDec2024}; US Army COLD STAR balloon SIGINT, press-attested~\cite{fortuneColdStar2022,visColdStar}. & Functional parity with airborne SIGINT (RQ-4 ASIP) at much longer time on station. \\
AIS / ADS-B & \cNotFlown & Theoretical only; geometry attractive but no validated flight. & Spire (LEO) and Saildrone (sea-surface) are references. \\
Broadband comms relay & \cFlown & Sunglider LTE link Sep~2020 with TYO--SV--NM live video~\cite{avSunglider2020PR}. AALTO Zephyr 4G D2US Kenya March~2025~\cite{aaltoFG2025Kenya}; Aerostar SATCOM relay Nov.--Dec.~2024~\cite{aerostarAirbusDec2024}; Sceye 12-day, 6{,}400-mile \emph{long-duration traverse} in March--April~2026 (a point-to-point traverse with station-keeping segments of $\sim$88~h, not a persistent-service demonstration)~\cite{sceye12day2026}. Loon (2011--2021, discontinued) is historical precedent only; see \S\ref{sec:forward}. & Technology demonstrations only; carrier-grade persistence, certification, and large-scale operations have not been demonstrated. Complement to Iridium / Starlink, not replacement. ITU Res.~165~\cite{itu_Res165}; 3GPP TR 38.811~\cite{tgpp38811}. \\
\bottomrule
\end{tabular}
\end{table*}

\subsection{Status Summary}

Of nineteen functions, five are \flown\ with attested stratospheric data return: optical EO, VNIR/SWIR hyperspectral, methane imaging, RF/SIGINT, and broadband comms relay. We qualify this headline plainly, because the five \flown\ functions are not five independent demonstrations. First, the evidence is predominantly operator-attested rather than peer-reviewed. Operator-reported figures are flagged \est\ in Table~\ref{tab:lens} and are ranked below peer-reviewed data by the evidence-weighting order of this section. Because the \flown\ definition of Table~\ref{tab:evidence} admits operator-attested sources that the weighting of \S\ref{sec:lens}-B ranks at categories (ii)--(v), we report two counts. The operator-attested count is five of nineteen. A peer-reviewed-only count, restricted to category (i) sources, is zero to one of nineteen: zero under a strict Earth-observation reading, and one if the peer-reviewed balloon-borne SuperBIT optical imaging~\cite{nasaSuperBIT2023} is admitted toward the optical-EO row. The gap between the two counts matters in its own right, since none of the headline HAPS capability claims has so far been validated in peer-reviewed work. Second, two of the five, VNIR/SWIR hyperspectral and methane imaging, are the \emph{same} Sceye HySpex instrument on the \emph{same} August~2024 flight: methane is the methane-tuned retrieval channel of that hyperspectral payload, not a separate sensor or a separate sortie. Third, RF/SIGINT and broadband relay are both RF-communications-payload demonstrations that partly overlap, with the Aerostar/Airbus SATCOM relay of November--December~2024 counted toward both rows. The five functions therefore rest on four flight programs and, more conservatively, on the order of three to four genuinely independent demonstrations. We report 5/1/2/3/8 as a measure of \emph{function coverage}, not as a count of distinct flights. Methane imaging is counted as a separate category from generic hyperspectral because the operational standard is set by mission-class GHG retrievals (TROPOMI, GHGSat, MethaneSAT, Carbon Mapper) and a HAPS implementation must match that retrieval, not just cover the SWIR band. One is \demonstrated: SAR (DLR HAPSAR is ground-tested; PHASA-35 has Aloft Sensing identified for future flights~\cite{dlr-hapsar,dlr-hapalpha-tests,av-phasa-aloft}). Two are \partialtag: thermal IR (announced and partially flown but no published thermal-IR scene from a HAPS at $\geq$~18~km) and student-grade GPS-RO. Three are \concept: GNSS/PNT broadcast, atmospheric trace-gas DOAS (band coverage exists, but no retrieval has been published), and atmospheric LiDAR. We tag atmospheric LiDAR \concept\ rather than \demonstrated\ deliberately: the Honeywell HALAS instrument sometimes read as a HAPS payload is a ground-based stratospheric profiler used for HAPS flight planning~\cite{honeywellHALAS2023}, and no HAPS-borne payload program has been published. Eight remain \notyet: topographic / bathymetric LiDAR, ocean / cryo altimetry, precipitation radar, soil moisture, atmospheric sounders, gravimetric / magnetic, AIS / ADS-B, and lightning mapping. Of these eight, only two (AIS/ADS-B and lightning mapping) are favorable for HAPS: these are the evidence-bound functions of \S\ref{sec:lens}-D, where the receiver SWaP is small and the geometry already favors a station-kept platform. The other six are functions where the satellite or UAV tier is the natural fit and HAPS competes only marginally.

\subsection{The Eight Open Functions} The eight \notyet\ rows fall into three classes. First, \emph{geometry-bound} functions whose science depends on altitudes or baselines a HAPS cannot reach: ocean and cryo altimetry require sub-cm orbit knowledge that is built on tracked LEO trajectories; gravimetry needs either inter-asset baselines (GRACE-FO) or single-platform attitude stability beyond current HAPS; atmospheric IR/MW sounders need cross-track scanning over wide swath that LEO geometry favors. Second, \emph{aperture-bound} functions whose minimum payload still exceeds the HALE 5--15~kg envelope: L-band soil-moisture radiometry needs a 6~m-class antenna (SMAP flies a 6~m reflector); Ku-band precipitation radar carries a kW-class transmitter and antenna; topographic LiDAR with the laser power and pointing needed at 20~km slant range is at the edge of stratospheric SWaP. Third, \emph{evidence-bound} functions where a HAPS implementation is straightforward in physics but has not yet been integrated and flown: AIS and ADS-B (small receivers, geometry favors HAPS), and lightning mapping (a wide-FOV optical or VHF imager would fit comfortably). The three classes have very different prognoses. Geometry-bound functions are the least likely to migrate to HAPS in a realistic timeframe, even with continued investment, because the underlying geometry requirements are very difficult to meet from a single station-kept platform. Aperture-bound functions are conceivable only at airship class, with a several-times larger SWaP budget than current solar HALE provides, and even there the payload integration remains an open engineering question. Evidence-bound functions, by contrast, should close within the next two to three years if a sponsor with a regional mission emerges, since the underlying hardware is small and the geometry is already favorable.

The obvious counter-question is why two decades of HAPS development have not produced these supposedly easy functions, and the answer is structural rather than technical. First, for most of those two decades there was no platform to manifest a payload on: routine, persistent stratospheric flight dates only from the 2020--2026 wave (\S\ref{sec:background}), and a function cannot accumulate flight evidence before its platform class flies. Second, the few payload slots that early flights offered were consumed by the functions that fund platform programs, namely ISR, optical EO, and communications relay, so small receiver-class payloads were displaced from the manifest. Third, both functions face entrenched satellite incumbents (Spire and Aireon for AIS/ADS-B, the GOES-R Geostationary Lightning Mapper for lightning~\cite{noaa-goes}), which caps the willingness to pay for a HAPS implementation and confines it to a secondary, ride-along role at near-zero marginal cost. The two-to-three-year forecast is therefore conditional on manifest capacity rather than on engineering progress, and its non-realization would falsify the screening rule of \S\ref{sec:why}.

Three caveats apply to the table. The Sceye tagging follows the rule as applied in \S\ref{sec:lens}-A: only the two HySpex hyperspectral instruments have publicly attested flight data, and the remaining advanced-payload-suite entries are \announced. Stratospheric Platforms Ltd's 2024 5G ``HAPS'' trial was performed on a tropospheric Britten-Norman Islander aircraft, not on a HAPS~\cite{splIslander2024}. The Zephyr 18~cm GSD figure is operator-attested but not peer-verified and is flagged \est. Read against Fig.~\ref{fig:radar} and Table~\ref{tab:lens}, the analysis shows that satellites are the only class with an operational implementation of every function in the catalog (atmospheric LiDAR, briefly without an operational spaceborne instrument after CALIPSO and Aeolus ended in 2023, regained one when EarthCARE ATLID completed commissioning at the end of 2024, with public data from January~2025~\cite{wehr2023earthcare}), the HAPS column is filled by the five \flown\ rows plus one \demonstrated\ row pending flight validation, and closing the eight \notyet\ rows is the central R\&D opportunity. The satellite reference is not static either: over the same 2020--2026 window the commercial tier pushed very-low-Earth-orbit optical toward 10~cm (Albedo Clarity-1, launched March~2025 and lost before completing its 10~cm demonstration~\cite{albedoClarity}), daily hyperspectral at 5~m (Pixxel Firefly~\cite{pixxelFirefly}), and sub-meter SAR~\cite{commercialSAR}. These advances move along the two axes an orbit can always improve, resolution and viewing geometry, by flying lower or enlarging the aperture; the slant-range ratios of \S\ref{sec:functions}, referenced to a 500~km LEO, accordingly narrow to about~16 against a $\sim$320~km platform. The axis they leave untouched is time on station, because no orbit can remain over a single point. Persistence, not GSD, is what delimits the HAPS tier, and the comparisons in this paper are drawn against this current satellite frontier.

% ----------------------------------------------------------------------------
\section{Engineering Constraints}\label{sec:why}\label{sec:parameters}
% ----------------------------------------------------------------------------

The function-level lens makes a pattern visible. The functions that have crossed over to HAPS are those whose engineering obstacles are most compatible with the structural constraints of stratospheric platforms, rather than those with the most marketing investment. The constraints fall into four engineering domains, bounded by a two-part operational envelope. The domains are coupled rather than strictly orthogonal, and we name each by its \emph{binding} physics: \emph{aperture} is the diffraction and antenna-footprint constraint that sets achievable angular resolution and footprint, whereas \emph{SWaP} is the realizable mass-and-power envelope. A larger aperture consumes SWaP, so the two interact. We treat aperture as the physical requirement and SWaP as the budget it must fit within, and we flag this coupling wherever it governs (most sharply for L-band radiometry and large optics). Table~\ref{tab:domains} summarizes how each domain maps onto the function tags of \S\ref{sec:lens} before the subsections below develop them in turn.

Two clarifications keep the taxonomy honest. First, it is an organizing classification constructed from the observed cross-over pattern, not a theory derived independently of it. Its value is that it compresses nineteen outcomes into four binding constraints and yields a screening test for any new function: a candidate should cross over when its minimal payload fits the relevant sub-class SWaP envelope, it requires no aperture beyond roughly half a meter, it does not depend on meter-class repeat-pass geometry, and persistence at close range adds mission value. AIS/ADS-B and lightning mapping pass all four screens, and \S\ref{sec:lens} accordingly predicts that they should close within two to three years once payload-manifest capacity exists (\S\ref{sec:lens} also confronts why they have not closed in two decades). Their fate is the falsification test of the scheme, conditional on a regional sponsor emerging, since the four screens establish only that no engineering barrier remains (\S\ref{sec:lens}). Second, the four domains are not co-equal. SWaP and station-keeping bind most of the open functions, aperture binds the resolution- and antenna-driven ones, and above-cloud viewing is the weakest of the four: its downward-looking benefit is modest, as the corresponding subsection shows, and it earns its place mainly for the upward- and limb-viewing functions.

\begin{table*}[!tbp]
\centering
\caption{Synthesis: how the four engineering domains and the operational envelope map onto the function tags of \S\ref{sec:lens}.}
\label{tab:domains}
\footnotesize
\rowcolors{2}{gray!8}{white}
\renewcommand{\arraystretch}{1.1}
\setlength{\tabcolsep}{4pt}
\begin{tabular}{@{}>{\raggedright\arraybackslash}p{2.7cm} >{\raggedright\arraybackslash}p{3.3cm} >{\raggedright\arraybackslash}p{5.1cm} >{\raggedright\arraybackslash}p{5.1cm}@{}}
\toprule
\textbf{Domain} & \textbf{What it bounds} & \textbf{Favours (crossed over)} & \textbf{Blocks or does not help (open)} \\
\midrule
SWaP envelope (5--15~kg, 50--250~W HALE) & Payload mass and power & Optical EO, hyperspectral, methane, SDR/SIGINT, broadband relay & L-band soil moisture, precipitation radar, large-aperture LiDAR \\
Station-keeping (km-class loiter) & Repeat-pass geometric coherence & Single-pass imaging and sensing (EO, HSI, ISR) & InSAR, repeat-pass LiDAR, radar altimetry, gravimetry \\
Aperture (diffraction limit; antenna size) & Angular resolution and antenna footprint & Sub-decimeter EO at small aperture; HAPS-SAR via the $1/R^{4}$ margin & L-band radiometry (meter-scale antenna) \\
Above-cloud viewing geometry & Sky-path scatter, not the downward column & Upward/limb sensing: atmospheric LiDAR, occultation, DOAS & Downward optical/HSI/SAR, whose gain is from short range and persistence, not altitude \\
Operational envelope (stability + payload operability) & On-station persistence and continuous data return & SAR (day/night, all-weather) & Daylight-clear-sky-biased EO/HSI/methane \\
\bottomrule
\end{tabular}
\end{table*}

\subsection{SWaP Envelope}
A typical solar HALE aircraft (Zephyr~S, PHASA-35, DLR HAP-alpha, Sunglider) carries 5 to 15~kg at 50 to 250~W. Stratospheric balloons and airships (Stratollite, Sceye, Aerostar) carry tens of kilograms at higher power, with larger balloon classes capable of substantially more (e.g., $\sim$57~kg on Aerostar Thunderhead, third-party-reported via~\cite{inflection-haps-economics}, while the AeroVironment Sunglider commercial line is rated up to $\sim$75~kg, with the Horus~A military variant at $\sim$68~kg). The flown or near-flown HAPS sensing payloads all sit inside this envelope: Zephyr OPAZ EO (flown within Zephyr's $\sim$5~kg payload envelope)~\cite{airbusZephyr2021PR}, Honeywell HALAS atmospheric LiDAR (mass not publicly disclosed)~\cite{honeywellHALAS2023}, Sceye HySpex Mjolnir V-1240~+~SWIR-640 ($\sim$4~kg~+~$\sim$5~kg catalog)~\cite{hyspexSceye2024}, PHASA-35 SDR ISR ($>$9~kg)~\cite{ustPHASA2024}, NOAA POPS ($\sim$0.6~kg)~\cite{noaaPOPSStratollite}, DLR MACS-HAP optical (5~kg)~\cite{dlr-macshap}, and DLR HAPSAR (S-band, $<$~0.7~m resolution, 5~kg, $<$~250~W)~\cite{dlr-hapsar}. The communications side has a comparable flown reference: the Abside Networks LTE eNodeB flown on Loon (7.5~kg, $4\times5$~W transmit, LTE bands 20/28) is vendor-attested to have served over 500{,}000 users~\cite{absideHAPS}. The \notyet\ functions are those whose smallest realistic payload still exceeds the HALE envelope. Examples include a topographic LiDAR with the laser power and PRF needed at 20~km slant range, an L-band soil-moisture radiometer with adequate antenna footprint, a Ku-band precipitation radar, and a gravimetric instrument with the required thermal stability and Analog-to-Digital Converter (ADC) chain.

\textit{Sub-class comparison.} The three HAPS sub-classes occupy distinct points in the SWaP--endurance--station-keeping space. Solar HALE aircraft (Zephyr, PHASA-35, HAP-alpha, Sunglider, Evenstar) achieve multi-week to multi-month endurance at a 5--15~kg, 50--250~W payload envelope and station-keep to within a kilometer-class loiter circle. Super-pressure balloons (Stratollite, Loon-class) carry tens of kilograms at hundreds of watts with multi-day to multi-month endurance and station-keep probabilistically through wind-layer steering rather than active propulsion. Stratospheric airships (Sceye, Stratobus, Thunderhead) carry tens-to-hundreds of kilograms at kilowatt-class power but are still maturing on endurance and station-keeping. The L-band soil-moisture and Ku-band precipitation-radar payloads of \S\ref{sec:functions} fit only the airship sub-class envelope. The optical, hyperspectral, methane, SAR, and broadband-relay payloads fit all three. This sub-class structure is the reason a single-platform-class HAPS taxonomy misleads: a useful-resolution L-band soil-moisture radiometer requires a meters-scale antenna whose mass alone exceeds the 5--15~kg HALE envelope by roughly an order of magnitude (\S\ref{sec:why}, \emph{Aperture}), so the function must be assessed against the airship sub-class rather than against HALE, not because the HALE programs are immature but because the antenna-area requirement is set by the physics of L-band footprint geometry.

\textit{Propulsion variants: hydrogen-electric aircraft and eVTOL.} The aircraft sub-class is predominantly solar-electric today, but its powertrain is not fixed. A solar HALE airframe flies on a balanced day/night energy budget, which holds payload power to the 50--250~W range and ties year-round operation to solar-favorable latitudes and seasons (\emph{Operational Envelope}, below). Hydrogen-electric propulsion, either a fuel cell or a hydrogen-fueled generator charging a battery bus, relaxes both limits. Compressed or liquid hydrogen stores roughly an order of magnitude more energy per unit mass than current lithium-ion cells, so a hydrogen aircraft can devote far more mass and power to the payload, and it does not depend on the diurnal solar cycle, which in principle extends operation to high latitudes and the polar winter where the solar budget fails. The flown precedent is the AeroVironment Global Observer, a liquid-hydrogen HALE designed for $\sim$20~km (65{,}000~ft), a $\sim$170~kg payload, and week-class endurance; it completed nine test flights before the demonstrator was lost in April~2011, without reaching those design targets in flight~\cite{avGlobalObserver}, together with the Boeing PhantomEye, which completed nine liquid-hydrogen flights to $\sim$16.5~km~\cite{boeingPhantomEye}. Among current programs, Stratospheric Platforms Ltd targets a hydrogen-fuel-cell HAPS at $\sim$18~km (60{,}000~ft) sized for a $\sim$140~kg payload and $\sim$20~kW of antenna power, against the $\sim$200~W of a solar platform, although its public flight trials to date have been on a tropospheric testbed~\cite{splIslander2024,raesHydrogenHaps}. Two advantages distinguish this variant: a payload mass and power approaching the airship sub-class but on a conventional runway-launched aircraft, and the reuse of existing airport and aviation refueling infrastructure for takeoff, landing, and turnaround, rather than the dedicated launch site a balloon or airship requires. The trade is endurance and logistics: onboard fuel bounds a hydrogen aircraft to days-to-weeks rather than the multi-month flights of a solar platform, and cryogenic liquid-hydrogen ground handling is a first-order operational cost that weighed on the PhantomEye procurement case (\S\ref{sec:lessons}). We therefore read hydrogen-electric HAPS as an emerging high-payload, high-latitude variant of the aircraft class rather than a flight-validated stratospheric capability, since no current program has yet returned station-kept data from the target band.

A further forward-looking option is the electric vertical-takeoff-and-landing (eVTOL) airframe. Hydrogen-electric eVTOLs are a low-altitude urban-air-mobility technology today, yet recent flights show the range a HAPS role would demand: a Joby--H2FLY liquid-hydrogen eVTOL covered 523~miles ($\sim$840~km) on a single $\sim$40~kg hydrogen load in 2024~\cite{jobyH2flight2024}, and the AMSL Aero Vertiia targets a $\sim$1{,}000~km hydrogen range~\cite{amslVertiia2025}. A hydrogen-eVTOL HAPS would trade the aerodynamic efficiency of a high-aspect-ratio solar wing for runway independence, launching and recovering vertically from a confined site and dispensing even with the airport infrastructure that the fixed-wing hydrogen variant relies on. However, no eVTOL has operated as a stratospheric platform, and sustaining rotor-borne lift through the climb to 20~km is a substantial open problem. We therefore treat the hydrogen-eVTOL HAPS as conceptual, included to complete the powertrain-and-airframe design space rather than as a near-term path.

A further variant inverts the persistence premise altogether. Spartan Space's TRITON is a lightweight, modular stratospheric drone that is lofted by a high-altitude balloon and released autonomously, reaching the stratosphere without runway or launch-site infrastructure and returning after a short mission rather than station-keeping; a first flight test on 31~July~2026 validated the balloon-launch, controlled-separation, telemetry, and onboard-imaging chain with operation demonstrated at $\sim$24~km~\cite{spartanTriton2026}. We read this class as a rapid-reach, short-dwell complement to the platforms of this paper: it shares the stratospheric operating band and the payload-miniaturization constraints of this section, but trades time on station for deployment speed, so it sits outside the persistence-defined HAPS tier that our function-level analysis evaluates. Its relevance here is twofold: it extends the launch-infrastructure axis of the design space beyond runway (fixed-wing), launch-site (balloon and airship), and vertical (eVTOL) options to balloon-deployed release, and it supplies an early flight datum for the rapid-test-and-training role that the defense literature assigns to stratospheric platforms~\cite{hapsAllianceGoldenDome2026}. The evidence is a single operator press release, category~(iv) of the weighting of \S\ref{sec:lens}, and we weight it accordingly.

\subsection{Station-Keeping}
Solar HALE aircraft and balloon platforms hold position to within a few kilometers of a target, sometimes drifting in loiter circles up to about 100~km across when winds are unfavorable. Most single-pass sensing functions tolerate this comfortably, because they form their image or measurement without revisiting a precise earlier position: optical EO, hyperspectral imaging, atmospheric LiDAR, and SDR-based ISR all depend on pointing accuracy, not on the platform returning to a precise earlier position. The exceptions are functions built on repeat-pass interferometry, where the science signal lives in the geometric coherence between two visits. These include Interferometric SAR (InSAR)~\cite{rosen2000insar}, which needs baselines stable to within meters; repeat-pass LiDAR for ground deformation; centimeter-class ocean radar altimetry; and gravimetry. Closing the gap from kilometer-class drift to meter-class station-keeping is roughly three orders of magnitude of control-system performance, and no HAPS demonstrator has yet shown it.

\subsection{Aperture}
The diffraction-limited Ground Sample Distance (GSD), defined by the Rayleigh criterion as $\text{GSD} = 1.22\,\lambda R / D$ for an operating wavelength $\lambda$, slant range $R$, and aperture diameter $D$, sets the physical floor on the GSD an optical imager can achieve, that is, the finest detail it can resolve. It depends only on geometry and the wavelength of light. In practice, sampling pitch, pointing jitter, and detector noise may dominate before the diffraction limit is reached, so the diffraction-limited GSD is a best case, the finest GSD the optics can deliver, rather than a guaranteed value. The operational GSD is coarser. At 20~km the diffraction-limited GSD for a 0.5~m aperture at 550~nm is $\sim$2.7~cm. A 0.5-m HAPS optical telescope (an aperture class flown on SuperBIT in upward-looking science configuration~\cite{nasaSuperBIT2023}) can in principle compete with sub-meter commercial satellite EO. The required pointing stability and payload mass, however, exceed the current solar HALE 5--15~kg lift envelope. A 0.5~m downward-pointing telescope remains aspirational rather than near-term, and UC4 (\S\ref{sec:uc4}) uses a more credible 30~cm aperture for the projected 5~cm GSD. Aperture is why the best HAPS optical GSD (Zephyr OPAZ 18~cm~\est) trails the best commercial EO. It is also why HAPS-SAR must exploit the slow ground speed for long coherent-integration windows and the close-range SNR margin to operate at lower transmit power than LEO SAR. The relevant margin for imaging is the distributed-target NE$\sigma^{0}$ scaling of $1/R^{3}$ ($\approx$42~dB). The steeper $1/R^{4}$ ($\approx$56~dB) relief applies only to isolated point targets, so the transmit-power saving for scene imaging is the more modest of the two. The minimum physical azimuth aperture remains range-independent~\cite{moreira2013sar}. L-band soil-moisture radiometry is aperture-bound on HAPS for the same SWaP reason: scaling SMAP's 6~m mesh reflector and 36~km conical-scan IFOV at 685~km to a 4~m antenna at 20~km gives a $\sim$1.6~km HAPS footprint~\cite{nasa-smap,entekhabi2010smap,esa-smos}, but fitting a 4~m antenna into a 5--15~kg, 50~W payload remains the obstacle. SAR phase fidelity adds further constraints: oscillator coherence over the integration window, structural vibration of the wing or airship envelope, and atmospheric decorrelation across repeat passes all bound usable interferometric performance.

\subsection{Above-Cloud Viewing Geometry}
A platform at 20~km sits above approximately 95\,\% of the atmospheric mass~\cite{wallace2006atmosci}. For a downward-looking sensor the integrated water-vapor column is the same to within a few percent as for a LEO sensor, since $>$99\,\% of water vapor sits below $\sim$12~km. Both view through the same tropospheric H$_{2}$O, and an optically thick cloud between platform and ground blocks LEO and HAPS equally per single observation. The genuine HAPS advantages from this geometry are therefore (i) for downward-viewing sensors, a modest reduction in upper-leg aerosol and Rayleigh scatter and in residual cirrus contamination of atmospheric correction, and (ii) for upward- or limb-viewing geometries (atmospheric LiDAR, occultation, ozone-DOAS), a much shorter atmospheric path that directly reduces molecular and aerosol scatter on both outbound and return legs. Per-observation cloud blockage itself is not reduced by altitude: the cloud sits between platform and ground in either case. The probability of catching a clear line of sight over a multi-day mission window is reduced for HAPS instead through persistence, an argument that belongs to the persistence-layer framing of \S\ref{sec:layer} and is independent of viewing geometry. For SWIR methane retrievals the dominant geometry benefit is sky-background and residual cirrus mitigation, not column-water-vapor reduction. The crossover of optical EO, hyperspectral, and SWIR methane imaging is explained primarily by persistence and the close-range $1/R^{2}$ SNR gain rather than by the platform sitting above the cloud column. Only atmospheric LiDAR and occultation crossover is genuinely sky-geometry-driven. The same geometry mechanism explains less of the SAR story (microwaves are minimally absorbed regardless of altitude), almost none of the gravimetric story (gravity does not care about the air column), and none of the radar-altimetry story (Ku/Ka altimeters are limited by orbit knowledge, not atmosphere).

\subsection{Operational Envelope}
The four engineering domains describe what HAPS can carry, point, and see. They do not describe whether a given platform can stay on station and whether its payload keeps producing data through that period. The operational envelope adds two intertwined dimensions: \emph{platform stability} (the airframe must remain on station under realistic atmospheric and reliability conditions) and \emph{payload operability} (the sensor must keep producing data through the day/night thermal cycle, local weather, and the lower-stratospheric environment). Persistent operation is only as useful as the worst-affected of the two.

\textit{(i) Platform stability.} Three constraints bound stability. (a)~The solar HALE seasonal-and-latitude envelope is real: a Zephyr-class aircraft flies on a balanced day/night energy budget, so polar-night and high-latitude winter operation is not yet within reach, and the longest demonstrated flights have been at low-to-mid latitude in solar-favorable seasons~\cite{aaltoFG2025Kenya,amprius2025Aalto67d}. (b)~Lower-stratospheric winds are not uniformly benign: subtropical-jet remnants, mountain-wave events, and the seasonal Brewer--Dobson circulation drive headwind episodes well above the kilometer-class loiter-circle drift typical of station-keeping flight~\cite{wallace2006atmosci}, so that drift figure is the favorable case rather than the annual average. (c)~Platform-availability statistics matter for any persistent-service mission: the discontinued-program record (Zephyr~S~8~\cite{zephyrLoss2022}, Loon, PhantomEye, Aquila) is consolidated in \S\ref{sec:forward} (\emph{Lessons from Discontinued Programs}). The relevant 24/7-service metric is mean time between platform losses multiplied by the fleet-multiplier required to maintain coverage through scheduled and unscheduled out-of-service windows.

\textit{(ii) Payload operability and survivability.} A platform that stays on station does not automatically deliver continuous data. Two payload-side constraints qualify the persistence promise.

\textit{(a) Day/night and weather coverage.} The flown sensing modalities cover different subsets of conditions. Optical EO and VNIR/SWIR hyperspectral instruments are daylight-and-clear-sky payloads: they need solar illumination, and optically thick cloud blocks the line of sight (the Sceye HySpex methane retrieval~\cite{hyspexSceye2024} depends on both). Thermal IR imagers operate continuously day and night (passive emissive radiation), but cloud and heavy precipitation attenuate the upwelling thermal signal. SAR is the only modality that is genuinely day/night and all-weather, since cm/dm wavelengths penetrate cloud and precipitation with negligible attenuation. The flown HAPS portfolio is consequently daylight-clear-sky-biased, and dedicated HAPS-SAR systems such as DLR HAPSAR~\cite{dlr-hapsar} are the principal route to all-weather, around-the-clock HAPS sensing.

\textit{(b) Thermal cycling and stratospheric environment.} The ambient stratospheric environment itself is more extreme than payload-internal envelopes suggest: MIL-HDBK-310 1\%-worst-case values for the 20--24~km band give ambient pressures of 6.5--2.0~kPa and ambient temperatures of $-29^{\circ}$C to $-87^{\circ}$C~\cite{hapsAlliancePayloadStrato}. The MACS-HAP thermal-simulation envelope of $-30^{\circ}$C to $+50^{\circ}$C is therefore the payload-internal survivability envelope after thermal management, not the ambient condition~\cite{dlr-macshap}. The thin air precludes forced-convection cooling: heat dissipation is handled passively through black-anodized radiation plates~\cite{dlr-macshap}. Glass-lensed optical payloads face the additional risk of lens fracture from steep temperature gradients. The MACS-HAP mitigation is heat mats activated when the optical-cage temperature drops below $+10^{\circ}$C~\cite{dlr-macshap}. The HAPSAR S-band radar specifies a $-20^{\circ}$C to $+40^{\circ}$C operating envelope at minimum 15~mbar ambient pressure~\cite{dlr-hapsar}. Payloads that cannot be encapsulated and thermally managed against the day/night cycle do not fly, regardless of how stable the host platform is.

Together, these platform- and payload-side constraints bound the operational footprint of any individual HAPS class and motivate the multi-class fleets considered in \S\ref{sec:layer}. The platform staying on station is necessary but not sufficient for persistent data return.

\subsection{Radiation Environment and Onboard Compute}\label{sec:radiation}
A frequently overlooked advantage of the stratospheric tier is the benign radiation environment it presents to electronics, which relaxes the onboard-compute ceiling that constrains satellites. The distinction is qualitative, not merely one of magnitude. A platform at 20~km does not sit in a low-dose location: the cosmic-ray secondary cascade peaks near this altitude at the Regener--Pfotzer maximum, where a recent balloon campaign measured an ambient dose rate of $\sim$3~$\mu$Gy/h at 18.6~km~\cite{stratRadMexico2025}. Integrated over even a multi-month flight, however, the cumulative dose is of order 1~rad, four to five orders of magnitude below the tens-to-hundreds of krad a multi-year satellite mission accrues, and to which radiation-hardened parts are qualified~\cite{berger2001rad750}. Equally important, the destructive components of the space environment are absent at this altitude: the trapped Van~Allen particles are confined to higher orbits and never reach 20~km, while the primary heavy galactic-cosmic-ray ions responsible for single-event latch-up and burnout are attenuated by the overhead atmosphere and deflected by the geomagnetic field. What remains at 20~km is the atmospheric-neutron single-event environment already characterized for avionics, in which soft upsets are handled by error-correcting memory, watchdog timers, and scrubbing rather than by special silicon~\cite{normand1996avionics}.

The consequence for the sensing and ISAC payloads of this paper is direct. A HAPS can fly the same commercial edge accelerators used on the ground (e.g., the NVIDIA Jetson AGX Orin at up to 275~TOPS or the Hailo-8 at $\sim$26~TOPS~\cite{nvidia-orin,hailo-8}), whereas a satellite is restricted to radiation-hardened processors that trail commercial parts by roughly two orders of magnitude in throughput, such as the 133~MHz, 240-MIPS RAD750~\cite{berger2001rad750}, or must accept commercial-off-the-shelf parts under heavy mitigation and a shortened service life~\cite{bruhn2020gpuSpace}. Recoverability compounds the gap: a HAPS is retrieved and re-equipped with current silicon between flights, while a satellite is frozen to its launch-era hardware for five to twenty years. Two caveats keep the advantage bounded. First, the atmospheric-neutron upset rate is itself elevated near the Pfotzer maximum, so error-correction and watchdog mitigation remain mandatory. Second, onboard-compute throughput on a HAPS is ultimately limited by heat rejection in thin air (\emph{Operational Envelope}, above), not by radiation.

\subsection{Carrier-Grade KPIs beyond the Flight Demonstrations}\label{sec:cgkpi}
The flown HAPS communication demonstrations (AALTO Zephyr 4G D2US from Kenya, BAE PHASA-35 SDR-ISR, AeroVironment Sunglider LTE link) have established that an air interface can terminate at altitude, but none has been operated long enough or against enough users to characterize the metrics that decide commercial deployment. Carrier-grade service is defined by a four-tuple of Key Performance Indicators (KPIs): (i) Block Error Rate (BLER) under load (10\,\% BLER at full traffic-channel duty cycle for the worst-case scheduled user), (ii) BLER under mobility (re-establishment latency and throughput continuity as users handover between beams or platforms), (iii) Mean Time Between platform Losses (MTBPL) and the fleet multiplier required for 24/7 regional coverage, and (iv) station-keeping precision adequate for repeat-pass interferometry (meter-class, three orders of magnitude tighter than the current km-class drift~\cite{kurt2021haps}). Each public flight to date has measured at most one of these: the AALTO trial sustained a 4G link for a single demonstration period without published BLER-under-load curves, the PHASA-35 SDR established receive-side path-loss margin but not user-plane KPIs, and Sunglider's LTE link demonstrated a single downlink path without a fleet-multiplier study. Closing the carrier-grade gap is the single largest empirical question for HAPS comms over the next three years.

\subsection{Security and Resilience}\label{sec:security}
HAPS deployed as a comms relay, SIGINT, or PNT-broadcast platform inherits the cross-layer threat model already characterized for non-terrestrial networks~\cite{yue2023leoSecurity}, but the stratospheric tier introduces several HAPS-specific differences that warrant a dedicated treatment.

\textit{Physical-layer security.} Over the inter-HAPS optical mesh and the regenerative service link, eavesdropping margins differ from LEO. The narrower-beamwidth optical inter-platform link is inherently directional and hard to eavesdrop without line-of-sight cooperation, whereas the wide-area service link (50--100~km cell radius) is more easily intercepted by a ground or airborne adversary. Physical-layer-security techniques developed for LEO NTN (artificial noise, secrecy beamforming, cooperative jamming~\cite{yue2023leoSecurity}) adapt naturally to HAPS with the modification that the long on-station time allows an adversary to integrate any received signal for hours rather than the LEO pass-length of minutes. Secrecy capacity calculations should be recomputed with this longer adversary integration window.

\textit{Anti-jamming.} The close-range geometry cuts both ways. The $\sim$28~dB lower free-space path loss derived in Table~\ref{tab:gains} benefits the service link, but the same advantage accrues to a directional ground-based jammer reaching a 20~km HAPS rather than a 500~km satellite. A regional adversary with a few hundred watts of effective isotropic radiated power can therefore deny service over the inner HAPS cell unless null-steering, frequency-hopping, or spatial-diversity countermeasures are employed. Frequency-hopping and beam-nulling anti-jam techniques of the kind studied for LEO NTN are a starting point, but the longer HAPS exposure window lets an adversary characterize the hop sequence faster.

\textit{Anti-spoofing.} A HAPS PNT-broadcast or SBAS-augmentation payload (\S\ref{sec:functions}) is more vulnerable to ground-based spoofing than a GEO PNT source because of the smaller geometric diversity in the elevation angle. Authentication codes (Galileo OSNMA-class signatures) and multi-platform cross-correlation are the natural mitigations.

\textit{Supply chain and adversarial ISR.} Because HAPS payloads sit inside the SWaP envelope of consumer/industrial parts (Jetson AGX Orin edge processors, commercial Li-ion cells, COTS SDR modules), supply-chain integrity is a first-order risk for any defense-aligned deployment. Provenance attestation and trusted-execution-environment payload partitioning are open work. Counter-ISR from peer or near-peer adversaries is a parallel concern: a HAPS with a wide-FOV optical or thermal payload is itself observable from the ground, and persistent stratospheric flight at low-to-mid latitude is geometrically vulnerable to both ground-based directed-energy and air-launched effects. Hardening and dispersal across the fleet, rather than per-airframe stealth, is the architectural response we expect to dominate the next five years~\cite{haps-defense-trl-2025}.

% ----------------------------------------------------------------------------
\section{Eight Use Cases}\label{sec:gains}
% ----------------------------------------------------------------------------

The four engineering domains of \S\ref{sec:why} are qualitative. The quantitative argument rests on a small set of physics-based scaling laws (Table~\ref{tab:gains}) that systematically favor a 20~km platform over a LEO satellite at approximately 500~km at the same carrier frequency, antenna, and receiver. We apply them to eight concrete use cases: the first three (\S\S\ref{sec:uc1}--\ref{sec:uc3}) are flight-validated or near-flight as of August~2026, and the remaining five (\S\S\ref{sec:uc4}--\ref{sec:uc8}) are physics-justified projections for which the relevant payload has not yet been integrated and flown. Together, the eight cases show what HAPS already delivers and what it could deliver if the appropriate payload programs were funded. For seven of the eight cases we also provide a same-sensor scaling-law visualization, generated by physics-based forward simulation, with the identical instrument placed on a satellite at 500~km (left panel) and a HAPS at 20~km (right): Figs.~\ref{fig:sim-uc1}--\ref{fig:sim-uc3} for the flight-validated cases and Fig.~\ref{fig:sim-proj} for the projected ones. Each panel comes from a scripted chain of a procedural ground-truth scene, a forward model of the shared instrument, and per-platform observation parameters in which only the range-dependent terms differ; the \hyperref[sec:appendix-sim]{Appendix} gives the sensor, plume, speckle, photon-budget, and Planck-radiance models with a full parameter table (Table~\ref{tab:simparams}). Holding the sensor fixed makes the satellite-to-HAPS difference a consequence of geometry and physics, namely proximity, footprint, and time on station, rather than of a better instrument. The HAPS panels are accordingly a best case: they bound from above the resolution, sensitivity, and photon-budget gains and omit the platform-level penalties (pointing jitter, station-keeping error, detector or thermal SNR limits) that \S\ref{sec:why} identifies as the gating constraints, while UC3's coverage radius and link budget are exact consequences of range. The figures visualize the scaling laws of Table~\ref{tab:gains}; they are not measured HAPS data and do not independently validate those laws.

\begin{table*}[!tbp]
\centering
\caption{Physics-based gain factors for a HAPS at 20~km relative to a LEO satellite at approximately 500~km at the same carrier frequency, aperture, and receiver.}
\label{tab:gains}
\footnotesize
\rowcolors{2}{gray!8}{white}
\renewcommand{\arraystretch}{1.06}
\setlength{\tabcolsep}{4pt}
\begin{tabular}{@{}>{\raggedright\arraybackslash}p{3.8cm} >{\raggedright\arraybackslash}p{2.4cm} >{\raggedright\arraybackslash}p{4.0cm} >{\raggedright\arraybackslash}p{6.0cm}@{}}
\toprule
\textbf{Mechanism} & \textbf{Scaling law} & \textbf{HAPS gain over LEO} & \textbf{Mission-design implication} \\
\midrule
Free-space path loss (one-way) & $20\log_{10}(R_{1}/R_{2})$ dB at fixed $f$ & $\approx 28$~dB & Comms link budget; passive RF reception sensitivity~\cite{hapsAlliance6G2026} \\
Passive optical / IR irradiance at pupil & $\propto 1/R^{2}$ & $25^{2}=625\times$ ($\approx 28$~dB)~\cite{inflection-haps-economics} & The pupil-irradiance ratio is a geometric upper bound, not the usable SNR; it collapses from $625\times$ toward $25\times$ in the shot-noise-limited regime (derived in \S\ref{sec:tutorial}). Equivalently, at fixed SNR the required integration time drops by up to $625\times$ \\
SAR received power (point target) & $\propto 1/R^{4}$ & $\approx 56$~dB & Lower TX power, smaller swath antenna, or longer coherent integration \\
SAR distributed-target NE$\sigma^{0}$ (after SAR averaging) & $\propto 1/R^{3}$ & $\approx 42$~dB & One factor of $R$ is recovered because per-pixel resolution-cell area scales linearly with range~\cite{moreira2013sar}. Minimum azimuth aperture is range-independent and $\geq 2\rho_{az}$ \\
Diffraction-limited GSD & $\propto R$ & $25\times$ finer & 0.3~m aperture at 550~nm gives $\sim$1.1~m GSD at LEO and $\sim$4.5~cm at HAPS \\
Sensor-side atmospheric path & $\sim$95\,\% mass below 20~km; $>$99\,\% H$_{2}$O below 12~km~\cite{wallace2006atmosci} & Same downward cloud and water-vapor column as LEO; HAPS gain is lower upper-leg sky-background and reduced residual cirrus contamination of atmospheric correction~\est & Improves SWIR methane and DOAS retrievals through reduced sky-background and cirrus contamination, not through column-water-vapor reduction; per-observation cloud blockage itself is unchanged; the multi-day blockage advantage comes from persistence (\S\ref{sec:layer}) \\
Ground velocity (SAR azimuth integration) & platform-dependent & $\sim 237\times$ slower ($\sim 30$~m/s HALE vs.\ $\sim$7.1~km/s LEO ground-track speed) & Long coherent-integration synthetic apertures; persistent ground-deformation InSAR; weak-signal coherent processing \\
Round-trip latency ($2R/c$) & $\propto 2R/c$ & $25\times$ lower than LEO at $\sim$500~km; $\sim$1{,}800$\times$ lower than GEO~\cite{hapsAlliance6G2026} & Closed-loop ISAC, low-latency C-V2X, real-time disaster response \\
Revisit cadence over a fixed target & platform-dependent & continuous vs.\ minutes (LEO constellation) to hours (single LEO) & 1--3 orders of magnitude more samples per day for change detection \\
\bottomrule
\end{tabular}
\end{table*}

\begin{figure*}[!t]
\centering
\includegraphics[width=0.88\textwidth,keepaspectratio]{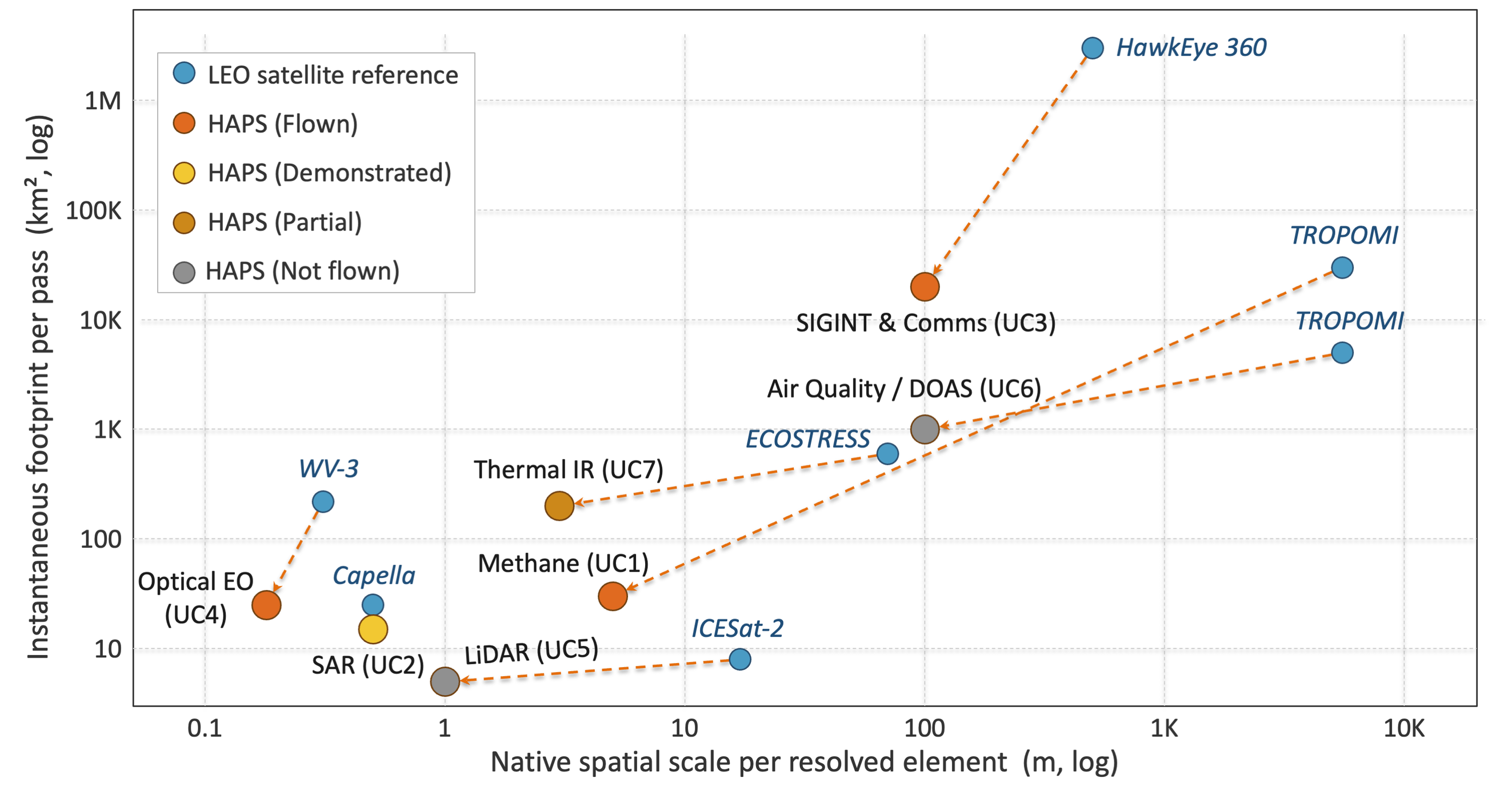}
\caption{Seven use cases on a common spatial-scale-versus-footprint trade space, with arrows showing the LEO-to-HAPS migration. Endpoints tagged \emph{Flight-validated} are operator-attested measurements; all other endpoints are upper bounds from the scaling laws of Table~\ref{tab:gains}, and the two are not equivalent. The x-axis is the smallest resolved ground element (imaging GSD for UC1, UC2, UC4--UC7; RF geolocation CEP for UC3). The UC3 CEP placement is notional: no published analysis or flight measurement supports a specific single-HAPS geolocation accuracy (\S\ref{sec:uc3}). The UC4 \emph{Flown} tag refers to the flown 18~cm endpoint, not the projected 5~cm operating point of \S\ref{sec:uc4}. The \emph{Not flown} marker class covers both the \emph{Conceptual} and \emph{No flight evidence} rows of Table~\ref{tab:lens}. UC8 appears separately in Fig.~\ref{fig:uc8-fso}.}
\label{fig:uc-summary}
\end{figure*}

Table~\ref{tab:flown-uc} summarizes the three flight-validated use cases (UC1--UC3). The remaining five (UC4--UC8) lack a comparable HAPS measurement and are best read as prose. Fig.~\ref{fig:uc-summary} places seven of the eight cases in a common trade space of native spatial scale against \emph{instantaneous} footprint per pass and renders the LEO-to-HAPS migration as an arrow for each use case. Both axes are instantaneous, so the HAPS persistence advantage (the ability to revisit the same instantaneous footprint continuously and to sweep the gimbal across a much larger steerable-access area over hours-to-days on station) is not encoded by the y-axis. It is the orthogonal dimension that motivates the entire perspective and is discussed in the per-case prose and in \S\ref{sec:layer}. The eighth case (UC8, the HAPS-to-HAPS FSO mesh) is an inter-platform link rather than a downward-looking observation and is shown separately in Fig.~\ref{fig:uc8-fso}.

\begin{table*}[!tbp]
\centering
\caption{Summary of the three flight-validated HAPS use cases (UC1--UC3).}
\label{tab:flown-uc}
\footnotesize
\rowcolors{2}{gray!8}{white}
\renewcommand{\arraystretch}{1.06}
\setlength{\tabcolsep}{4pt}
\begin{tabular}{@{}>{\raggedright\arraybackslash}p{2.6cm} >{\raggedright\arraybackslash}p{4.0cm} >{\raggedright\arraybackslash}p{4.0cm} >{\raggedright\arraybackslash}p{2.4cm} >{\raggedright\arraybackslash}p{2.6cm}@{}}
\toprule
\textbf{Use case} & \textbf{LEO reference} & \textbf{HAPS deployment (flight evidence)} & \textbf{Headline gain} & \textbf{Status} \\
\midrule
UC1: methane monitoring & TROPOMI 5.5$\times$7~km daily~\cite{esa-tropomi}; GHGSat-C $\sim$25~m hours--days~\cite{ghgsat-mission} & Sceye + HySpex SWIR-640, Aug 2024~\cite{hyspexSceye2024} & meter-class sampling; continuous observation; intermittent-event capture & Flight-validated \\
UC2: SAR pathfinder & Capella / ICEYE X-band~\cite{capella-stringham,eoportal-iceye} & DLR HAPSAR S-band, ground-tested~\cite{dlr-hapsar}; PHASA-35 + Aloft~\cite{av-phasa-aloft} & $\approx$56~dB $1/R^{4}$ point-target gain (Table~\ref{tab:gains}), of which $>$50~dB is recoverable as link-budget margin after subtracting realistic antenna-pattern and atmospheric-loss terms; minutes-class coherent integration & Hardware-mature, awaiting flight \\
UC3: SIGINT \& comms relay & HawkEye 360~\cite{eoportal-he360}; Iridium / Starlink & PHASA-35 SDR-ISR Dec 2024~\cite{ustPHASA2024}; AALTO 4G D2US Mar 2025~\cite{aaltoFG2025Kenya} & $\approx$28~dB lower receiver path loss; substantially lower propagation latency than LEO; D2US link margin & Flight-validated \\
\bottomrule
\end{tabular}
\end{table*}

\subsection{UC1: Persistent Methane Monitoring}\label{sec:uc1}

\begin{figure}[!t]
\centering
{\small\bfseries Methane Monitoring over a Remote Industrial Facility}\\[2pt]
{\footnotesize\bfseries\makebox[0.5\linewidth]{Satellite (500~km)}\makebox[0.5\linewidth]{HAPS (20~km)}}\\[1pt]
{\setlength{\fboxsep}{1pt}\fcolorbox{black!35}{white}{\includegraphics[width=\dimexpr\linewidth-2.8pt\relax]{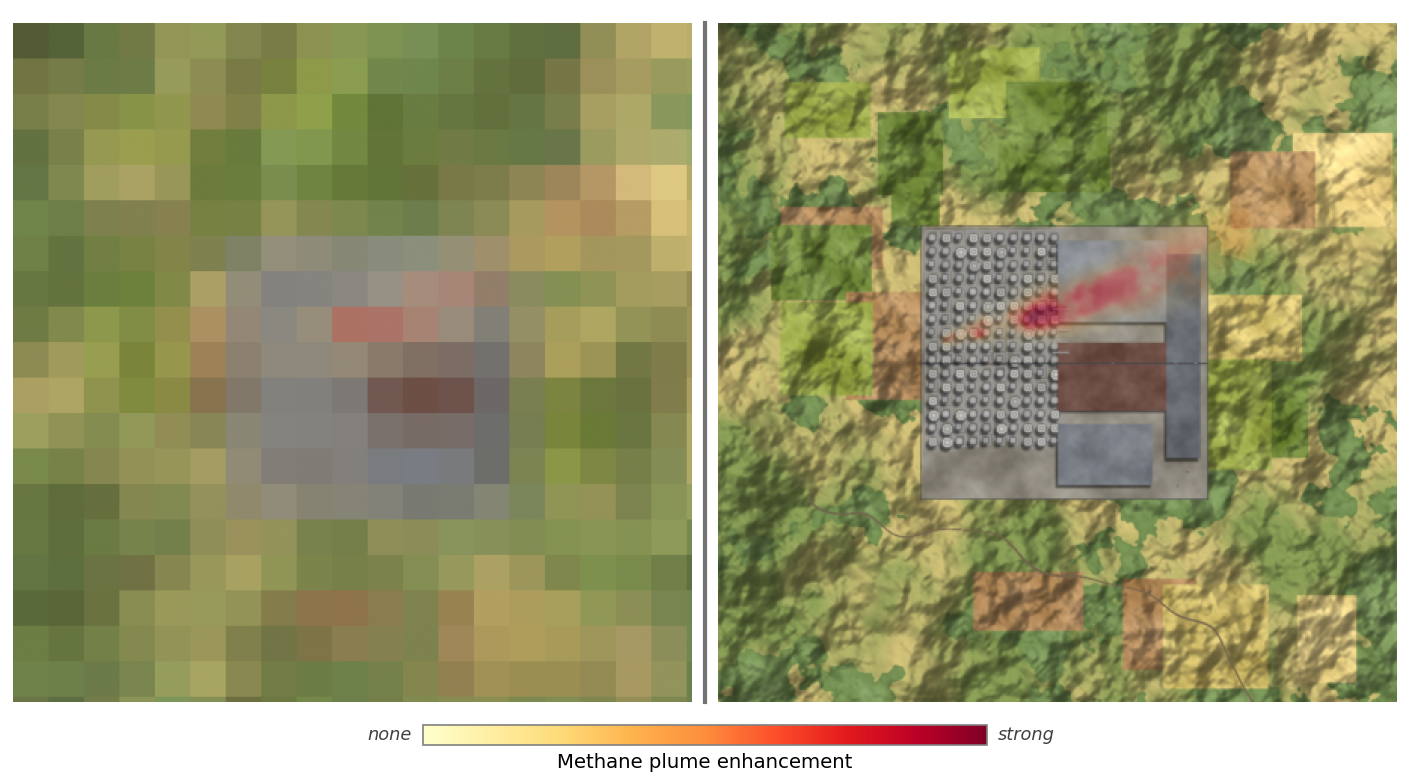}}}
\caption{Methane monitoring over a remote industrial facility, forward-simulated with the same SWIR imager at both ranges (ground sample 5~m versus 125~m). The satellite dilutes the plume into a few pixels; the HAPS traces it to a single tank.}
\label{fig:sim-uc1}
\end{figure}

Methane monitoring is the highest-impact HAPS use case today, filling a clear gap in the satellite portfolio: time-resolved attribution of intermittent emission events to specific facilities. The closest point-source-attribution references are GHGSat-C~\cite{ghgsat-mission,jervis2021ghgsat}, at 25~m GSD with a $\sim$100--200~kg/h detection threshold on hours-to-days revisit, and the now-retired MethaneSAT~\cite{methanesat} (loss of contact 20~June~2025), at $\sim$100~m GSD with a few-hundred~kg/h threshold. TROPOMI~\cite{esa-tropomi,veefkind2012tropomi} at 5.5$\times$7~km daily maps basin-scale flux, detects individual point sources only above a scene-dependent single-overpass limit of roughly 4{,}000--25{,}000~kg/h, and is not a point-source attribution instrument. We cite it for the global baseline rather than the operational comparator. The satellite methane-retrieval literature quantifies this split between area-flux mapping and point-source attribution, together with the plume-inversion methods that any finer-sampling platform inherits~\cite{jacob2022methane,varon2018methane}. Against this set, the Sceye + HySpex SWIR-640 flight in August~2024~\cite{hyspexSceye2024,sceyeDiurnalPR2024} delivers meter-class sampling, continuous days-to-weeks coverage, and an operator-projected detection limit on the order of tens of kg/h~\est\ (Sceye marketing reports figures in the 10--100~kg/h range, though the lower end is unverified and treated as upper-bound by this paper). Carbon Mapper~\cite{carbonmapper} provides a tasking-driven satellite reference, and SeekOps SeekIR~\cite{seekops-seekir-spec} the UAV reference. Against this set, a HAPS resolves plume morphology drifting downwind from individual tank clusters and wellpads at meter-class sampling rather than the $\sim$20~km$^{2}$ TROPOMI-class pixel of a column-average flux, making source attribution unambiguous. The compounding gain is spatial resolution sharp enough to render the plume plus a persistent stare that catches intermittent venting events between LEO overpasses. The downward water-vapor column and per-observation cloud blockage are identical to LEO (\S\ref{sec:why}). UC1's position in Fig.~\ref{fig:uc-summary} ($\sim$5~m, $\sim$30~km$^{2}$ single-swath footprint, \emph{Flight-validated}) shows the consolidated regime. The same-sensor contrast in Fig.~\ref{fig:sim-uc1} isolates this resolution gain, which follows from proximity rather than from a better instrument.

\subsection{UC2: HAPS-SAR Pathfinder}\label{sec:uc2}

\begin{figure}[!t]
\centering
{\small\bfseries SAR Imaging over an Urban District}\\[2pt]
{\footnotesize\bfseries\makebox[0.5\linewidth]{Satellite (500~km)}\makebox[0.5\linewidth]{HAPS (20~km)}}\\[1pt]
{\setlength{\fboxsep}{1pt}\fcolorbox{black!35}{white}{\includegraphics[width=\dimexpr\linewidth-2.8pt\relax]{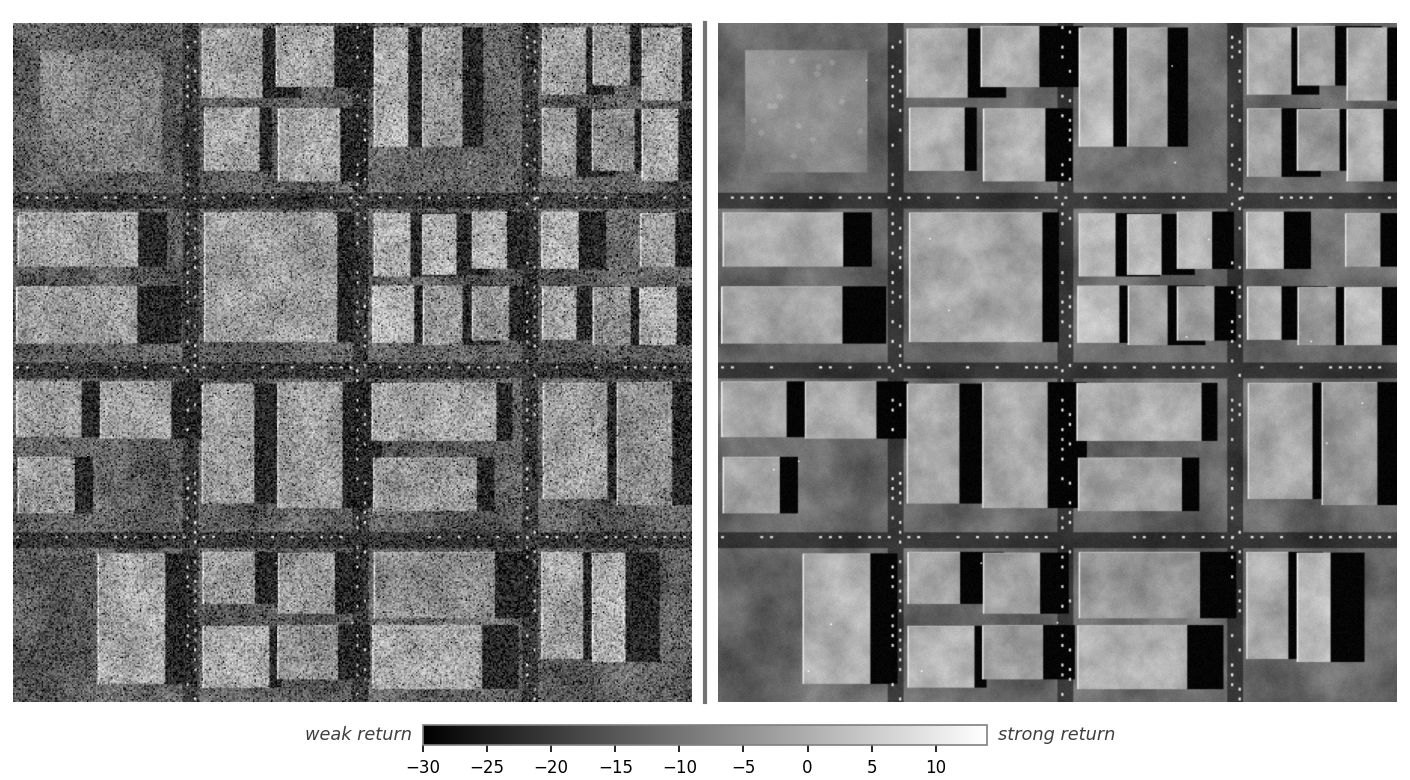}}}
\caption{SAR over an urban district. Resolution is range-independent, so both panels share one pixel scale; the HAPS gain is sensitivity (NE$\sigma^{0}\propto R^{3}$, $\approx$42~dB) plus multi-look integration. Single-look speckle and the raised noise floor bury the streets for the satellite; the HAPS recovers a clean scene. Grayscale: backscatter (dB).}
\label{fig:sim-uc2}
\end{figure}

The HAPS-SAR gap is the most consequential opportunity where the hardware exists but flight-validated stratospheric imagery does not. DLR HAPSAR (S-band)~\cite{dlr-hapsar} is ground-tested on the HAP-alpha airframe, with first stratospheric flights targeted for 2026~\cite{dlr-hapalpha,dlr-hapalpha-tests}, and BAE/Prismatic PHASA-35 has Aloft Sensing identified as its X-band SAR supplier~\cite{av-phasa-aloft}. The close-range SAR scaling of Table~\ref{tab:gains}~\cite{moreira2013sar} ($1/R^{4}$ for point targets and $1/R^{3}$ for distributed targets after speckle averaging) can be spent as shorter antennas at fixed swath, watts-class transmit power against the hundreds-of-watts LEO smallsats~\cite{capella-stringham,eoportal-iceye} or the kW-class TerraSAR-X carry, finer resolution at the same SWaP, or longer coherent-integration windows supported by the slow HAPS ground velocity. Revisit shifts from sub-hourly or daily LEO passes to continuous coverage of a fixed target, repositioning the mission from wide-area screening toward persistent ground-deformation, maritime surveillance, and slow-moving-target detection. The open obstacles (\S\ref{sec:why}) are antenna integration in stratospheric SWaP and station-keeping precision for repeat-pass interferometry. UC2 sits in Fig.~\ref{fig:uc-summary} at the sub-meter end of the spatial-scale axis at $\sim$0.5~m (DLR HAPSAR S-band ground-tested) and a $\sim$15~km$^{2}$ ground-test scene, \emph{Ground-demonstrated}. The migration arrow is essentially vertical because the ground-tested S-band resolution matches Capella's X-band spotlight class within a slightly smaller instantaneous scene. Fig.~\ref{fig:sim-uc2} contrasts the identical radar on the two platforms and shows that, for SAR, the geometric advantage is sensitivity and integration time rather than resolution.

\subsection{UC3: RF/SIGINT and Comms Backhaul}\label{sec:uc3}

\begin{figure}[!t]
\centering
{\small\bfseries Communication Coverage over a Remote Rural Area}\\[2pt]
{\footnotesize\bfseries\makebox[0.5\linewidth]{Satellite (500~km)}\makebox[0.5\linewidth]{HAPS (20~km)}}\\[1pt]
{\setlength{\fboxsep}{1pt}\fcolorbox{black!35}{white}{\includegraphics[width=\dimexpr\linewidth-2.8pt\relax]{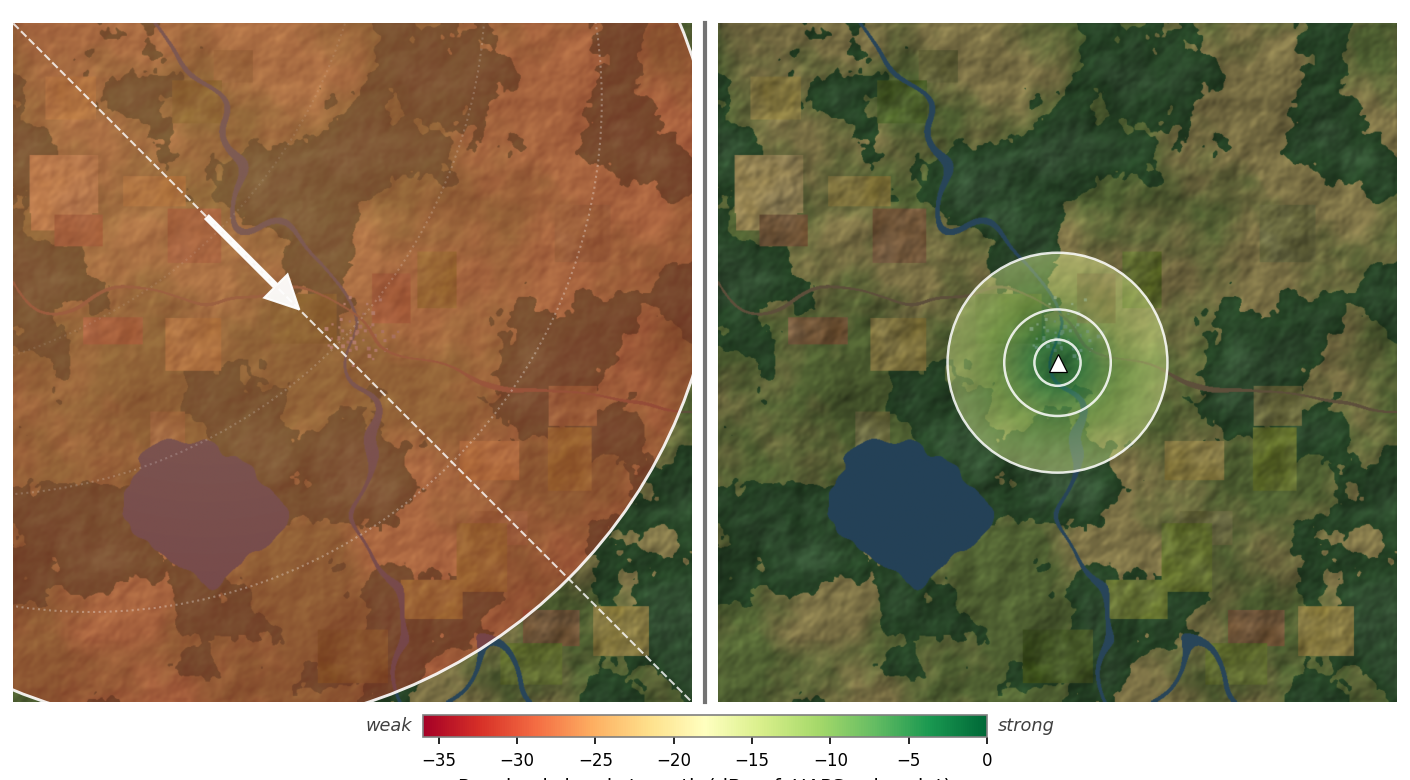}}}
\caption{Coverage of a remote village with the same radio payload. The satellite blankets a wide area but sweeps past at $\sim$7.1~km/s with a $\sim$28~dB weaker link; the HAPS holds a small footprint ($r\approx$113~km at $\varepsilon=10^{\circ}$) continuously. Colour: received signal strength.}
\label{fig:sim-uc3}
\end{figure}

UC3 combines two flight-validated capabilities, PHASA-35 SDR ISR (Dec~2024)~\cite{ustPHASA2024} and the AALTO Zephyr 4G D2US trial from Kenya (March~2025)~\cite{aaltoFG2025Kenya}, against LEO references HawkEye~360~\cite{eoportal-he360,sfl-he360} and the Iridium/Starlink constellations. In the HAPS Alliance defense taxonomy (C3, ISR, EW, ALE, and other effects~\cite{haps-defense-trl-2025}), UC3 spans the C3 and ISR functions. EW and air-launched effects are out of scope for this paper. The $\approx 28$~dB lower receiver path loss at 1~GHz (Table~\ref{tab:gains})~\cite{ustPHASA2024,hapsAlliance6G2026,eoportal-he360} admits lower-noise-figure receivers. Geolocation differs in kind: HawkEye~360 uses TDOA across a 3-satellite formation~\cite{sfl-he360}, while a single HAPS can exploit its kilometer-class loiter circle as a synthetic baseline, accumulating FDOA measurements over the trajectory during its long loiter. The geometry makes accuracy approaching the constellation-TDOA class plausible in principle, but no published analysis or flight measurement yet supports a specific single-HAPS geolocation figure, and we do not assert one. The principal asymmetry is coverage radius: $\sim$50--100~km per HAPS~\cite{hapsAlliance6G2026} against continental Iridium and thousands-of-km Starlink beams. The Alliance 6G target user-experienced data rates are $\sim$500~Mbps downlink and $\sim$50~Mbps uplink, with 50--100~Mbps per beam (peak 200~Mbps) and on the order of $2{,}000$ concurrent voice calls per beam~\cite{hapsAlliance6G2026,hapsAllianceAdvantages2025}. A LEO satellite sweeps a hurricane-affected coastal region in minutes at elevation-angle-dependent quality, while a station-kept HAPS holds a smaller circular footprint at uniform high quality for days. The HAPS gain is in link margin, uniform high-elevation-angle coverage within the served footprint, and persistent fixed-cell service rather than in coverage area. D2US itself is also offered from orbit (Starlink Direct, AST SpaceMobile, Lynk~\cite{hapsAlliance6G2026}). The persistent fixed-cell regime is a particularly strong match for MCX, the 3GPP push-to-talk, data, and video profiles (MCPTT/MCData/MCVideo) that national public-safety networks are migrating to over LTE/5G~\cite{tgpp22179,eenaMCX2024}. MCX demands priority and preemption, group communication, and guaranteed availability, and its highest-stress case, a disaster that damages the terrestrial network, is precisely where a station-kept HAPS holds a uniform high-elevation cell over the affected area while ground sites are down. The spectrum is already adjacent: U.S.\ public safety operates on Band~14 at 700~MHz~\cite{firstnetBand14}, inside the 694--960~MHz range that carries the WRC-23 HIBS identification (Table~\ref{tab:wrc}), so a regenerative HAPS base station could serve public-safety broadband on identified spectrum, subject to the carrier-grade KPIs of \S\ref{sec:cgkpi}. UC3 occupies the upper-right region of Fig.~\ref{fig:uc-summary} ($\sim$100~m geolocation accuracy as a notional placement pending analysis, $\sim$20{,}000~km$^{2}$ cell, \emph{Flight-validated} for the comms and SDR-ISR payloads, not for the geolocation figure). Fig.~\ref{fig:sim-uc3} contrasts the two coverage regimes over a remote rural area and makes the large-but-sweeping versus small-but-fixed trade explicit.

\subsection{UC4: Sub-Decimeter Optical EO}\label{sec:uc4}

\begin{figure*}[!t]
\centering
{\setlength{\fboxsep}{1pt}\fcolorbox{black!35}{white}{\includegraphics[width=0.92\textwidth]{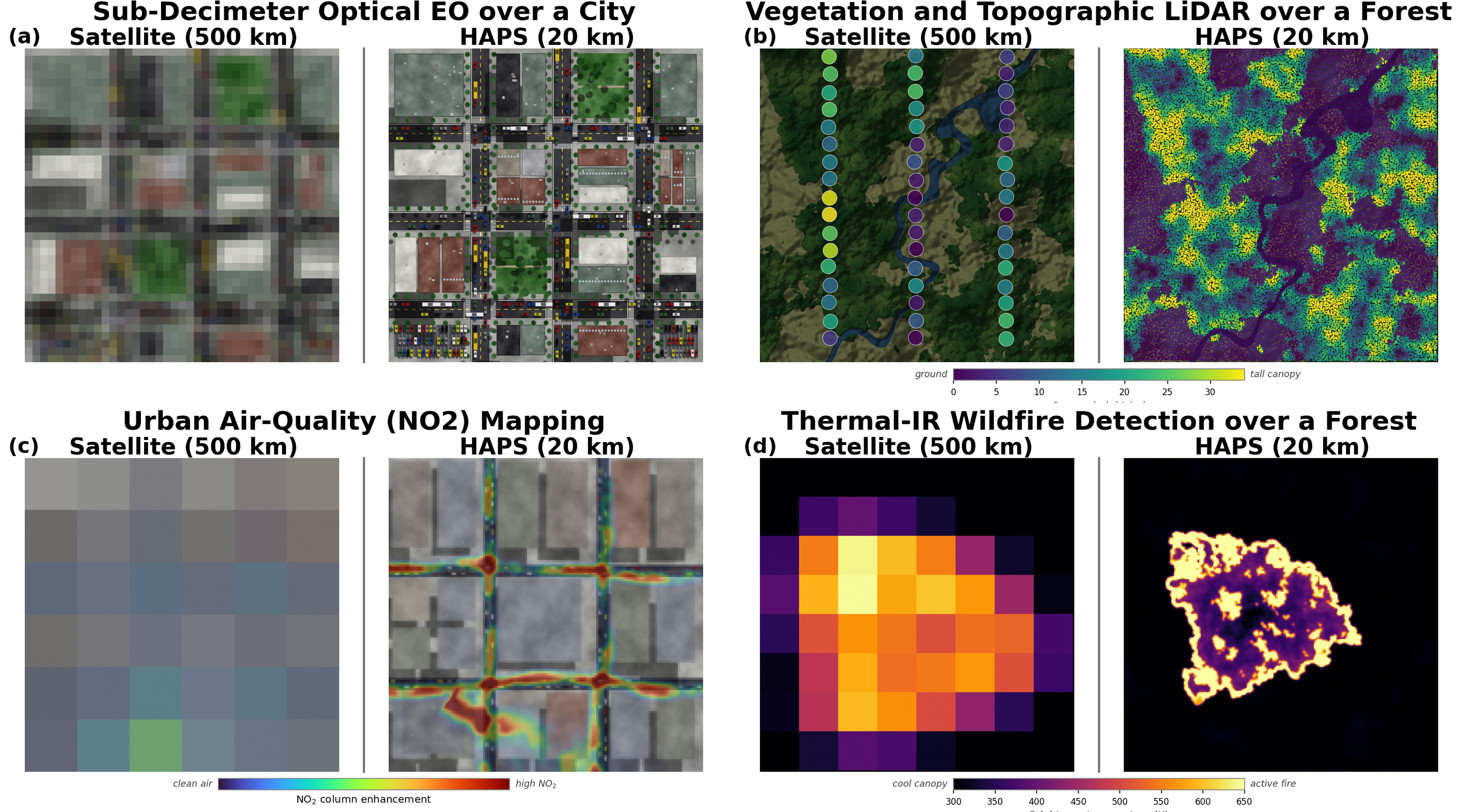}}}
\caption{Forward-simulated comparisons for the four projected use cases, satellite at 500~km (left of each pair) versus HAPS at 20~km (right): (a)~optical EO over a city (0.2~m versus 5~m); (b)~forest LiDAR (photons $\propto 1/R^{2}$, footprint $\propto R$); (c)~urban NO$_2$ (2~m versus 50~m); (d)~thermal-IR wildfire (15~m versus 375~m, Planck-space averaging). Synthetic scenes, not flight data; models in the \hyperref[sec:appendix-sim]{Appendix}.}
\label{fig:sim-proj}
\end{figure*}

The diffraction-limited GSD scaling of Table~\ref{tab:gains} opens a HAPS regime in advanced development: \emph{sub-decimeter optical EO from a small payload}. Zephyr OPAZ has flown 18~cm GSD~\cite{airbusZephyr2021PR}, and DLR MACS-HAP is peer-published at 15~cm GSD with a 90~mm aperture~\cite{dlr-macshap}. A 30~cm aperture in a 5--10~kg HALE envelope can in principle deliver $\sim$5~cm persistent GSD against $\sim$1~m for the same aperture at LEO. The geometric ratio is an upper bound: many flagship LEO sensors are sampling- or jitter-limited (Sentinel-2 at 10~m in its VNIR bands, PlanetScope at $\sim$3~m). The R\&D obstacle is pointing stability at $\sim$10~$\mu$rad, with HALE wing flex of $10^{-3}$--$10^{-2}$~rad in stratospheric turbulence requiring the inboard gimbal to slave to an optical-bench attitude reference. UC4 sits at the finest end of Fig.~\ref{fig:uc-summary} ($\sim$18~cm, $\sim$25~km$^{2}$ single optical frame, \emph{Flight-validated}); the tag refers to the flown 18~cm endpoint (Zephyr OPAZ), while the $\sim$5~cm operating point of this subsection remains a projection. The wider $\sim$1{,}200~km$^{2}$ steerable access area accessible across one station-kept session is a persistence-driven gain not captured by an instantaneous-footprint axis. Fig.~\ref{fig:sim-proj}(a) contrasts the identical telescope on the two platforms and isolates the resolution gain from proximity.

\subsection{UC5: Topographic and Vegetation LiDAR}\label{sec:uc5}

Topographic and vegetation LiDAR is \notyet\ on HAPS but physics-favored: at 20~km it needs $\sim$$625\times$ less laser energy than ICESat-2~\cite{icesat2-mission,markus2017icesat2} for the same ground photon rate. The ATLAS strong-beam transmitted energy is 48--172~$\mu$J per pulse, with $\sim$100~$\mu$J at the energy setting flown for most of the mission~\cite{martino2023atlas}, which scales to $\sim$0.16~$\mu$J at $\sim$1.6~mW average power at the 10~kHz pulse rate. The $\sim$11~m measured on-orbit spot diameter~\cite{martino2023atlas} shrinks to sub-meter, and slow ground speed compresses along-track sampling to millimeters. Coverage shifts from 91-day polar tracks to continuous coverage of a 1{,}000--5{,}000~km$^{2}$ basin, enabling sub-canopy biomass, ice-sheet thickness, and shallow-water bathymetry. The unsolved obstacle is station-keeping precision for repeat-pass deformation work. UC5 sits in Fig.~\ref{fig:uc-summary} at $\sim$1~m, $\sim$5~km$^{2}$ per-pass illuminated area, \emph{No flight evidence}. The 1{,}000--5{,}000~km$^{2}$ basin coverage comes from continuous raster scanning over days on station, which is a persistence-driven gain. Fig.~\ref{fig:sim-proj}(b) contrasts the identical lidar on the two platforms and shows the sampling-density and footprint gain that the photon budget dictates.

\subsection{UC6: Urban Air-Quality Monitoring}\label{sec:uc6}

Urban air-quality monitoring from a HAPS is not yet flown and is presented here as a physics-justified projection. The geostationary trace-gas constellation (TEMPO~\cite{nasa-tempo}, GEMS~\cite{kari-gems}, Sentinel-4 in cal/val) gives hourly continental NO$_{2}$/O$_{3}$/SO$_{2}$/HCHO at 2--10~km, and LEO TROPOMI~\cite{esa-tropomi} samples at 5.5$\times$7~km daily. A HAPS UV-Vis DOAS over a single megacity occupies a different trade-space point: hourly-or-finer cadence at $\sim$0.1~km ground sampling, enabling sub-neighborhood emission attribution. The R\&D risk is detector-driven spectral SNR, not altitude. UC6 sits in Fig.~\ref{fig:uc-summary} at $\sim$100~m, $\sim$1{,}000~km$^{2}$ single push-broom frame, \emph{Conceptual} (the Table~\ref{tab:lens} tag for trace-gas DOAS). Megacity-wide coverage comes from sweeping the gimbal over the observation session. Fig.~\ref{fig:sim-proj}(c) contrasts the identical spectrometer on the two platforms and isolates the resolution gain in the column field.

\subsection{UC7: Thermal-IR Fire Detection}\label{sec:uc7}

Wildfire monitoring exploits the $1/R^{2}$ thermal-IR SNR scaling of Table~\ref{tab:gains} and a persistent stare over a fire-prone region. Against Landsat~8/9 TIRS~\cite{nasa-landsat9} (100~m native, 8-day combined revisit), ECOSTRESS~\cite{jpl-ecostress} (38$\times$69~m native, 1--5-day revisit), and GOES-R ABI~\cite{noaa-goes} (2~km Band-7, minute-class CONUS/mesoscale cadence), against the MODIS/VIIRS active-fire products that define the operational detection algorithms~\cite{giglio2016modisfire}, scaling ECOSTRESS optics to 20~km gives $\sim$2--3~m geometric sampling~\est. HAPS thermal IR complements rather than competes with GOES (sub-meter resolution where ABI flags a 2~km hot spot). Matching ECOSTRESS thermal SNR at the meter scale requires longer integration than per-pixel scaling alone provides. UC7 sits in Fig.~\ref{fig:uc-summary} at $\sim$3~m, $\sim$200~km$^{2}$ single thermal scene, \emph{Partially demonstrated}. Fig.~\ref{fig:sim-proj}(d) contrasts the identical thermal imager on the two platforms and shows how a sub-pixel fire front is recovered only at close range.

\subsection{UC8: HAPS-to-HAPS FSO Mesh}\label{sec:uc8}

\begin{figure}[!t]
\centering
\includegraphics[width=\linewidth,height=4.2cm,keepaspectratio]{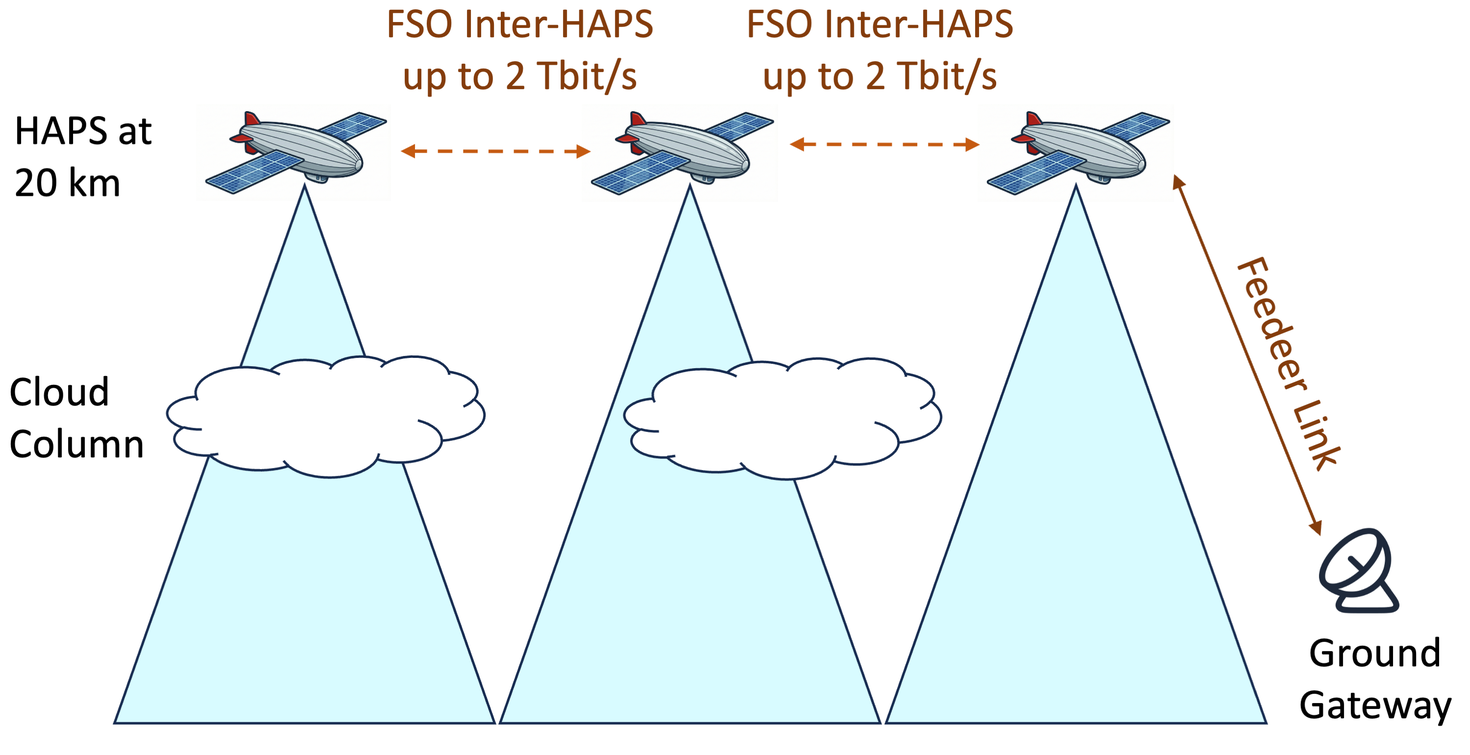}
\caption{HAPS-to-HAPS FSO mesh at 20~km with ground-gateway reach-back. The inter-HAPS links sit above the cloud column and the boundary-layer $C_{n}^{2}$ (projected $>$99\,\% availability); the feeder leg is weather-bounded. Conceptual illustration; no HAPS-to-HAPS link has been flown.}
\label{fig:uc8-fso}
\end{figure}

The HAPS-to-HAPS free-space-optical (FSO) mesh is not yet flown and is presented here as a projection. FSO between two HAPS has a structural advantage: both endpoints sit above the cloud column and above the boundary-layer-concentrated $C_{n}^{2}$. Ground-to-LEO/GEO FSO link availability is dominated by ground-site cloud cover and ranges from $\sim$50\,\% at a single mid-latitude ground site to $\sim$90\,\% across a globally diverse optical-ground-station network~\cite{ata2023hapsFSO,priyadarshani2024earthHapFSO}; HAPS-to-HAPS at 20~km is projected above 99\,\% because both endpoints sit above the cloud column~\est. Performance analyses with atmospheric turbulence, pointing error, and adaptive-optics correction for HAPS-borne FSO links are given in~\cite{ata2023hapsFSO}, and an Earth-to-HAP FSO uplink channel model with spatial diversity is developed in~\cite{priyadarshani2024earthHapFSO}. Terminals span the 10~Gbps Mynaric CONDOR~Mk3 / Tesat ConLCT class~\cite{mynaric-condor,tesat-conlct} and NICT's December~2025 2~Tbit/s WDM demonstration (a 7.4~km ground-to-ground link with HAPS/satellite-mountable terminals)~\cite{nict2tbps2025}. No HAPS-to-HAPS link has been flown. Pointing, tracking, and handover between two drifting endpoints is an open integration question. Because the FSO mesh is an inter-platform link rather than a downward-looking sensing payload, it does not fit the spatial-scale-vs-footprint axes of Fig.~\ref{fig:uc-summary} and is shown separately in Fig.~\ref{fig:uc8-fso}.

\subsection{What the Eight Cases Show}\label{sec:uc-summary}

The eight cases fall into three structural gain types: a \emph{close-range physics gain} (UC2, UC4, UC5) following the $1/R^{2}$, $1/R^{3}$, $1/R^{4}$ and diffraction-limited scaling laws of Table~\ref{tab:gains}; a \emph{cloud-geometry gain} that applies only to links whose optical path stays above the cloud column (UC8 and upward-looking atmospheric LiDAR), \emph{not} to downward-viewing optical, hyperspectral, or SWIR methane sensing; and a \emph{persistence gain} (UC1, UC3, UC6, UC7) that converts continuous presence into 1--3 orders of magnitude more samples per day. UC1--UC3 are the strongest publicly attested cases as of August~2026; UC4--UC8 are physics-derived projections. The same-sensor visualizations in Figs.~\ref{fig:sim-uc1}--\ref{fig:sim-uc3} and Fig.~\ref{fig:sim-proj} make these gain types explicit on a scene: because the instrument is held fixed and only the platform range changes, each satellite-to-HAPS difference is the geometric and radiometric consequence of proximity, footprint, and time on station. Across all eight, the common architectural primitive is persistence at close range, which \S\ref{sec:layer} develops as the unifying middle tier in the multi-tier NTN.

% ----------------------------------------------------------------------------
\section{Persistence Layer in a Multi-Tier Architecture}\label{sec:layer}
% ----------------------------------------------------------------------------

The eight use cases of \S\ref{sec:gains} instantiate the same structural pattern, which we call the \emph{persistence layer}. Satellites continue to provide wide-area screening, and UAVs provide on-demand high-resolution survey at the meter-to-centimeter scale. HAPS adds a persistent middle layer over the high-priority area, exploiting the dimension neither tier serves well. A regional fleet of order 50 platforms can cover populated land at LEO-SAR-equivalent revisit~\cite{kurt2021haps}. However, blanket global daily coverage remains a satellite-tier role. Recent inter-platform and cross-tier studies cover GEO-to-HAPS backhaul~\cite{grieco2024sat-hap-link}, HAPS-altitude tuning for downlink capacity~\cite{jang2025altitude}, and joint user association and beamforming across integrated satellite-HAPS-ground networks~\cite{liu2024satHapsBeamforming}.

\begin{figure}[!t]
\centering
\includegraphics[width=\linewidth]{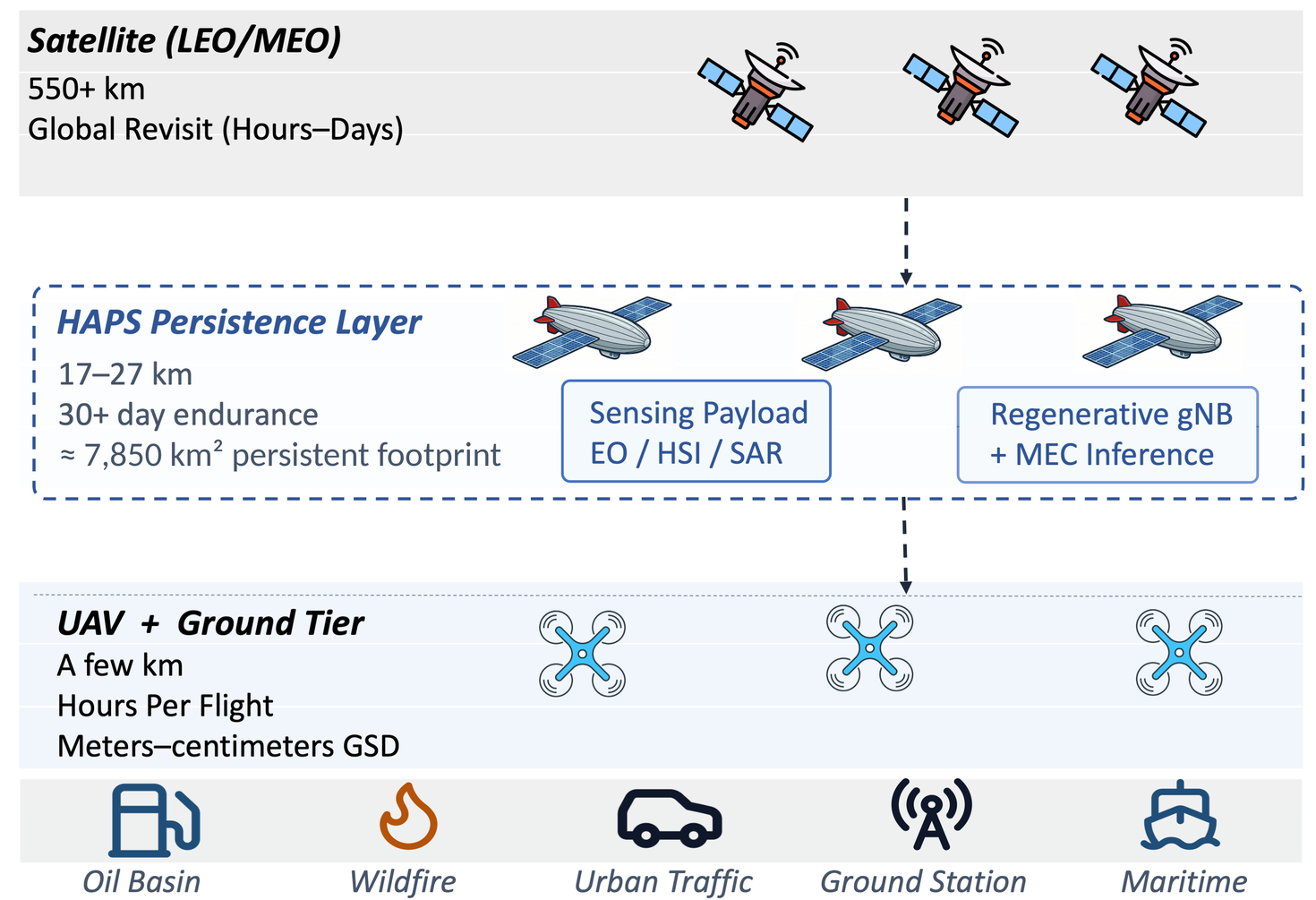}
\caption{HAPS as a regenerative edge node in the multi-tier NTN architecture. The persistent service footprint shown on the HAPS layer is the 50~km-radius case ($\sim$7{,}850~km$^{2}$); the operational range is 50--100~km cell radius giving 7{,}850--31{,}400~km$^{2}$ (\S\ref{sec:layer}; Table~\ref{tab:bg}).}
\label{fig:arch}
\end{figure}

Fig.~\ref{fig:arch} maps the resulting three-tier stack.  LEO/MEO satellites at $550+$~km sit at the top with global revisit on the hours-to-days time scale. The HAPS persistence layer at 17--27~km in the middle holds 30+~day endurance over a $\sim$7{,}850--31{,}400~km$^{2}$ footprint (50--100~km cell radius~\cite{hapsAlliance6G2026}) and hosts both a sensing payload (EO/HSI/SAR) and a regenerative gNodeB-DU plus on-board MEC inference. The UAV-plus-ground tier at the bottom delivers meter-to-centimeter GSD over kilometer-scale areas in flights of hours per sortie.  Inter-tier data flows are color-coded by source tier. The application domains the persistence layer serves (oil-and-gas basins, wildfires, urban traffic, ground stations, maritime) run along the bottom of the figure.

\subsection{Infrastructure Resilience} The LEO failure modes that current EO and connectivity infrastructure depends on (Kessler-cascade collision risk, anti-satellite weapons, severe space weather, and mega-constellation economic fragility) are correlated with each other but uncorrelated with HAPS failure modes. The Outer Space Institute CRASH clock~\cite{osi-crash-clock} has fallen by more than an order of magnitude since 2018, and major LEO operator maneuver counts now reach the hundreds of thousands per year~\cite{spacex-conjunction-report,esaSpaceEnv2024,lewis-kessler-2025}. HAPS at 20~km sit below the debris shell, are atmosphere-shielded, and (when spare airframes are pre-positioned) can be re-launched on the order of days. The airframe production lead time itself is months and is the dominant determinant of fleet-replacement schedule. A multi-tier architecture including a stratospheric layer therefore accommodates a real probability of LEO disruption that a satellite-only architecture implicitly assumes away.

\subsection{Application Case: Public-Safety Mission-Critical Services}
\begin{figure}[!t]
\centering
\includegraphics[width=\linewidth]{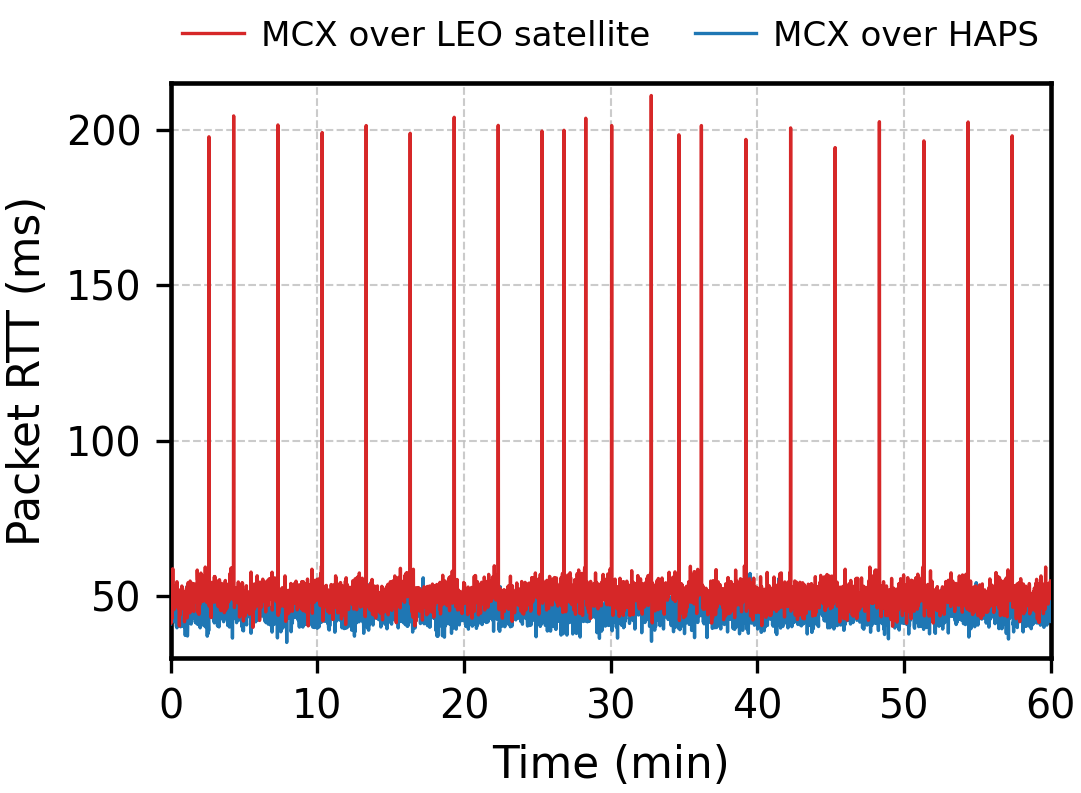}
\vspace{-7pt}
\caption{Packet RTT over a one-hour mission-critical incident for a stationary responder (handover times and propagation exact from a 550~km Walker-1584 geometry; spike magnitudes representative). MCX over a LEO satellite (red) spikes from a $\sim$50~ms baseline to $\sim$200~ms at each satellite handover ($\sim$22 per hour), where a RAN handover and an MCX edge-server migration coincide; MCX over a HAPS (blue) holds one fixed cell and one anchored MC server, staying flat with no spikes.}
\label{fig:mcx}
\end{figure}

The clearest near-term demand for the persistence layer is public-safety MCX, the push-to-talk, data, and video profiles that national responder networks are migrating to from legacy narrowband systems over LTE/5G~\cite{tgpp22179,eenaMCX2024}. MCX is defined less by peak throughput than by a set of guarantees: priority and preemption over commercial traffic, group and broadcast communication, end-to-end latency on the order of a few hundred milliseconds, and availability that must survive the very events, storms, earthquakes, and wildfires, that destroy terrestrial infrastructure. Each maps onto a property developed above. Priority, preemption, and network slicing are enforced by the regenerative gNodeB-DU of \S\ref{sec:ntn-integration}; stable QoS and uniform high-elevation coverage follow from the fixed station-kept cell rather than a sweeping orbital beam; and the disaster-availability requirement is the uncorrelated-failure-mode argument made just above, since a stratospheric cell is untouched by the ground-level damage that takes terrestrial sites offline. Sovereignty reinforces the fit, as responder traffic can be terminated and routed on national soil under national regulation (\S\ref{sec:hae}), a first-order requirement for both the European BroadEU.net initiative and U.S.\ FirstNet~\cite{eenaMCX2024,firstnetBand14}; and the spectrum is already at the band edge, with U.S.\ public safety on Band~14 at 700~MHz~\cite{firstnetBand14} inside the 694--960~MHz WRC-23 HIBS identification (Table~\ref{tab:wrc}). An analogy from edge computing makes the cost concrete. EdgeWarp~\cite{edgewarp} showed that when a stateful application's edge server migrates during a client handover, the packet round-trip time (RTT) spikes sharply before recovering, and that avoiding the migration removes the spike. MCX over a moving satellite pays this penalty twice over: at each satellite handover a stationary responder undergoes a radio (RAN) handover \emph{and}, for an edge-hosted MC server, an MCX session migration to the new ground anchor. We simulate this over a one-hour incident, comparing a 550~km LEO Walker-1584 constellation with a HAPS for a stationary responder. The serving satellite changes every $\sim$2.7~minutes (sticky minimum-handover association at a $25^{\circ}$ service elevation, computed from orbital propagation); the baseline RTT is the geometry-exact propagation plus a representative processing-and-queueing term, and each LEO handover adds a RAN-handover interruption ($\sim$50~ms) plus an MCX migration ($\sim$100~ms). Fig.~\ref{fig:mcx}, generated by modifying the open-source EdgeWarp code to the satellite-versus-HAPS geometry, shows the result: the LEO link spikes from a $\sim$50~ms baseline to $\sim$200~ms at every handover, about twenty-two times an hour, whereas the HAPS holds one fixed cell and one anchored MC server and stays flat at $\sim$45~ms with no spikes. The handover times and propagation are exact; only the spike magnitudes are representative. Edge-hosting the MC server is the configuration that the LEO MCX latency targets themselves motivate, so it is the latency-appropriate baseline rather than a worst case. A satellite operator can avoid the migration spike only by surrendering that edge latency, and even then the placement of the central anchor is constrained in the disaster regime that defines MCX. Anchoring it on the space segment keeps the dependence on a moving node, so the spike returns whenever the anchoring satellite itself hands over. Anchoring it on the ground either exposes it to the same terrestrial damage the responder service must survive, or, if it is moved to an undamaged distant site, imposes a persistent ground round-trip on every packet rather than only at handover. The RAN handover, set purely by orbital motion, remains in all three cases. The HAPS holds the MC server at the stratospheric edge directly over the affected footprint, at once close to the responders, survivable against ground-level damage, and stationary, so it escapes all three; what it still owes, platform reliability across a fleet, is the carrier-grade question of \S\ref{sec:cgkpi} rather than a latency penalty.

What MCX does not yet have from a HAPS is flight evidence at carrier grade. The flown demonstrations of \S\ref{sec:lens} establish that an air interface terminates at altitude, but none has measured the MCX-defining metrics: priority handling and group-call setup under load, throughput continuity as responders hand over between a HAPS cell and terrestrial sites, and the mean time between platform losses and fleet multiplier an always-available responder service requires (\S\ref{sec:cgkpi}). Off-network operation, the direct device-to-device mode mandated for responders when no infrastructure is reachable, is also outside what a relay or base-station HAPS provides. MCX therefore sits exactly where this paper places broadband relay as a whole: flight-validated as a link, unproven as a carrier-grade service, and bounded by the carrier-grade-KPI gap rather than by geometry, which for this application is favorable.

% ----------------------------------------------------------------------------
\section{Lessons from Discontinued Programs}\label{sec:lessons}
% ----------------------------------------------------------------------------

The engineering envelope of the field is most sharply defined by the programs that did not survive. Seven HAPS programs of the past two decades have been discontinued or lost in flight (Table~\ref{tab:lessons}). A brief review of each clarifies what the current generation must avoid~\cite{inflection-haps-economics}.

\begin{table*}[!htbp]
\centering
\caption{Discontinued or lost HAPS programs of the past two decades: the factual record from primary or contemporaneous sources, and the lesson each holds for the post-2020 cohort.}
\label{tab:lessons}
\footnotesize
\rowcolors{2}{gray!8}{white}
\renewcommand{\arraystretch}{1.12}
\setlength{\tabcolsep}{4pt}
\begin{tabular}{@{}>{\raggedright\arraybackslash}p{2.2cm} >{\raggedright\arraybackslash}p{1.1cm} >{\raggedright\arraybackslash}p{2.0cm} >{\raggedright\arraybackslash}p{5.6cm} >{\raggedright\arraybackslash}p{6.0cm}@{}}
\toprule
\textbf{Program} & \textbf{Years} & \textbf{Cause class} & \textbf{What happened} & \textbf{Lesson for current-generation programs} \\
\midrule
Helios (NASA / AeroVironment)~\cite{nasaHeliosNoll2004} & 2003 & Aeroelastic instability & Pitch--phugoid divergence after a fuel-cell pod added wing flexibility and shifted mass distribution; airframe exceeded structural airspeed and broke up over the Pacific. & Any configuration change on a high-aspect-ratio HALE airframe must trigger a full re-derivation of the flexible-airframe aeroelastic and stability envelope before flight; legacy rigid-body envelopes do not transfer. \\
SolarEagle / Vulture~II (Boeing/QinetiQ for DARPA) & 2010--2012 & Energy budget & DARPA cancelled the airframe element before flight after battery and regenerative-fuel-cell specific energy were judged insufficient to sustain a 100+\,m wingspan through a mid-latitude winter night. & Solar HALE feasibility is gated by overnight energy storage; year-round mid-to-high-latitude operation requires battery specific energy, hydrogen-fuel-cell architectures, or polar-winter avoidance, not just photovoltaic-area scaling. \\
PhantomEye (Boeing)~\cite{boeingPhantomEye} & 2012--2014 & Procurement / ISR market & Nine LH$_{2}$-powered flights from Edwards AFB up to $\sim$16.5~km; demonstrator placed in storage in 2014 without transition to a program of record as ISR demand consolidated on MQ-9 derivatives. & Persistent-ISR HAPS programs must demonstrate a procurement case differentiated from existing UAS classes, not just flight; cryogenic-fuel ground-handling is a first-order operational cost that must close in the business case. \\
Aquila (Facebook)~\cite{ntsbAquila2016,facebookAquila} & 2014--2018 & Structural / business & First full-scale flight (28~June~2016) ended in right-outboard-wing structural failure on final approach in gusty conditions, attributed by NTSB to airspeed-envelope exceedance from wind gusts beyond the autopilot, with insufficient approach drag; program closed June~2018 without 24/7 station-keeping ever publicly demonstrated. & Solar-electric HALE airframes need gust-tolerant flight-control envelopes verified across the full atmospheric-disturbance spectrum, not just calm-air pathfinder flight; commercial viability requires demonstrated round-the-clock service. \\
Project Loon (Google~X / Alphabet)~\cite{bloombergLoon2021,zhang2022LoonLessons} & 2013--2021 & Unit economics & Over one million stratospheric flight hours and forty million km of navigated balloon track, with single flights up to 312~days, a stratospheric endurance record that stood until Aerostar's 336-day HBAL684 flight of 2024--25~\cite{aerostarHBAL684}; closed in January~2021 when cost-per-connected-user could not beat terrestrial 4G expansion plus emerging LEO direct-to-cell. & Stratospheric connectivity must compete on cost-per-user against parallel-evolving terrestrial and LEO tiers, not on technical feasibility alone; flight maturity is necessary but not sufficient for market viability. \\
Zephyr~S serial~8 (Airbus)~\cite{zephyrLoss2022} & 2022 & Storm turbulence / propulsion & An engine component failed during unusual high-altitude storm turbulence below its $\sim$20~km stratospheric cruise altitude on 19~August~2022, ending a 64-day, 18-hour record flight over Yuma; the component has since been redesigned. & Lower-stratospheric weather (subtropical jets, mountain waves, mesoscale convective storms) is not uniformly benign at HAPS altitudes; propulsion must be qualified against the 1\,\% worst-case turbulent envelope of MIL-HDBK-310, not the standard atmosphere. \\
HAPSMobile (SoftBank / AeroVironment) & 2018--2023 & Strategic / market & HAPSMobile JV merged into SoftBank Corp.\ in 2023 (absorption-type merger); SoftBank has continued HAPS airframe development with AeroVironment (an improved Sunglider reached the stratosphere in Aug.\ 2024) while the LEO direct-to-cell market matured faster than HAPS unit costs. & A HAPS airframe program competes for capital against rival NTN architectures (LEO direct-to-cell, GEO IoT, terrestrial 5G); strategic alignment with the operator-customer's total-NTN-cost view matters as much as airframe maturity. \\
\bottomrule
\end{tabular}
\end{table*}

\textit{Helios (NASA/AeroVironment, June 2003).} The prototype was lost over the Pacific when, following a configuration change that added a fuel-cell pod and increased wing flexibility, the vehicle entered a divergent pitch--phugoid oscillation and exceeded its airspeed limit~\cite{nasaHeliosNoll2004}. The aeroelastic finding subsequently informed every solar HALE design.

\textit{SolarEagle / Vulture II (Boeing/QinetiQ, 2012).} The airframe element of DARPA's Vulture~II program was terminated prior to first flight after battery and regenerative-fuel-cell specific energy were judged insufficient to sustain the planned 120~m wingspan through a mid-latitude winter solstice.

\textit{PhantomEye (Boeing, 2012--2014).} The liquid-hydrogen-fuelled demonstrator completed nine flights but did not transition to a program of record. Liquid-hydrogen ground-handling logistics and the consolidation of persistent intelligence, surveillance, and reconnaissance demand around MQ-9 derivatives removed the procurement case~\cite{boeingPhantomEye}.

\textit{Aquila (Facebook, 2014--2018).} The program was discontinued on 26~June~2018, two years after its first full-scale flight (28~June~2016) ended in an in-flight structural failure of the right outboard wing on final approach. The National Transportation Safety Board attributed the failure to airspeed exceedance in gusty conditions beyond the autopilot's envelope~\cite{ntsbAquila2016}. Round-the-clock solar station-keeping was never publicly demonstrated, and the parent organization transitioned to a partnership-only role~\cite{facebookAquila}.

\textit{Project Loon (Google~X / Alphabet, 2013--2021).} Loon was closed in January~2021 for commercial rather than technical reasons: cost per connected user could not be reduced to a level competitive with terrestrial 4G expansion or with emerging low-Earth-orbit direct-to-cell services~\cite{bloombergLoon2021,zhang2022LoonLessons}. The program nonetheless produced the largest empirical record of the lower-stratospheric operating environment in the literature, accumulating over one million stratospheric flight hours and over forty million kilometers of navigated track~\cite{hapsAllianceCert2024,zhang2022LoonLessons}, with single flights of up to 312 days, the stratospheric endurance record until Aerostar's 336-day HBAL684 flight of 2024--25~\cite{aerostarHBAL684}. This dataset directly quantifies several of the operational-envelope constraints cataloged in \S\ref{sec:why}: the seasonal stratospheric-wind statistics that bound passive-balloon station-keeping, the day/night battery cycle that bounds solar-electric persistence, and the platform-loss rate that drives the fleet-multiplier required for 24/7 regional service. The HAPS Alliance has identified the Loon record as a principal input to forthcoming HAPS environmental design and certification standards~\cite{hapsAllianceCert2024}, and the post-mortem of Zhang et al.~\cite{zhang2022LoonLessons} translates the commercial failure modes into a 6G research roadmap that this perspective builds on directly.

\textit{Zephyr~S serial~8 (Airbus, August 2022).} The airframe was lost over Yuma, Arizona on 19~August~2022 following a propulsion-system failure in storm turbulence below its stratospheric cruise altitude, ending a 64-day, 18-hour record flight~\cite{zephyrLoss2022}.

\textit{HAPSMobile (SoftBank--AeroVironment, 2023).} The HAPSMobile joint venture that developed the Sunglider platform was merged into SoftBank Corp.\ in 2023 (absorption-type merger), consolidating the HAPS effort under the parent company rather than ending it: SoftBank has continued airframe development with AeroVironment, and an improved Sunglider reached the stratosphere in a U.S.\ Department of Defense field trial in August~2024. The strategic backdrop is a market in which LEO direct-to-cell economics have matured faster than HAPS unit costs, so the lesson here is one of capital competition between NTN architectures rather than of a failed airframe.

These closures share a common root cause: an unclosed stratospheric energy balance. The solar photovoltaic areal power and battery specific energy available during the design horizons of these programs were collectively insufficient to sustain, at the same time, propulsion against seasonal stratospheric winds, payload operation, avionics thermal management, and overnight reserve at useful latitudes and coverage footprints. The resulting weight-minimization pressure produced ultralight high-aspect-ratio airframes, and thin balloon envelopes, with limited structural margin against gusts, turbulence, and aeroelastic excitation, rendering ascent and descent more failure-prone than steady-state cruise. Station-keeping was a coupled second-order constraint: passive balloons relied on probabilistic wind-layer steering, while solar aircraft expended a substantial fraction of their energy budget on position hold, leaving limited margin for payload growth or adverse seasonal winds. Multi-year mean time between failures for motors, batteries, power electronics, and envelope materials remained insufficiently demonstrated, and communications performance was further constrained by long slant ranges, platform drift, and limited aperture size. The enabling technologies (photovoltaic efficiency, battery specific energy, lightweight durable structures, and autonomous gust-tolerant flight control) were collectively about one technology generation short of supporting economically viable persistent stratospheric operation.

The current solar-electric generation (Zephyr, PHASA-35, HAP-alpha, Sceye, Stratobus, Sunglider) shares solar-electric propulsion, dedicated station-keeping, and integrated airframe--battery design. The most recent entrant, the Seattle-based Radical Evenstar, completed a full-scale battery-powered first flight in fall~2025 (announced October~2025) and targets solar-wing stratospheric testing in 2026. Hydrogen-electric architectures have also re-emerged: Stratospheric Platforms Ltd has continued to develop a hydrogen-fuel-cell HAPS targeting high-latitude operation, where the diurnal solar energy budget fails through the polar winter, although its public flight demonstrations to date have been on tropospheric aircraft~\cite{splIslander2024}.

% ----------------------------------------------------------------------------
\section{R\&D Roadmap}\label{sec:forward}
% ----------------------------------------------------------------------------

The function-level lens identifies eight \notyet\ rows plus one \demonstrated\ row (SAR) that awaits flight validation. The four-domain engineering frame from \S\ref{sec:why} sequences the next R\&D moves. The tiers below are grouped by the class of gating obstacle, not ordered strictly by expected time-to-flight-validation (TTFV; per-tier estimates in Table~\ref{tab:roadmap}), by mission priority, or by HAPS-Alliance Technology Readiness Level alone: Tiers 1--3 are payload-engineering gaps with identified programs or hardware paths, Tiers 4--6 have a clear technical path but no current sponsor, Tier 7 collects low-priority functions where the satellite or UAV tier is the natural fit, and Tier 8 (ISAC) is a forward research direction whose payload has not yet been integrated. Within a group TTFV varies, so tier number does not imply temporal order.

\subsection{Technology-Readiness Framework} A useful separation when reading the roadmap below is: (1)~\emph{physics-feasible}: the scaling law admits a useful HAPS implementation; (2)~\emph{subsystem-feasible}: a complete payload exists and has been benched against a representative load; (3)~\emph{flight-feasible}: the payload has been integrated to a stratospheric platform and returned operationally relevant data at $\geq$~18~km; (4)~\emph{operationally scalable}: the system supports continuous service at carrier-grade KPIs across a fleet, through seasonal weather and platform-loss windows. Most of the cross-over functions of \S\ref{sec:lens} sit at level~(3). None has yet credibly reached level~(4). On the HAPS Alliance nine-level HAPS-specific TRL scale~\cite{haps-defense-trl-2025}, the four-domain frame above corresponds approximately to HAPS-TRL 1--4 (physics-feasible), 5--6 (subsystem-feasible, including thermal-vacuum testing), 7 (flight-feasible: actual stratospheric demonstration), and 8--9 (operationally scalable). The Alliance identifies the TRL~6 to TRL~7 transition as the dominant cost and schedule discontinuity in HAPS development, since difficult-to-replicate stratospheric conditions can only be characterized in flight, and tracks platform and payload TRLs separately because each must be flown in its operational configuration to reach TRL~8 or 9.

\begin{table}[!htbp]
\centering
\caption{Roadmap tiers and gating obstacles, grouped by gating-obstacle class, with expected time-to-flight-validation (TTFV) per tier.}
\label{tab:roadmap}
\footnotesize
\rowcolors{2}{gray!8}{white}
\renewcommand{\arraystretch}{1.06}
\setlength{\tabcolsep}{3pt}
\begin{tabular}{>{\raggedright\arraybackslash}p{0.45cm} >{\raggedright\arraybackslash}p{1.65cm} >{\raggedright\arraybackslash}p{2.35cm} >{\raggedright\arraybackslash}p{1.0cm} >{\raggedright\arraybackslash}p{1.6cm}}
\toprule
\textbf{T} & \textbf{Function} & \textbf{Gating obstacle} & \textbf{TTFV} & \textbf{Sponsor pull} \\
\midrule
1 & HAPS-SAR flight validation & Antenna integration in HALE SWaP; flight-attested NE$\sigma^{0}$ & 1--2~yr & DLR; BAE Aloft \\
2 & L-band soil moisture & 6~m-class antenna on airship SWaP & 3--5~yr & Drought-mgmt agencies \\
3 & Topographic LiDAR & Pointing stability; laser power & 2--3~yr & Forestry / cryo science \\
4 & Lightning mapping & Sponsor pull (no SWaP gap) & 2~yr & Regional storms; wildfire precursor \\
5 & AIS / ADS-B & Sponsor pull (no SWaP gap) & 1--2~yr & ATM; EEZ surveillance \\
6 & PNT broadcast & Regulatory; payload integration & 3--5~yr & GPS-denied regional \\
7 & Low-priority (altimetry, precip.\ radar, sounders, gravimetry) & Geometry- or aperture-bound & n/a & Satellite tier is the natural fit \\
8 & ISAC at altitude & Shared-aperture phased array; joint waveform & 3--5~yr & 6G; defense ISR+comms \\
\bottomrule
\end{tabular}
\end{table}

\subsection{Tier 1: HAPS-SAR Flight Validation}
The hardware is mature. DLR HAPSAR is ground-tested and waits on its host platform HAP-alpha to begin stratospheric flights post-2026~\cite{dlr-hapsar,dlr-hapalpha,dlr-hapalpha-tests}. PHASA-35 has Aloft Sensing X-band SAR identified for future flights~\cite{av-phasa-aloft}. Slant-range geometry at 20~km is favorable ($25\times$ shorter range than from a LEO orbit at $\sim$500~km), and the relevant SWaP envelope is now demonstrated to be feasible in a 5~kg payload. A near-term priority is to publish flight-attested SAR imagery from one of these hardware programs and characterize image quality (NE$\sigma^{0}$, range resolution, azimuth coherence) against LEO commercial SAR. This converts SAR from a hardware demonstration to an operational capability.

\subsection{Tier 2: L-band Soil Moisture}
SMAP and SMOS established the global coverage at $\sim$36~km footprint~\cite{nasa-smap,entekhabi2010smap,esa-smos}. A persistent HAPS L-band passive radiometer over an agricultural basin could resolve sub-field-scale variability that LEO does not, enabling drought-management applications that LEO temporal sampling misses. The obstacle is antenna size. An aperture-synthesized array on a Sceye-class airship is the credible path.

\subsection{Tier 3: Topographic LiDAR}
A smaller SWaP gap than SAR but a larger pointing-stability gap. A HAPS-borne single-photon LiDAR demonstration with a UAV-class instrument (Riegl, YellowScan)~\cite{riegl-vux1uav-spec,yellowscan-mapper-plus-spec} flown on a station-kept airship is plausibly closable on a 2--3-year horizon if pointing-stability and laser-power R\&D is resourced. Atmospheric LiDAR sits further back than topographic LiDAR: no HAPS-borne payload program has been published, and the Honeywell HALAS instrument occasionally read as one is a ground-based stratospheric profiler used for HAPS flight planning~\cite{honeywellHALAS2023}, so the function enters this roadmap only as a conceptual candidate.

\subsection{Tier 4: Lightning Mapping}
A wide-FOV optical or VHF lightning imager is small, low-power, and trivially within the SWaP envelope of any HAPS. The reason it has not crossed over is opportunity cost on payload manifests, not difficulty. Closure is most likely when a sponsor with a regional-storms or wildfire-precursor mission emerges.

\subsection{Tier 5: AIS and ADS-B}
A low-engineering-risk gap. AIS and ADS-B receivers are small and well understood. The technical case is persistent geographic specificity (port surveillance, Exclusive Economic Zone (EEZ) monitoring) that LEO AIS does not deliver. Recent prototype-design work for an ADS-B-equipped HAPS targeting Air Traffic Management (ATM) surveillance over oceanic and remote regions~\cite{rezo2026hapsAdsb}, motivated by the coverage limits of ground-based ADS-B and the cost of space-based ADS-B (annual surveillance cost on the order of EUR~0.14--2.19 per km$^{2}$ per year for the satellite reference, with HAPS positioned as a regional complement), illustrates the practical pull on this tier from the ATM community. GNSS-reflectometry receivers (\S\ref{sec:functions}) share the same small-SWaP, signal-of-opportunity profile and would ride along on the same manifest.

\subsection{Tier 6: PNT Broadcast}
HAPS as alternative-PNT (APNT) broadcast is in concept stage~\cite{kurt2021haps}. The engineering is closer to a comms-relay payload than to a sensing payload, and the regulatory path under 3GPP NTN~\cite{tgpp38811} is plausible. The case is most defensible in GPS-denied regional scenarios. An explicit HAPS-aided GNSS architecture for urban canyons has recently been analyzed in~\cite{zheng2023hapsGNSS}.

\subsection{Tier 7: Low-Priority Functions}
The remaining \notyet\ rows are low priority on a HAPS roadmap. Ocean altimetry is well served by Jason / Sentinel-6 / SWOT~\cite{nasa-jason3,nasa-swot}, and HAPS regional persistence does not provide the global-mean Sea Surface Height (SSH) product that altimetry primarily delivers. Precipitation radar~\cite{nasa-gpm} would benefit from HAPS persistence over a basin, but the Ku/Ka antenna SWaP is at the edge of the HAPS envelope. Gravimetry inherently requires either inter-asset baselines (GRACE-FO~\cite{nasa-gracefo}) or extreme single-platform attitude stability, and is a poor match. Atmospheric sounders are likewise a poor match for HAPS' single-point geometry.

\subsection{Tier 8: ISAC at Altitude}
Returning to the center of the trefoil (Fig.~\ref{fig:trefoil}), Integrated Sensing and Communication (ISAC) is the joint-services result that closes the function-level argument, but no HAPS demonstrator has yet flown a shared-aperture, shared-waveform ISAC payload. Three open research directions define the stratospheric-ISAC agenda.

\emph{Shared-aperture phased-array hardware.} A single steerable aperture must serve both the gNodeB downlink and the sensing radar at a fractional bandwidth and power compatible with the 5--15~kg solar-HALE envelope. Digital phased arrays with one ADC per element are the natural fit in the solar HALE thermal budget because their amortized power dominates analogue-hybrid alternatives at the moderate array sizes a HAPS will carry, and self-interference cancellation between the co-resident transmit and receive chains is the principal architectural problem.

\emph{Joint waveform design and the persistence-enabled integration frontier.} Dual-function OFDM-based radar-comm waveforms with embedded radar pilots, and OTFS variants for the high-Doppler edge of the stratospheric wind envelope, are the leading candidates. The genuinely stratospheric question, however, is the coherent-integration versus scheduling frontier that the long on-station time creates, and it has no terrestrial or LEO analogue. At a $\sim$30~m/s platform velocity, a 0.5~m azimuth resolution at X-band requires a synthetic aperture of roughly $T=\lambda R/(2v\rho_{az})\approx26$~s at a 25~km slant range~\est, an order of magnitude longer than the 2--3~s aperture of a LEO SAR at the same resolution, while the clutter Doppler band shrinks to $\pm2v/\lambda\approx\pm2$~kHz against hundreds of kHz from orbit. A shared-aperture HAPS ISAC node must therefore schedule communication service through sensing apertures measured in tens of seconds rather than milliseconds. Either the radar aperture is fragmented into comms-interleaved bursts and recombined coherently, which transfers the burden to oscillator stability and platform-motion compensation across the gaps, or the aperture is held continuous and the cell accepts periodic capacity outages. Mapping this radar-resolution versus comms-capacity frontier as a function of fragmentation ratio, oscillator Allan deviation, and HALE wing-flex spectra is, in our assessment, the first stratosphere-specific ISAC problem that merits dedicated treatment. Neither terrestrial 6G nor LEO ISAC encounters second-class apertures at kHz-class clutter bandwidths.

\emph{Joint sensor fusion at the MEC layer.} On-board inference that combines the ISAC radar return with co-resident EO, HSI, or SAR payloads can produce range-resolved methane plumes, fire-line motion vectors, or ADS-B/ELINT refinements that no single modality delivers in isolation. A near-term demonstrator is a HAPS-borne SDR plus a small steerable antenna in a $\leq 5$~kg payload exercising a low-power dual-function NR-plus-radar waveform during a single stratospheric sortie, exploiting the PHASA-35 SDR backbone already qualified at altitude~\cite{ustPHASA2024}. The HAPS Alliance 6G targets that frame the KPI envelope for such a demonstrator are sub-10~cm positioning accuracy at sub-10~ms latency, sub-5~ms control loops for low-altitude drone management, and sub-0.5~ms near-field sensing~\cite{hapsAlliance6G2026}.

\subsection{A HAPS-Native Frontier: In-Situ and Acoustic Sensing}
The function catalog of \S\ref{sec:functions} is, by construction, a catalog of electromagnetic remote sensing: every row observes a distant target through the spectrum, with the gravimetric and magnetic potential-field pair the only exception. A persistent stratospheric platform additionally enables a class of functions that no satellite can perform, because their value is \emph{presence in the medium} rather than a view of it, and two of these already carry flown or peer-reviewed heritage. \emph{In-situ sampling} measures the air the platform flies through: the NOAA POPS aerosol spectrometer returned stratospheric aerosol size distributions directly from a station-kept Stratollite~\cite{noaaPOPSStratollite}, extending the decades-long balloon-sonde record of in-situ composition to persistent operation, and trace-gas, water-vapor, and ionization probes are natural ride-along candidates of the same small-SWaP class. \emph{Acoustic (infrasound) sensing} exploits the low-noise stratospheric waveguide, where the signal-to-noise ratio is roughly an order of magnitude better than on the ground: balloon-borne microbarometers have recorded a buried chemical explosion more clearly than ground arrays~\cite{bowman2021infrasound}, the acoustic signature of a magnitude-4.2 earthquake~\cite{brissaud2021balloonquake}, and the infrasound of the 2022 Hunga eruption from thousands of kilometers away~\cite{podglajen2022hunga}, with direct relevance to explosion, earthquake, and volcanic-hazard monitoring. Neither class has an operational satellite implementation, so both sit outside the cross-over count of \S\ref{sec:lens}; they are the purest expression of the persistence-at-close-range thesis, in which the platform observes by being present in the medium it measures. The R\&D path is integration and endurance rather than new physics, and the natural first demonstrators are ride-along payloads on the platforms already flying (\S\ref{sec:background}).

\subsection{Cross-Cutting Investments}
Several investments cut across the function tiers. \emph{Station-keeping precision} from km-class to m-class is the gating control-system problem for HAPS-InSAR, repeat-pass LiDAR, and altimetry. \emph{Onboard edge processing and optical inter-platform links} are the gating data-architecture problems for persistent Hyperspectral Imaging (HSI) and high-rate Wide-Area Motion Imagery (WAMI), where raw rates of GB/s class exceed practical RF downlink budgets. Standard compressed ISR video at the tens-of-Mbps class is handled comfortably by RF. \emph{Reconfigurable Intelligent Surface (RIS) integration} is a third cross-cutting investment whose stratospheric application has been treated in depth by~\cite{ye2022nonterrestrialRIS}: a passive aerial RIS payload fits the HAPS SWaP envelope without the active-payload power budget of a regenerative gNB, and offers a coverage-extension tool for the dense-urban and obstructed regimes that neither a satellite nor a single terrestrial cell serves well. \emph{Regulatory pathways} are the gating non-technical risk. The FAA and EASA have no HAPS-specific airworthiness category as of August~2026~\cite{hapsAllianceCert2024}, though both regulators offer generic special-class pathways (FAA Part~21.17(b) in the U.S., EASA Special Condition for Light UAS in Europe) that current HAPS programs are pursuing. Case law under these pathways is thin and timelines remain operator-attested rather than regulator-published. The HAPS Alliance has proposed third-party-centric quantitative risk targets to fill the gap: $5\times10^{-9}$ mid-air collisions per exposed manned-aircraft flight hour (drawn from ICAO SASP work) and a $10^{-4}$--$10^{-6}$ per-person-per-year tolerable-risk band on the ground (drawn from the UK HSE ALARP framework)~\cite{hapsAllianceRisk2024}. The operational layer is the emerging Higher Airspace Operations (HAO) framework, built on Cooperative Zones above FL500 ($\sim$15~km) and serviced through ETM in the U.S.\ and ECHO in Europe, in which HAPS operators self-separate under a UTM-like deconfliction regime~\cite{hapsAllianceHAO2025}. Spectrum allocations are summarized in Table~\ref{tab:wrc} (which consolidates ITU-R Resolutions 122, 145, 150, 165--168, 213, 218, and 221). \emph{Business-model and deployment economics.} The Frontex HAPS market study~\cite{frontexHAPSMarket2024} ranks HAPS lowest on regulatory-framework maturity but highest on sustainability among HAPS, UAS, satellites, and traditional aircraft, and behind only UAS on application and coverage. The unit-economics gap relative to satellites is narrowest where HAPS persistence is the operational differentiator (methane monitoring, regional broadband over underserved territory, persistent SIGINT) and widest where the satellite tier already amortizes its cost across a global revisit (wide-area screening, ocean altimetry, gravimetry). A HAPS perspective that does not engage with this economics layer will mis-rank the use cases of \S\ref{sec:gains}. The asymmetry between strong technical and sustainability scores and weak regulatory scores is the single largest non-engineering risk to operational HAPS deployment over the next five years.

\begin{table*}[!htbp]
\centering
\caption{HAPS frequency identifications under the ITU Radio Regulations as of the WRC-23 Final Acts. Entries are identifications within existing Fixed-Service (gateway) or Mobile-Service/IMT (HIBS) allocations, not separate allocations, and are subject to the conditions of the cited Resolution and Article 5 footnotes.}
\label{tab:wrc}
\footnotesize
\rowcolors{2}{gray!8}{white}
\renewcommand{\arraystretch}{1.1}
\setlength{\tabcolsep}{4pt}
\begin{tabular}{@{}>{\raggedright\arraybackslash}p{2.6cm} >{\raggedright\arraybackslash}p{1.4cm} >{\raggedright\arraybackslash}p{4.4cm} >{\raggedright\arraybackslash}p{2.7cm} >{\raggedright\arraybackslash}p{4.6cm}@{}}
\toprule
\textbf{Resolution} & \textbf{WRC adoption / last revision} & \textbf{Frequency band(s)} & \textbf{Service allocation \& link type} & \textbf{Geographic scope and principal conditions} \\
\midrule
Res.~150 (WRC-12) & 2012; no change at WRC-19 & 6440--6520~MHz (HAPS$\rightarrow$ground); 6560--6640~MHz (ground$\rightarrow$HAPS) & FS; HAPS feeder (gateway) links & \emph{Scope:} administrations listed in RR No.~5.457. \emph{Conditions:} antenna-pattern, e.i.r.p., GSO-arc pfd, and coastal-separation limits per the Resolution; explicit agreement of affected administrations under RR No.~5.457 \\
Res.~122 (Rev.~WRC-19) & WRC-97; last rev.\ WRC-19 & 47.2--47.5~GHz; 47.9--48.2~GHz & FS; HAPS broadband links (gateway and fixed terminal) & \emph{Scope:} worldwide (original WRC-97 HAPS bands); both link directions. \emph{Conditions:} RR No.~5.552A; WRC-19 updated the limits protecting FS and FSS incumbents \\
Res.~145 (Rev.~WRC-19) & WRC-2000 (footnote); Res.\ last rev.\ WRC-19 & 27.9--28.2~GHz & FS; HAPS broadband links & \emph{Scope:} administrations listed in RR No.~5.537A (China added at WRC-19). \emph{Conditions:} protection limits of the Resolution \\
Res.~165 (WRC-19) & WRC-19 (new) & 21.4--22~GHz (HAPS$\rightarrow$ground) & FS; HAPS broadband links (gateway and fixed terminal) & \emph{Scope:} Region 2 only. \emph{Conditions:} pfd and e.i.r.p.\ limits of the Resolution protecting in-band and adjacent services \\
Res.~166 (WRC-19) & WRC-19 (new) & 24.25--27.5~GHz & FS; HAPS broadband links (25.5--27~GHz gateway only) & \emph{Scope:} Region 2 only. \emph{Conditions:} sub-band link directions: 24.25--25.25~GHz HAPS$\rightarrow$ground; 25.25--27~GHz ground$\rightarrow$HAPS; 27--27.5~GHz HAPS$\rightarrow$ground \\
Res.~167 (WRC-19) & WRC-19 (new) & 31--31.3~GHz & FS; HAPS broadband links (gateway and fixed terminal) & \emph{Scope:} worldwide; both link directions. \emph{Conditions:} limits protecting the passive services in the adjacent band 31.3--31.5~GHz (EESS (passive), radio astronomy) \\
Res.~168 (WRC-19) & WRC-19 (new) & 38--39.5~GHz & FS; HAPS broadband links (gateway and fixed terminal) & \emph{Scope:} worldwide; both link directions. \emph{Conditions:} development of FSS, FS, and MS shall not be unduly constrained \\
Res.~213 (WRC-23) & WRC-23 (new) & 694--960~MHz, or portions thereof & MS (IMT identification); HIBS service (access) links to IMT user equipment & \emph{Scope:} regional identification in Regions 1 and 2; by country footnote in Region 3 (RR Nos.~5.312B and 5.314A). \emph{Conditions:} protection of broadcasting at GE06 pfd thresholds; agreement under RR No.~9.21 with respect to ARNS in the countries of RR Nos.~5.312 and 5.323; pfd limits protecting terrestrial IMT and FS of other administrations \\
Res.~218 (WRC-23) & WRC-23 (new) & 2500--2690~MHz, or portions thereof & MS (IMT identification); HIBS service (access) links to IMT user equipment & \emph{Scope:} Regions 1 and 2: 2500--2690~MHz (2500--2510~MHz receive-only); Region 3: 2500--2655~MHz (2500--2535~MHz receive-only), per RR No.~5.409A. \emph{Conditions:} pfd limits protecting terrestrial IMT, FS, and BSS (2520--2630~MHz); unwanted-emission limits protecting ARNS and radiolocation (2700--2900~MHz), MSS (2483.5--2500~MHz), and radio astronomy (2690--2700~MHz) \\
Res.~221 (Rev.~WRC-23) & WRC-2000; last rev.\ WRC-23 & 1710--1885~MHz (added at WRC-23); 1885--1980~MHz; 2010--2025~MHz; 2110--2170~MHz & MS (IMT identification); HIBS service (access) links to IMT user equipment & \emph{Scope:} 1710--1885~MHz identified globally; the pre-existing 2~GHz sub-bands follow the regional structure of RR No.~5.388A (Regions 1 and 3, with 1885--1980 and 2110--2160~MHz in Region 2). \emph{Conditions:} WRC-23 revised the operational, coexistence, and unwanted-emission limits \\
\bottomrule
\end{tabular}

\smallskip
\footnotesize HIBS = HAPS as IMT base stations at nominal 18--25~km altitude~\cite{hapsAllianceRegPos2024,wrc23hibs}. Identifications confer non-exclusive access within the existing allocation, do not establish priority in the Radio Regulations, and HIBS shall neither cause harmful interference to nor claim protection from existing primary services (e.g., RR No.~5.409A). Band edges, link directions, and conditions follow the WRC-19 and WRC-23 Final Acts and the cited Resolutions~\cite{itu_Res165,ituHapsBackgrounder}.
\end{table*}

% ----------------------------------------------------------------------------
\section{Toward a High Altitude Economy}\label{sec:hae}
% ----------------------------------------------------------------------------

The 2020--2026 stratospheric flight wave has produced more than a set of new technical capabilities. It has produced the seed of an industrial ecosystem. The flight evidence of \S\ref{sec:lens}, the regulatory substrate of \S\ref{sec:forward} and Table~\ref{tab:wrc}, and the capital flows of \S\ref{sec:hae-engines} no longer behave as a loose collection of platform programs. They now have the structure of an economic stratum, with a definable value chain, a set of industry verticals, a capital base, and a forming regulatory regime. We designate this stratum the emerging \emph{High Altitude Economy} (HAE): the civil and defense economic activity enabled by persistent operations at 17--27~km, per the HAPS scope of \S\ref{sec:perspective}. Treating the stratospheric tier as a single economic stratum allows its technical, regulatory, and economic dimensions to be analyzed together rather than program by program. We present this section as a forward-looking framing. Its market estimates are indicative context only, and the technical conclusions of \S\ref{sec:lens} through \S\ref{sec:layer} do not depend on them.

The HAE sits between two tiers that are already studied as economies in their own right. Above it is the established \emph{Space Economy} (SE), which the World Economic Forum and McKinsey value at \$630~billion in 2023 and project to reach \$1.8~trillion by 2035 at a $\sim$9\,\% CAGR~\cite{wefMckinseySpace2024}. Below it is the \emph{Low Altitude Economy} (LAE), recently named in national industrial policy~\cite{stateCouncilLAE2024} and developed in a growing academic and policy literature that sets out its scope, value chain, verticals, and enablers~\cite{wangShusenLAE2025,wangLAENetworks2025}. A complementary recent framing places HAPS as the stratospheric extension of the LAE itself, bridging the LAE's communication, computing, and regulatory layers~\cite{huangBelmekkiAlouini2026}. We take a different reading. Where that work positions HAPS within the LAE's layers, the present paper contributes the per-function flight evidence of \S\ref{sec:lens} and the engineering envelope of \S\ref{sec:why}, and uses them to justify treating the stratospheric tier as a distinct economic stratum rather than a sub-layer of the LAE. On that basis its value chain, verticals, capital structure, and regulatory substrate can be discussed in their own right, alongside the LAE and the SE.

\subsection{The Three Altitude-Stratified Economies}\label{sec:hae-trichotomy}
Table~\ref{tab:hae} compares the three altitude-stratified economies on a common set of dimensions. These cover altitude scope, representative platforms and verticals, midstream infrastructure, regulatory anchor, capital structure, market scale, and maturity. The HAE column is filled entirely from the technical and regulatory analysis of this paper. The three layers do not overlap in altitude, in platform class, or in primary use case. They are complementary rather than competitive. The HAE and the LAE also share structural features that set both apart from the SE. Both operate in nationally regulated airspace rather than under the international ITU space-services and UN COPUOS regime. Both avoid the orbital-mechanics launch barrier, use recoverable and reusable assets, and integrate directly with terrestrial networks. This holds even though the HAE is closer in altitude scale to the SE.

The market-scale figures in Table~\ref{tab:hae} come from different analysts and rest on non-harmonized accounting boundaries, so they indicate orders of magnitude rather than like-for-like market sizes. The SE figure is the most established and is drawn from the World Economic Forum and McKinsey~\cite{wefMckinseySpace2024}. The LAE figure is a national estimate~\cite{caacLAE2025}. For the HAE we deliberately report no figure: published HAPS-market estimates differ by roughly an order of magnitude depending on scope and methodology, none is archival or methodologically transparent enough to anchor a quantitative claim in this venue, and nothing in this section depends on one. We treat all longer-horizon addressable-opportunity projections as unverified upper bounds.

\begin{table*}[!htbp]
\centering
\caption{The three altitude-stratified civil-and-defense economies on common dimensions. The HAE column is built from the analysis of this paper. Market-scale figures come from non-harmonized sources and are indicative only (see text).}
\label{tab:hae}
\footnotesize
\rowcolors{2}{gray!8}{white}
\renewcommand{\arraystretch}{1.12}
\setlength{\tabcolsep}{4pt}
\begin{tabular}{@{}>{\raggedright\arraybackslash}p{2.8cm} >{\raggedright\arraybackslash}p{4.4cm} >{\raggedright\arraybackslash}p{4.4cm} >{\raggedright\arraybackslash}p{4.4cm}@{}}
\toprule
\textbf{Dimension} & \textbf{Low Altitude Economy (LAE)} & \textbf{High Altitude Economy (HAE)} \textit{(this paper)} & \textbf{Space Economy (SE)} \\
\midrule
Altitude scope & $<1$~km; corridors to 3~km & 17--27~km lower stratosphere & $>100$~km (LEO/MEO/GEO and beyond) \\
Representative platforms & Drones, eVTOLs, light aircraft, agri-UAVs & Solar HALE aircraft (Zephyr, PHASA-35, HAP-alpha, Sunglider, Evenstar), hydrogen-fuel-cell aircraft (Stratospheric Platforms), super-pressure balloons (Stratollite, Thunderhead), stratospheric airships (Sceye, Stratobus) & LEO/MEO/GEO satellites, launchers, in-space services \\
Representative verticals & Logistics, eVTOL UAM, agriculture, infrastructure inspection, tourism, public safety & Persistent EO and climate (incl.\ methane), stratospheric connectivity (D2US, IoT, backhaul), defense and sovereign ISR, regional PNT augmentation, on-demand industrial sensing, atmospheric science (\S\ref{sec:hae-verticals}) & Global broadband, EO, PNT, science and exploration, launch services, in-space manufacturing \\
Midstream infrastructure & U-space (EU, Reg.\ 2021/664), UTM (US), 5G-integrated low-altitude airspace management (China); vertiports, charging stations & Higher Airspace Operations framework (HAO; ETM, ECHO)~\cite{hapsAllianceHAO2025}; ITU spectrum identifications (Table~\ref{tab:wrc}); FSO inter-platform mesh (\S\ref{sec:uc8}); ground gateways; on-board MEC & Launch infrastructure, orbital tracking, optical/RF ground-station networks, frequency-coordination registries \\
Regulatory anchor & EASA U-space, FAA Part 107/108, China LAE Standards System 2025 & WRC-23 Final Acts (Table~\ref{tab:wrc}); HAO; HAPS Alliance Reference Architecture and Risk targets~\cite{hapsAllianceRefArch2024,hapsAllianceRisk2024}; FAA Part~21.17(b) / EASA Special Condition for Light UAS~\cite{hapsAllianceCert2024} & UN COPUOS; ITU RR space-service and coordination provisions (Arts.~9, 11, 22); national space agencies and authorities \\
Capital structure & eVTOL OEM raises (Joby, Archer, Vertical, EHang/Geely); municipal infra (Shenzhen \$1.7~B over two years)~\cite{caacLAE2025} & National R\&D programs (DRDO, JAXA, KARI, DLR, ESA); strategic OEMs (Airbus/AALTO, BAE/Prismatic, AeroVironment, SoftBank, Thales); venture-backed (Sceye, Stratospheric Platforms Ltd, Radical, Kea Aerospace); early-stage EU entrants (StratoSyst~\cite{stratosyst2026}) & National space agencies; mega-constellation operators; commercial launch and analytics; defense procurement \\
Indicative market scale (non-harmonized sources; see text) & $\sim$\$220~B (China 2025 est.); $\sim$\$500~B by 2035~\cite{caacLAE2025} & Early-stage; published analyst estimates disagree by about an order of magnitude and lack methodological transparency, so no figure is reported (see text) & $\sim$\$630~B (2023); $\sim$\$1.8~T by 2035 at $\sim$9\,\% CAGR~\cite{wefMckinseySpace2024} \\
Maturity & Operational and scaling rapidly; recently named in national industrial policy (2024) & Early operational; 5 of 19 functions flight-validated (\S\ref{sec:lens}); regulatory framework forming under WRC-23 and HAO & Mature; well-defined investor, regulator, operator, and analytics ecosystems \\
\bottomrule
\end{tabular}
\end{table*}

\subsection{The HAE Value Chain and Industry Verticals}\label{sec:hae-verticals}
The HAE value chain has three layers. At the \emph{upstream} layer it comprises the platform programs cataloged in \S\ref{sec:ntn-integration} and \S\ref{sec:why}, across the three sub-classes: solar HALE aircraft, super-pressure balloons, and stratospheric airships. The named programs and their lead jurisdictions are listed in Table~\ref{tab:jurisdictions} and Table~\ref{tab:hae}. Subsystem suppliers anchor this tier. They provide III-V solar cells, silicon-anode and lithium-sulfur cells, ultra-light CFRP airframes, COTS and aerospace-grade SDR, and the inference accelerators of \S\ref{sec:forward}.

At the \emph{midstream} layer the HAE consists of the airspace, spectrum, and ground infrastructure that the stratospheric tier needs in order to operate as a service rather than as a sequence of test flights: (i) the ITU-R spectrum identifications consolidated in Table~\ref{tab:wrc}, which span the Fixed-Service gateway bands and the Mobile-Service HIBS bands under the WRC-23 Final Acts; (ii) the Higher Airspace Operations framework built on Cooperative Zones above FL500 ($\sim$15~km) and serviced through ETM in the U.S.\ and ECHO in Europe~\cite{hapsAllianceHAO2025}, in which HAPS operators self-separate under a UTM-like regime; (iii) the HAPS Alliance Reference Architecture~\cite{hapsAllianceRefArch2024}, third-party safety-risk targets~\cite{hapsAllianceRisk2024}, and payload-environment guidelines~\cite{hapsAlliancePayloadStrato} that constitute the de-facto industry-standard substrate; (iv) ground-segment infrastructure (gateway terminals, optical ground stations, RF backhaul); and (v) the FSO inter-platform mesh of \S\ref{sec:uc8} that converts isolated stratospheric assets into a continental-scale network.

The \emph{downstream} layer is the set of operational services the value chain delivers. These map onto the function-level cross-over of \S\ref{sec:lens}: the five flight-validated functions plus the two demonstrated and two partial define the initial addressable surface, and the eight not-yet-flown functions define the expansion path. Six operationally identifiable HAE verticals are:

\emph{(i) Persistent Earth observation and climate monitoring.} Methane attribution to specific facilities (the Sceye SWIR hyperspectral flight of August~2024~\cite{hyspexSceye2024} is the operational seed), sub-meter optical EO, fire-line tracking (\S\ref{sec:uc7}), and atmospheric LiDAR. The NASA and USGS procurement signal in this vertical~\cite{sceyeNASAUSGS2024} is the most mature commercial demand-pull anchor in the HAE today.

\emph{(ii) Stratospheric connectivity.} D2US 4G/5G in rural and underserved regions and in disaster zones, IoT relay, broadband backhaul, and the public-safety MCX of \S\ref{sec:layer}. The AALTO 4G demonstration in Kenya~\cite{aaltoFG2025Kenya} is the operational seed.

\emph{(iii) Defense and sovereign ISR.} Persistent surveillance, SIGINT and ELINT, maritime-domain awareness, and theater-scale communications relay. Defense-and-surveillance applications presently account for the largest single share of HAPS market activity and are the primary operational driver for many platform programs (Zephyr, PHASA-35, Sunglider, the DRDO airship). The most recent industry articulation of this vertical is the HAPS Alliance Defense Applications Working Group's \emph{HAPS in the Golden Dome} white paper and round table of July~2026~\cite{hapsAllianceGoldenDome2026}, which positions the stratospheric tier inside layered defensive architectures, including the U.S.\ Golden Dome initiative, through six roles: persistent mid-layer sensing of low-flying threats (cruise missiles, UAS) that terrain-limited ground radar and orbital revisit cover poorly, communications extension and relay into the C2 layer, contested-EW support, left-of-launch force projection, maritime-domain awareness and gap filling, and rapid test and training. We read the mid-layer-sensing role as the defense-sector restatement of the persistence-layer thesis of \S\ref{sec:layer}: a repositionable, station-kept sensor node placed where coverage geometry, not sensor quality, is the binding constraint. The same document also highlights an unsettled legal substrate, observing that no modern multilateral defense treaty explicitly considers HAPS and citing a 2023 U.S.\ INDOPACOM legal position that places high-altitude balloons and HAPS in sovereign airspace rather than in ``near space,'' which reinforces the sovereign-control differentiation of \S\ref{sec:hae-diff}. As an industry position paper its capability claims are advocacy rather than measured KPIs, and we weight them accordingly under the evidence hierarchy of \S\ref{sec:lens}.

\emph{(iv) Regional Positioning, Navigation, and Timing augmentation.} HAPS as an alternative-PNT broadcast layer for GPS-denied or jammed regions, including urban canyons and contested airspace~\cite{zheng2023hapsGNSS}. Conceptually mature, operationally pre-flight.

\emph{(v) On-demand industrial sensing services.} Sensing-as-a-service for oil and gas, mining, agriculture, and infrastructure-asset monitoring. The HAE differentiator is persistence at meter-class resolution that satellite EO does not match and that UAV endurance does not reach.

\emph{(vi) Atmospheric science and meteorological platforms.} Persistent in-situ measurement (POPS aerosol, ozone profiling, occultation), limb sounding, and gravity-wave characterization. The NOAA Stratollite-borne POPS aerosol payload~\cite{noaaPOPSStratollite} is the flown reference.

\subsection{Technology, Policy, and Capital Enablers}\label{sec:hae-engines}
Three convergent enablers explain why the HAE has reached its current threshold. We adopt the technology--policy--capital framework that has emerged in the recent emerging-aerospace-economy literature~\cite{wangShusenLAE2025} as a useful general framework. The evidence below is the HAE's own.

\emph{Technology.} The HAE's technological foundation is the III-V solar, lithium-ion, and edge-compute convergence quantified in \S\ref{sec:perspective} and \S\ref{sec:background}. These advances have crossed the threshold at which autonomous flight control, useful sensing payloads, and on-board ISR processing fit inside the solar-HALE 5--15~kg payload envelope.

\emph{Policy.} The HAE's policy substrate is the WRC-23 Final Acts (Table~\ref{tab:wrc}) for spectrum, the HAO framework~\cite{hapsAllianceHAO2025} for airspace, the FAA Part~21.17(b) special-class and EASA Special Condition for Light UAS pathways for airworthiness~\cite{hapsAllianceCert2024}, and the HAPS Alliance Reference Architecture~\cite{hapsAllianceRefArch2024} and Risk targets~\cite{hapsAllianceRisk2024} as the de-facto industry-coordination layer. The Frontex market study~\cite{frontexHAPSMarket2024} scores HAPS lowest among aviation classes on regulatory-framework maturity. Closing this gap is the most consequential non-engineering action available to the HAE community.

\emph{Capital.} The HAE's capital structure rests on three pools: (a) national R\&D programs (the DRDO stratospheric airship, JAXA HAPS, KARI EAV, DLR HAP-alpha, ESA), (b) strategic OEM investment (Airbus's AALTO subsidiary, BAE Systems / Prismatic, AeroVironment, the former SoftBank HAPSMobile joint venture, Thales Stratobus), and (c) a growing tier of venture-backed dedicated entrants (Sceye, Stratospheric Platforms Ltd, Radical, Kea Aerospace). The coming years will determine whether a dedicated venture-and-infrastructure capital pool joins this strategic-OEM and national-program mix. Such a pool would be needed to take the HAE through its first operational deployments at fleet scale.

\subsection{How the HAE Differs from the LAE and the SE}\label{sec:hae-diff}
The engineering envelope of \S\ref{sec:why} translates into four economic-layer differentiations against the flanking tiers, none of which the LAE or SE delivers natively.

\emph{Capital intensity per persistent-coverage unit} is, by our estimate, one to two orders of magnitude below the SE~\est. The figure follows from order-of-magnitude arithmetic rather than from analyst reports. Persistent regional coverage from LEO is bought with constellations: revisit scales with spacecraft count (HawkEye~360, for example, operates in three-satellite clusters and needs multiple clusters for dense temporal coverage~\cite{sfl-he360}), every spacecraft pays \$1{,}000--5{,}000 per kg to reach orbit, and no asset is recoverable. The HAPS equivalent of a 50~km-radius persistent service cell ($\sim$7{,}850~km$^{2}$; \S\ref{sec:layer}) is a single station-kept, recoverable, refurbishable, re-deployable airframe with zero launch cost. The order of magnitude in asset count, compounded by launch cost and non-recoverability, yields the one-to-two-order figure; a sharper estimate awaits published HAPS unit economics.

\emph{Coverage geometry} is intermediate between LAE (block-to-neighborhood scale, kilometer-class footprints) and SE (regional-to-global, but with revisit gaps for non-GEO and slant-range penalties for GEO), addressing the city-to-province-scale persistent-service gap that neither flanking tier serves natively. This is the geometry behind the methane, persistent SAR, fire-line, and D2US use cases of \S\ref{sec:gains}.

\emph{Sovereign control} is structurally stronger than for LEO because a HAPS fleet can be operated from national soil under national regulation, without the international-coordination overhead, dependence on foreign launch capacity, or shared-orbit constraints of orbital systems. This differentiator is particularly relevant for defense, public-safety, and critical-infrastructure customers, and it is the structural reason national programs (DRDO, JAXA, KARI) are anchor investors in the HAE today.

\emph{Direct integration with terrestrial networks} is structurally easier because the platform is reachable by conventional ground-segment hardware at much lower link margin than orbital nodes ($\sim$28~dB less free-space path loss; Table~\ref{tab:gains}), and the WRC-23 HIBS spectrum identifications (Table~\ref{tab:wrc}) place the air interface squarely inside existing IMT bands. The result is 3GPP-NTN unmodified-handset service without the per-satellite handover cost and Doppler-tracking overhead characteristic of LEO direct-to-cell.

\subsection{Risks and Open Questions}\label{sec:hae-risks}
The technical analysis bounds the HAE framing with five risks that deserve explicit treatment.

\emph{Carrier-grade KPI closure} remains unproven. No flown HAPS comms demonstration has yet measured the carrier-grade KPI four-tuple defined in \S\ref{sec:cgkpi}, and that four-tuple is what decides operator deployment.

\emph{Regulatory maturity is weak.} The Frontex study~\cite{frontexHAPSMarket2024} ranks HAPS lowest among aviation classes on regulatory-framework maturity. The HAO framework is being built~\cite{hapsAllianceHAO2025} but has not been operationalized at scale. The FAA and EASA HAPS-specific airworthiness category does not yet exist, and HAO equivalents in Asia and the Middle East are even earlier.

\emph{Platform-loss rates} drive a fleet multiplier on top of nominal coverage budgets. The discontinued-program record of \S\ref{sec:lessons} shows that unit economics depend on Mean Time Between Platform Losses, and the current generation has not yet had time to demonstrate this at scale.

\emph{Cross-tier substitution.} A pessimistic scenario is that the LAE absorbs the urban-persistent-sensing market from below (eVTOL-borne wide-FOV cameras, drone-swarm SAR-equivalent) and that LEO direct-to-cell economics absorb the rural-broadband market from above before the HAE establishes itself. An early instance of this pressure from below is Kelluu, a Finnish operator of hydrogen-lift autonomous ISR airships that fly \emph{below the clouds} at 50~m--2~km and raised a EUR~15~M Series~A led by the NATO Innovation Fund in April~2026~\cite{kelluu2026}, pursuing the same persistent ``between satellites and drones'' gap from the low-altitude tier rather than the stratosphere. The HAE niche must be earned: the technical analysis of this paper supports it, but it requires demonstrated execution within the next three to five years.

\emph{Cross-cutting investment dependence.} A credible five-year HAE forecast is conditioned on the cross-cutting investments of \S\ref{sec:forward} (edge compute, FSO mesh, ISAC, station-keeping precision) maturing on schedule. None of these are guaranteed.

\subsection{A Testable 2030 Outlook}\label{sec:hae-2030}
So that the HAE designation is a testable hypothesis rather than a label, we state what the stratum should look like by 2030 under two scenarios, both verifiable from public records. The supportive scenario has four markers: (i) at least two operators each sustain regional fleets of order ten platforms through a full seasonal-wind cycle, (ii) methane monitoring and defense ISR anchor recurring revenue, (iii) at least one carrier-grade D2US service operates commercially in an underserved market, and (iv) published market estimates converge rather than continuing to disagree by an order of magnitude (\S\ref{sec:hae-trichotomy}). The pessimistic scenario has four markers as well: (i) fleets remain at one to three platforms flown campaign-style, (ii) the carrier-grade KPI four-tuple of \S\ref{sec:cgkpi} remains unmeasured, (iii) LEO direct-to-cell absorbs the rural-broadband vertical from above and the LAE absorbs urban persistent sensing from below, and (iv) the HAE remains a procurement niche inside defense budgets. Which scenario is realized is observable platform by platform, and the verticals of \S\ref{sec:hae-verticals} name where to look.

\medskip
The cross-over evidence of \S\ref{sec:lens} and the engineering taxonomy of \S\ref{sec:why} together map the addressable surface of the HAE. The persistent-tier architecture of \S\ref{sec:layer} and the roadmap of \S\ref{sec:forward} describe how that surface scales. The value chain, vertical structure, and triple-engine analysis of this section position the HAE as a coherent economic stratum. Whether it matures along the supportive or the pessimistic trajectory of \S\ref{sec:hae-2030} will be determined by the next three to five years of carrier-grade KPI closure, regulatory operationalization, and platform-loss-rate data. The function-level evidence as of WRC-23 and the 2026 platform cohort supports the former.

\FloatBarrier

% ----------------------------------------------------------------------------
\section{Conclusion}\label{sec:conclusion}
% ----------------------------------------------------------------------------

The HAPS literature has long oscillated between two extremes: HAPS as a perennial ``platform of the future,'' or HAPS as a near-term replacement for satellites and UAVs. The discipline this paper imposes, a strict and falsifiable per-function evidence rule, supports neither. Measured against the operational satellite and UAV reference, five of nineteen functions have credibly crossed over, and they rest on four flight programs with at most one peer-reviewed result. The advantage that survives this scrutiny is narrow but real: persistence at close range, not altitude, and not the above-cloud geometry that much of the literature assumes.

Read this way, HAPS is neither the platform of the future nor a replacement for the established tiers. It is a distinct persistent layer, a middle tier in the multi-tier non-terrestrial network (\S\ref{sec:layer}) and the seed of a \emph{High Altitude Economy} (\S\ref{sec:hae}), whose reach is bounded by four engineering domains rather than by altitude. The same evidence rule that deflates the headline count also sharpens what comes next: it separates the functions that are flight-validated from the few that are one integration away, namely SAR, AIS/ADS-B, and lightning mapping, and from those the geometry will never permit. The same-sensor forward models of the \hyperref[sec:appendix-sim]{Appendix} turn that separation into a measurement, so each future flight can be scored against the geometry-and-physics ceiling rather than against a brochure. Beyond the catalog lies a class of functions with no satellite analogue at all, in-situ stratospheric sampling and infrasound sensing, where the platform observes by being present in the medium it measures. This is persistence at close range taken to its limit, and the one capability the stratospheric tier alone provides.

If, over the next several years, regional fleets of order fifty platforms sustain carrier-grade SAR or broadband revisit through a full seasonal-wind cycle and into a polar-night latitude band, the persistent-tier reading holds. If instead the carrier-grade KPIs, Block Error Rate under load and mobility, mean time between platform losses, and meter-class station-keeping for repeat-pass interferometry, stall despite continued progress in solar PV, batteries, and edge compute, then the four-domain engineering envelope of \S\ref{sec:why} is tighter than we judge, and HAPS settles as a niche complement rather than a tier. We make these alternatives concrete and dated in the two 2030 scenarios of \S\ref{sec:hae-2030}, each verifiable from the public flight record. Either way, the next chapter of the stratosphere will be written in flight data, not in architecture papers, and the flight demonstrations of the coming years will decide which.

\FloatBarrier

% Appendix comes BEFORE references per IEEE/PIEEE style

% ============================================================================
\appendix[Forward Models behind the Same-Sensor Use-Case Figures]\label{sec:appendix-sim}
% ============================================================================

Each panel of Figs.~\ref{fig:sim-uc1}--\ref{fig:sim-uc3} and Fig.~\ref{fig:sim-proj} is produced by a scripted three-stage chain: (i) a procedurally generated synthetic ground-truth scene, (ii) a physical forward model of the shared instrument, and (iii) per-platform observation parameters in which only the range-dependent terms differ. The chain is deliberately simple. We state its components and assumptions here for reproducibility: the figures visualize the scaling laws of Table~\ref{tab:gains} on concrete scenes, and they do not constitute independent validation of those laws. Table~\ref{tab:simparams} collects the per-case parameters. The full generation scripts (Python with NumPy and SciPy, rendered with Matplotlib; one self-contained script per case) are available from the authors.

\subsection{Common Chain}
Every scene is built on a $900\times900$ ground-truth grid from geometric primitives (building footprints with assigned heights, storage tanks, roads and lane markings, vehicles, tree crowns, canopy patches) textured with multi-octave OpenSimplex noise. Cast shadows are offset by $h/\tan\theta_{e}$ for object height $h$ and solar elevation $\theta_{e}$ ($38$--$43^{\circ}$ depending on the scene). Platform ranges are fixed at $R_{\mathrm{sat}}=500$~km and $R_{\mathrm{HAPS}}=20$~km, a ratio $K=25$. The instrument model is identical on both platforms; only $R$ enters the observation stage.

\subsection{Passive Imaging (UC1, UC4, UC6)}
The optics are modeled as a Gaussian point-spread function with full width at half maximum equal to the diffraction-limited GSD, which scales as $R$ at fixed aperture, followed by $f\times f$ detector block-averaging, where $f$ is the GSD expressed in ground-truth pixels. The satellite panel adds a $0.7\sigma$ anisotropic along-track smear representing the single fast overpass. Extended-scene SNR is range-independent at fixed IFOV (\S\ref{sec:tutorial}), so no differential noise is injected: the panels differ in sampling, not in noise, which is exactly the claim under illustration. The gas overlays (methane in UC1, NO$_{2}$ in UC6) are steady-state Gaussian plumes with lateral spread growing linearly downwind, exponential downwind decay, multiplicative simplex-noise turbulence of order unity, and a simplex-displacement centerline meander, normalized to a relative column scale and sampled through the same PSF-plus-detector chain.

\subsection{SAR (UC2)}
Backscatter is assigned by scattering class: roads $-21$~dB (specular), roofs $-4$ to $+3$~dB, radar-facing walls $+8$ to $+13$~dB (double bounce), radar shadow $-29$~dB, park vegetation $-8$~dB, vehicles $+6$ to $+11$~dB, and isolated reflectors $+13$ to $+17$~dB. The observed intensity is $I=(\sigma^{0}+\mathrm{NESZ})\,s$ with multiplicative speckle $s\sim\Gamma(L,1/L)$. The noise-equivalent $\sigma^{0}$ follows the distributed-target $R^{3}$ law: $-22$~dB at 500~km and $-64$~dB at 20~km for the identical radar. The satellite is single-look ($L{=}1$, one overpass). In contrast, station-keeping admits $L{=}30$ temporally independent looks for the HAPS. Resolution is set by the synthetic aperture and chirp bandwidth and is range-independent, so both panels share one pixel grid.

\subsection{LiDAR (UC5)}
Returned photons scale as $1/R^{2}$ ($625\times$ fewer per shot at 500~km) and footprint diameter as $R$ (1.5~m versus 37~m). The forward model adopts a generic large-footprint full-waveform configuration (37~m footprints, 45~m along-track spacing; GEDI, the closest operational analogue, uses 25~m footprints at 60~m spacing) rather than the ICESat-2 11~m photon-counting geometry used for the photon-budget scaling in \S\ref{sec:uc5}: the large-footprint full-waveform class is the conservative satellite reference for canopy-height mapping at scene scale, whereas ICESat-2 is used in the text because its published per-pulse energies permit the direct $1/R^{2}$ energy calculation. The satellite panel samples three such ground tracks. Each shot reports the canopy height averaged over its footprint (Gaussian weighting, FWHM equal to the footprint) plus $\mathcal{N}(0,3~\mathrm{m})$ photon-limited height noise. The HAPS panel rasters 1.5~m footprints on a 4.5~m grid with negligible height noise.

\subsection{Thermal IR (UC7)}
The scene temperature field spans cool canopy ($\sim$300~K), a smouldering scar (330--425~K), ember hotspots (480--630~K), and the active front (720--900~K). Radiance is computed with the Planck function at 4~$\mu$m, passed through the PSF-plus-detector chain of the passive-imaging model \emph{in radiance space}, and inverted back to brightness temperature, so sub-pixel hot-target dilution is treated exactly rather than by averaging temperatures.

\subsection{Coverage and Link (UC3)}
The overlay is exact geometry rather than a bound: HAPS coverage radius $r=h/\tan\varepsilon$; LEO visible footprint from spherical-Earth elevation geometry; ground-track speed $v_{g}=\sqrt{\mu/(R_{E}+h)}\,R_{E}/(R_{E}+h)\approx7.1$~km/s; and received signal strength from $20\log_{10}R$ free-space path loss at equal EIRP, a 28~dB offset between the two platforms.

\subsection{Limitations}
The chain omits atmospheric radiative transfer (the downward column is nearly identical from 20~km and 500~km in nadir view, so it largely cancels in a same-sensor ratio; \S\ref{sec:why}), platform pointing jitter, station-keeping error, and detector-level noise on the passive cases. The HAPS panels are therefore the geometry-and-physics best case stated in \S\ref{sec:gains}, and the scenes themselves are synthetic and carry no real-world information.

\begin{table*}[!tbp]
\centering
\caption{Forward-model parameters of the same-sensor use-case figures (satellite at 500~km, HAPS at 20~km, identical instrument).}
\label{tab:simparams}
\footnotesize
\rowcolors{2}{gray!8}{white}
\renewcommand{\arraystretch}{1.2}
\setlength{\tabcolsep}{4pt}
\begin{tabular}{@{}>{\raggedright\arraybackslash}p{1.6cm} >{\raggedright\arraybackslash}p{1.9cm} >{\raggedright\arraybackslash}p{4.2cm} >{\raggedright\arraybackslash}p{3.6cm} >{\raggedright\arraybackslash}p{3.4cm} >{\raggedright\arraybackslash}p{1.9cm}@{}}
\toprule
\textbf{Case} & \textbf{Scene (extent)} & \textbf{Shared instrument model} & \textbf{Satellite observation} & \textbf{HAPS observation} & \textbf{Range law} \\
\midrule
UC1 methane & remote plant (2.4~km) & Gaussian PSF + block sampling; Gaussian-plume overlay & GSD 125~m + overpass smear & GSD 5~m & GSD $\propto R$ \\
UC2 SAR & urban district (0.6~km) & class-based $\sigma^{0}$; $I=(\sigma^{0}{+}\mathrm{NESZ})\,\Gamma(L,1/L)$ & NESZ $-22$~dB, $L{=}1$ & NESZ $-64$~dB, $L{=}30$ & NESZ $\propto R^{3}$ \\
UC3 comms & rural region (700~km) & FSPL at equal EIRP; footprint geometry & footprint to $\varepsilon{=}40^{\circ}$, sweeps at 7.1~km/s, $-28$~dB & $r{=}113$~km at $\varepsilon{=}10^{\circ}$, fixed, 0~dB ref & FSPL $\propto R^{2}$; $r=h/\tan\varepsilon$ \\
UC4 optical & downtown (0.2~km) & Gaussian PSF + block sampling & GSD 5~m + overpass smear & GSD 0.2~m & GSD $\propto R$ \\
UC5 LiDAR & forest (0.8~km) & footprint-averaged canopy height + height noise & 37~m footprints, 3 tracks, $\sigma_{h}{=}3$~m & 1.5~m footprints, 4.5~m raster & photons $\propto 1/R^{2}$; footprint $\propto R$ \\
UC6 NO$_{2}$ & city block (0.3~km) & Gaussian PSF + block sampling; plume + line sources & GSD 50~m + overpass smear & GSD 2~m & GSD $\propto R$ \\
UC7 thermal & forest fire (3~km) & Planck radiance at 4~$\mu$m, averaged in radiance space & GSD 375~m + overpass smear & GSD 15~m & GSD $\propto R$; Planck dilution \\
\bottomrule
\end{tabular}
\end{table*}

\FloatBarrier

% Acknowledgment and generative-AI use disclosure (IEEE policy: disclose AI use in the acknowledgement)
\section*{Acknowledgment}
During the preparation of this manuscript, the authors used Claude Opus~4.8 (Anthropic) to assist with language editing and proofreading and with implementing the Python forward-simulation scripts behind the same-sensor use-case figures (Figs.~\ref{fig:sim-uc1}--\ref{fig:sim-uc3} and Fig.~\ref{fig:sim-proj}), and ChatGPT (OpenAI) to assist with the rendering of selected conceptual illustration figures (Fig.~\ref{fig:bigpicture} and the platform renderings of Fig.~\ref{fig:platforms}). The physical models, parameters, and scaling laws that those scripts implement were specified by the authors and are stated in full in the \hyperref[sec:appendix-sim]{Appendix}, and every simulation output was checked by the authors against the closed-form scaling laws of Table~\ref{tab:gains}. These tools were not used to generate, analyze, or interpret any measured research data, and the evidence assessment, technical analysis, and conclusions are the authors' own. All AI-assisted text, code, and figures were subsequently reviewed and verified by the authors, who take full responsibility for the content of this article.

% References

% ----------------------------------------------------------------------------

\vspace{11pt}

% Author biographies (last)
\begin{IEEEbiography}[{\IfFileExists{MukhtiarAhmad.png}{\includegraphics[width=1in,height=1.25in,clip,keepaspectratio]{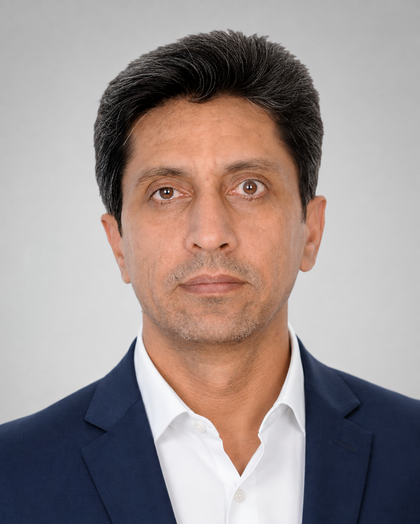}}{\includegraphics[width=1in,height=1.25in,clip,keepaspectratio]{example-image}}}]{Mukhtiar Ahmad}
is a Postdoctoral Researcher with the Computer, Electrical and Mathematical Sciences and Engineering (CEMSE) Division, King Abdullah University of Science and Technology (KAUST), Thuwal, Saudi Arabia. His research interests span Integrated Sensing and Communications (ISAC), High-Altitude Platform Station (HAPS)-enabled remote sensing, and the design of Ultra-Reliable Low-Latency Communications (URLLC) for Space, Ground, Air, and Integrated Non-Terrestrial (SGAIN) networks. He is currently advancing 3GPP-compliant Non-Terrestrial Network (NTN) experimentation on open-source 5G testbeds, alongside distributed digital twin frameworks orchestrated by agentic AI. Prior to KAUST, he served as Chief Technology Officer and Principal System Architect, leading R\&D teams delivering industrial automation, safety-critical systems, and the integration of AI into industrial remote monitoring and automation applications.
\end{IEEEbiography}

\begin{IEEEbiography}[{\IfFileExists{slim_alouini.jpg}{\includegraphics[width=1in,height=1.25in,clip,keepaspectratio]{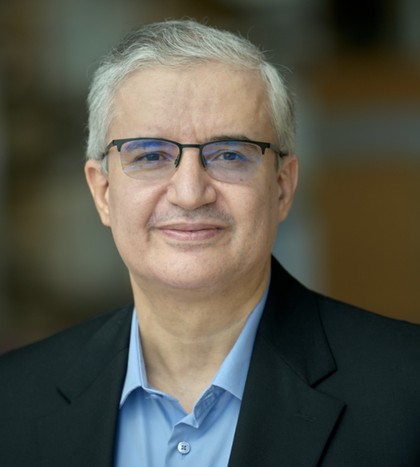}}{\includegraphics[width=1in,height=1.25in,clip,keepaspectratio]{example-image}}}]{Mohamed-Slim Alouini}
(Fellow, IEEE) was born in Tunis, Tunisia. He earned his Ph.D. from the California Institute of Technology (Caltech) in 1998 before serving as a faculty member at the University of Minnesota and later at Texas A\&M University at Qatar. In 2009, he became a founding faculty member at King Abdullah University of Science and Technology (KAUST), where he currently is the Al-Khawarizmi Distinguished Professor of Electrical and Computer Engineering and the holder of the UNESCO Chair on Education to Connect the Unconnected. Dr.\ Alouini is a Fellow of the IEEE, OPTICA, and SPIE, and his research interests span a broad range of topics in wireless and satellite communications. He is currently particularly interested in addressing the technical challenges associated with the uneven distribution, access to, and use of information and communication technologies in rural, low-income, disaster, and/or hard-to-reach areas.
\end{IEEEbiography}

\end{document}